\RequirePackage{fix-cm}
\documentclass[smallextended]{svjour3}       
\smartqed  
\usepackage{graphicx}

\usepackage{color}
\usepackage{amsmath}
\usepackage{amssymb}
\usepackage{color}
\usepackage{multicol}
\begin{document}
\title{A simulation technique for slurries interacting with moving parts and deformable solids with applications
}



\author{Patrick Mutabaruka \and
	Ken Kamrin
}


\institute{P. Mutabaruka \at
              Department of Mechanical Engineering,
              Massachusetts Institute of Technology, 
              Cambridge, MA 02139, USA\\
              \email{pmutabar@mit.edu}           
           \and
           K. Kamrin \at
              Department of Mechanical Engineering,
              Massachusetts Institute of Technology, 
              Cambridge, MA 02139, USA\\
              \email{kkamrin@mit.edu}           
}


\maketitle

\begin{abstract}
A numerical method for particle-laden fluids interacting with a deformable solid domain and mobile rigid parts is proposed and implemented in a full engineering 
system. The fluid domain is modeled with a lattice Boltzmann representation, the particles and rigid parts are modeled with a discrete element 
representation, and the deformable solid domain is modeled using a Lagrangian mesh. The main issue of this work, since separately each of these methods is a mature tool, is to develop coupling and 
model-reduction approaches in order to efficiently simulate coupled  problems of this nature, as occur in various geological and engineering applications. The lattice Boltzmann method incorporates a large-eddy simulation technique using the Smagorinsky 
turbulence model.  The discrete element method  incorporates spherical and polyhedral particles for stiff contact interactions. A neo-Hookean hyperelastic model is used for the deformable solid.
We provide a detailed description of how to couple the three solvers within a unified algorithm. The technique we propose for rubber modeling/coupling exploits a simplification that prevents having to solve a finite-element problem each time step. 
We also develop a technique to reduce the domain size of the full system by replacing certain zones with quasi-analytic 
solutions, which act as effective boundary conditions for the lattice Boltzmann method. The major ingredients of the routine are are separately validated. To demonstrate the coupled method in full, we simulate slurry flows in two kinds of piston-valve geometries. The dynamics of the valve and slurry are studied and reported over a large range of input parameters.

\end{abstract}

\section{Introduction}
\label{sec:intro}
For systems that involve grains, fluids, and deformable solids, a key challenge is to determine reasonable methodologies 
to couple very distinct numerical techniques. On their own, systems of dry grains are commonly simulated using the 
discrete element method (DEM), wherein each grain's position is evolved by Newton's laws applied by way of contact interactions 
with other grains.  For fluids, a variety of approaches exist including finite volume methods, finite difference methods, and 
the Lattice Boltzman Method (LBM), which are based on updating fluid data on an Eulerian background set.  
While the former two methods directly simulate Navier-Stokes, the latter utilizes a lattice discretization of the Boltzmann equation, 
which approaches Navier-Stokes under the proper refinement. As for solids, in the large deformation limit, finite-element methods 
are commonly used, typically based on a moving Lagrangian node set.  Systems that mix particles, fluids, and deformable solids 
require development of methods that allow proper momentum exchange between these disparate representations, which can be 
computationally quite complex if not reduced.  However, because the particles can enter a dense-packed state, we do not wish to 
reduce the particle-fluid mixture to a simplified dilute-suspension continuum model.

\begin{figure}[!ht]
\centerline{
\includegraphics[width=1.\textwidth]{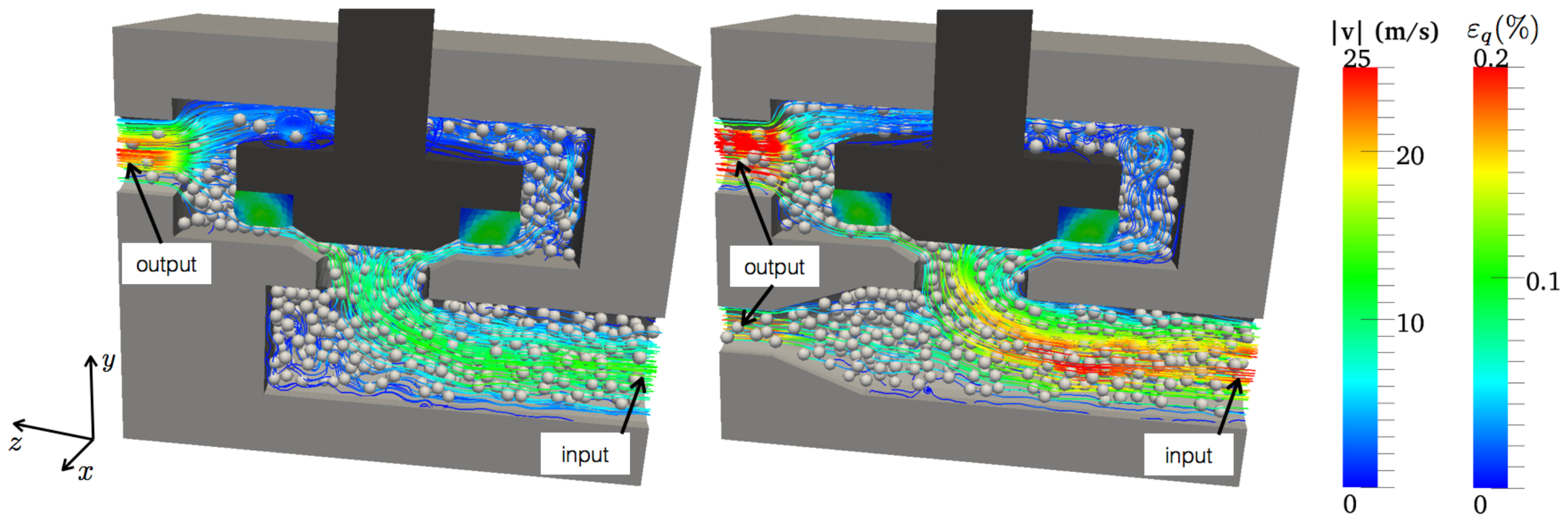}
}
\caption{Pressure valve (left) and safety valve (right) setups with particles (silver), fluid stream lines (colored according to fluid speed 
$|v|\ (m/s)$), deformable solid (colored according to equivalent shear strain $\varepsilon_q$), valve (dark gray), and frame (light gray). A spring (not shown) applies a downward force on the valve.}
\label{fig:opned_closed}
\end{figure}

The purpose of this paper is three-fold:
\begin{enumerate}
\item We introduce a reduced-order method that permits continuum deformable solid models, represented with finite-elements, 
to interact with both grains and fluid in a dynamic environment. The fluid-particle implementation is based on a joint LBM-DEM 
method similar to those used in \cite{Bouzidi2001,Guo2000,Feng2007,Han2007,Iglberger2008,hou1994}. LBM is well-suited to this problem because of its ease dealing with many moving boundaries. The solid interaction method 
introduced uses data interpolation to map deformed solid configurations from separate, individual solid deformation tests to 
the in-situ solid as it interacts with particles and fluid.
\\
\item Because of the inherent complexity in multi-material modeling, the ability to remove the need to simulate large zones of 
the computational domain can be advantageous, as long as the macro-physics in those zones can be properly represented otherwise. 
Herein, we introduce an LBM sub-routine that allows us to remove a large zone of the computational fluid domain and replace 
it with a global analytical form, which handshakes back to the LBM simulation domain appropriately.
\\
\item  As a key example of where these methods may come to use, we demonstrate their usage in two different piston-valve geometries.  
In both piston-valve geometries, a large piston pushes a particle-laden fluid through a passive valve.  The valve is spring-loaded, and opens when the slurry 
pressure beneath is large enough.  The deformable solid aspect comes into play because the valve has a rubber component along its bottom, which is intended 
to make a seal with the valve seat.  We conduct a systematic parameter study of valve behavior under variations in particle size, input packing 
fraction, and polydispersity as well as variations in fluid viscosity and piston speed.  We consider two types of valve setups:  
(1) A `pressure valve', in which the valve separates a  zone of pressurized slurry above it from a zone of low pressure below it.  Slurry pushed through the valve is hence pressurized as it passes through. 
(2) A `safety valve', whose goal is to ensure the pressure in a flowing conduit does not exceed a critical limit.  Here, the valve is placed 
adjacent to a flowing conduit and remains closed unless the pressure is high enough. Figure~\ref{fig:opned_closed} shows mid-simulation snapshots of both valve setups,
showing particles, fluid stream lines, rubber deformation, and the geometry of the valve and frame.  Note that we exploit symmetry about the $zy$-plane and simulate only half the domain.
\end{enumerate}

In testing our method, we provide numerical validations of the new techniques introduced.  We also perform validations of the LBM 
approach in the simulated valve geometry.  In analyzing valve simulation results, we provide physical commentary where possible to justify 
various observations.

\section{Numerical method}
\label{sec:num_method}

The discrete-element method (DEM) is already a mature tool that is applied in conjunction with experiments 
both for a better understanding of the micromechanics of granular materials and as a means of $virtual$ 
experimentation when laboratory experiments are unavailable. In a similar vein, the inclusion of a fluid 
at the subgranular scale in DEM simulations provides a powerful tool in the broad field of fluid-grain 
mixtures. Obviously, the available computational power and research time restrict considerably the 
number of  particles or the size of a physical system. 

In the case of dry granular materials, statistically representative samples are obtained and simulated 
with $O(10^4)$ of particles in 2D \cite{Voivret2007}. Despite enhanced kinematic constraints, 2D simulations often 
lead to novel physical insights and realistic behaviors that can be easily generalized to 3D configurations. 
However, with fluid in the pore space, 2D simulations are much less reliable in the dense regime since 
the pore space is discontinuous with zero permeability. This two-dimensional flaw can be partially 
repaired by adding artificially a permeable layer on the particles. But only 3D simulations may account 
for a realistic behavior of particle-fluid mixtures with their natural permeability. Moreover, complex 
geometries/boundaries relating to realistic engineering problems cannot be fully captured in 2D simulations 
or symmetric 2D extensions (e.g. axis-symmetry); only 3D approaches can handle such problems in full generality. 

We developed a 3D fluid dynamics algorithm based on the lattice Boltzmann method (LBM). This algorithm 
was interfaced with a DEM algorithm with a standard linear spring-dashpot-friction model of contact between 
particles. The combined LBM-DEM method for particle-laden fluid is then further coupled to a deformable solid 
domain using finite elements to model a rubber-like behavior.  The rubber coupling is intentionally simplified.

Within actual computer power, it is still a significant challenge to model the entirety of most engineering systems and problems. 
Certain sub-scale details and complex interactions are unnecessary to capture the macroscale system response for a given loading. 
We utilize symmetric boundaries (where possible) and a variety of techniques to shrink the system size and average-up
sub-scale phenomena. Specifically in this work: to handle sub-scale behavior in the fluid we use a Large-Eddy-Simulation (LES) technique 
(see Sect.~\ref{subsec:lbm}), to mimic a large fluid domain outside the focus region we have created a technique we denote 
\textit{Zoom-in with Effective Boundaries}  (ZIEB) (see Sect.~\ref{subsec:zoominlinktech}), and to reduce simulation time 
we developed a weak coupling to the rubber domain based on a \textit{Neo-Hookean} model developed in Abaqus. 
The last part is computed separately and only the result is imported into LBM-DEM simulation; the coupling and description 
of this part is expounded in Sec.~\ref{subsec:lbmdemrubbercoupling}.

\subsection{Discrete-element method}
\label{subsec:dem}
The DEM is based on the assumption of elastic solids with damping  and frictional contact behavior
\cite{Cundall1971,Allen1987,Jean1999,Moreau1993,Luding1996,Luding2008,Radjai2009,Brilliantov2002}. 
Newton's equations of motion are integrated for all degrees of freedom with simple force laws expressing the normal and friction forces 
as explicit functions of the elastic deflection defined from the relative positions and displacements  at contact points. We treat all quasi-rigid solids in the domain using this DEM description, including grains, the valve, and solid system boundaries.  Correspondingly, all solid-on-solid contact forces (e.g. grain on grain, grains on valve, grain on solid wall) are obtained using DEM contact laws. The valve and system walls are discretized as a kinematically constrained connected mesh of polyhedral solid `particles'.


To simplify contact interactions, we assume linear elastic normal and tangential contact forces characterized by a normal stiffness $k_n$ and tangential stiffness $k_t$. This is applied to all contact interactions, e.g. between spheres, polyhedra, or sphere-polyhedra, though the stiffnesses can vary depending on the two objects in contact. In additional to the elastic part, a dissipative part of the contact force is necessary \cite{Moreau1993,Luding2008,Hart_1988,Walton_1993}. In our model, we use a linear visco-elastic law for normal damping
and a linear visco-elasto-plastic law for tangential damping and friction forces where the plastic part uses a Coulomb law. The visco-elastic law is modeled by a parallel spring-dashpot model.
The contact normal force is defined as:
\begin{equation}
  \mbox{f\ }^n = \left \{ \begin{array}{ll} 
  k_n   - \gamma_n\ \vec{v}^r_n\cdot\vec{n} & \mbox{        if}\ \delta_n \leq 0 \\
 0 & \mbox{        otherwise}
  \end{array}\right.
  \label{eq:normal_force}
\end{equation}
where $\vec n$ is the contact normal vector and $\vec{v}^r_n$ is the relative velocity along the contact normal. $\gamma_n$ represents 
a viscosity parameter with a value that depends on the normal restitution coefficient between grains.   According to Coulomb's law 
the friction force is given by:
\begin{equation}
  \vec{f\ }^t = \left\{ \begin{array}{ll} 
k_t {\vec \delta}_t \  - \gamma_t\ \vec{v}^r_t 	& \mbox{        if} \ \ | \vec{f\ }^t | \leq \mu_s \mbox{f}^n \\
  - \mu_s f_n \frac{{\vec v}^r_t}{ | {\vec v}^r_t |}	& \mbox{        otherwise}
  \end{array}\right.
  \label{eq:friction_force}
\end{equation}
and 
\begin{equation}
 \vec {\delta}_t = \left\{ \begin{array}{ll} 
\int_t \vec{v}^r_t dt		& \mbox{        if} \ \ | \vec{f\ }^t | \leq \mu_s \mbox{f}^n \\ \\
 \frac{1}{k_t} \vec{f\ }^t	& \mbox{        otherwise}
 \end{array}\right.
  \label{eq:slingdisp_force}
\end{equation}
where $\mu_s$ is the friction coefficient, ${\vec v}^r_t = {\vec v}^r - {\vec v}^r_n$ is the tangential relative velocity, 
and $\gamma_t$ is a viscosity parameter, which depends on the tangential restitution coefficient. 


The equations of motion (both linear and angular momentum balance) are integrated according to a Velocity Verlet scheme \cite{Allen1987}.

\subsection{Lattice Boltzmann method}
\label{subsec:lbm}
The LBM is based on a material representation of fluids as consisting of particle distributions moving and colliding on a lattice 
\cite{Bouzidi2001,Feng2007,Feng2004,Yu:2007aa}. The partial distribution functions $f_{i}(\vec{r},t)$ are introduced 
to represent the probability density of a fluid particle at the position $\vec r$ with a velocity $\vec u = \vec {c_i}$ at time $t$ 
along discrete direction $i$. The three components of $\vec {c_i}$ are given in Tab.~\ref{tab:ciValue}.
\begin{table}[!ht]
\caption{The $\vec {c_i}$ components for a $D3Q19$ scheme (see Fig.~\ref{fig:SchemaLattice}).}
\label{tab:ciValue} 
\begin{tabular}{rrr rrr rrr rrr rrr rrr rr}
\hline\noalign{\smallskip}
$i$	& 0	& 1 & 2	& 3	& 4	& 5	& 6	& 7	& 8	& 9	& 10& 11& 12& 13& 14& 15& 16& 17& 18 \\
\noalign{\smallskip}\hline\noalign{\smallskip}
$x$	& 0	&-1	& 0	& 0	&-1	&-1	&-1	&-1	& 0 & 0	& 1	& 0	& 0	& 1	& 1	& 1	& 1	& 0	& 0	\\
$y$	& 0	& 0	&-1	& 0	&-1	&1	& 0	& 0 &-1 &-1	& 0	& 1	& 0	& 1	&-1	& 0	& 0	& 1	& 1	\\
$z$	& 0	& 0 & 0	&-1	& 0	& 0	&-1	& 1	&-1 & 1	& 0	& 0	& 1	& 0	& 0	& 1	&-1	& 1	&-1	\\
\noalign{\smallskip}\hline
\end{tabular}
\end{table}

The Lattice Boltzmann method is often non-dimensionalized when applied to physical problems. 
The governing macroscopic equations are given in terms of lattice units: 
\begin{equation}
\begin{array}{lr}
\mbox{Characteristic length scale} \ \  \ \ & \Delta x  \\
\mbox{Characteristic velocity}  \ \  \ \  & c \\
\mbox{Characteristic density}  \ \  \ \  & \rho_{_f} \\
\end{array}
\end{equation}
where $\Delta x$ is the lattice spacing, $c = \Delta x/\Delta t$ is the lattice speed with $\Delta t$ the time step, and 
$\rho_{_f}$ is  the fluid density at zero pressure. For the following, we will describe the method in lattice units.
 
\begin{figure}[!ht]
\centering
{
\includegraphics[width=.8\textwidth]{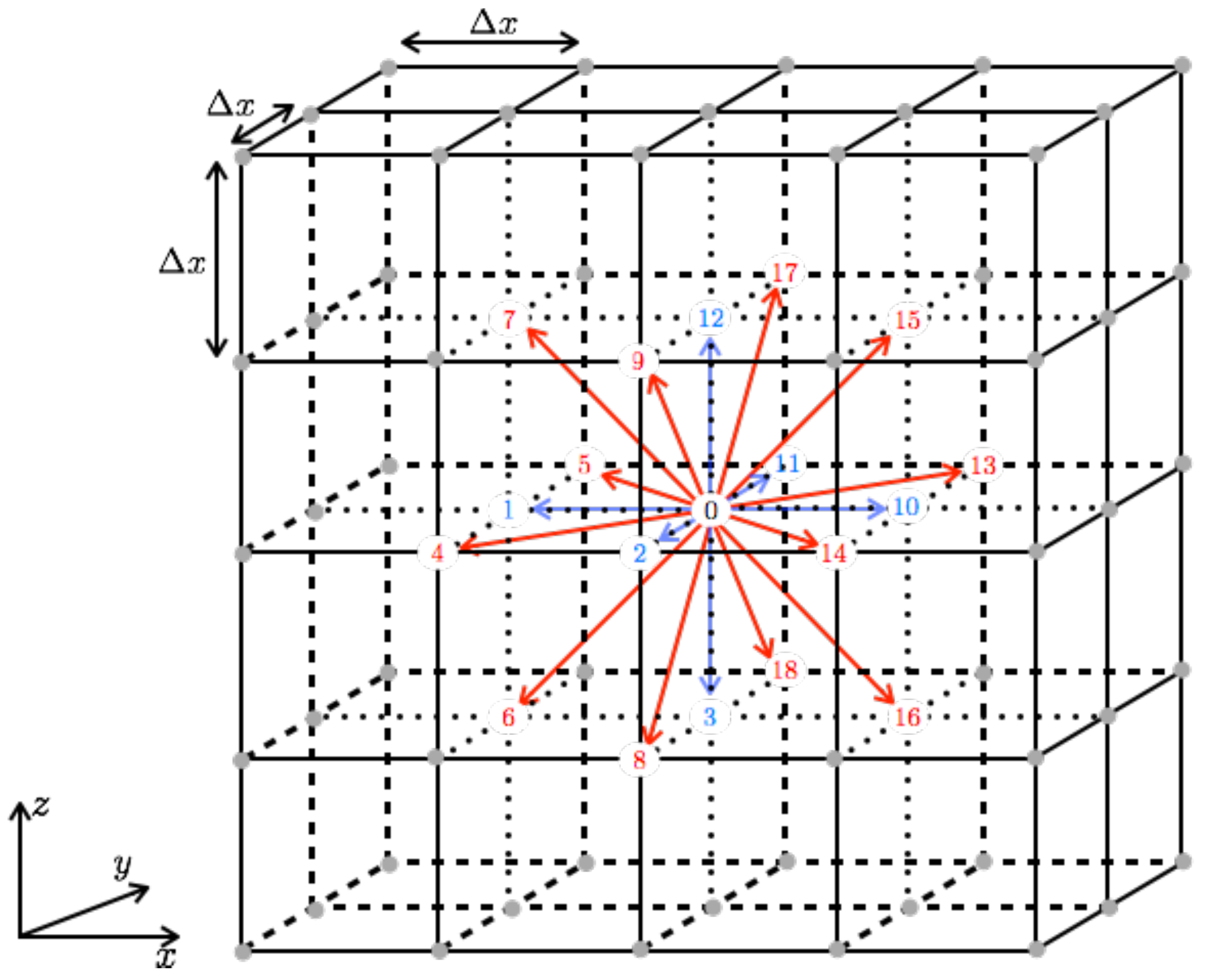}
}
\caption{3D lattice discretization with 18 directions (D3Q19).}
\label{fig:SchemaLattice}
\end{figure}

Figure~\ref{fig:SchemaLattice} shows a cartesian grid where the meshing scheme D3Q19, corresponding to 18 space directions 
in 3D used in our simulations, is represented. In LBM, the scheme D3Q19 is defined for each node where the distribution functions evolve 
according to a set of rules, which are constructed so as to ensure the conservation equations of mass, momentum and energy (with dissipation), 
so as to recover the Navier-Stokes equations \cite{He:1997aa}. This holds only when the wave lengths are small compared to the lattice 
spacing \cite{Chapman:1970aa}.

At each lattice node, the fluid density $\rho$ and momentum density $\rho \vec u$ are defined as    
\begin{equation}\label{dens}
\rho=\sum_{i}f_{i}.
\end{equation}
\begin{equation}\label{momentum}
\rho\vec{u}=\sum_{i}f_{i}\vec{c_{i}}.
\end{equation}
and the temperature is given by 
\begin{equation}
\frac{D}{2}kT=\sum_{i} \frac{1}{2} m (\vec{c_{i}}-\vec{u})^{2}\frac{f_{i}}{\rho} 
\end{equation}
where $D$ is the number of space dimensions, $m$ is particle mass, and $k$ is the Boltzmann constant. The equilibrium state 
is assumed to be governed by the Maxwell distribution: 
 \begin{equation}
f^{eq}(\vec{c})=\rho\left(\frac{m}{2\pi kT}\right)^{D/2}\exp\left[-\frac{m}{2kT}(\vec{c}-\vec{u}){}^{2}\right]
\label{eqn:maxw}
\end{equation}
where $\vec u$ is the mean velocity. By expanding (Eq.~\ref{eqn:maxw}) to order 2 as a function of $u/c_{s}$, which is 
the local Mach number with $c_s$ being the LBM sound velocity, a discretized form of the Maxwell distribution is 
obtained and used in the LBM:           
\begin{equation}
f_i^{eq}=\rho w_{i}\left[1 + \frac{\vec{c_{i}}\cdot\vec{u}}{c_{s}^{2}} - \frac{u^{2}}{2 c_{s}^{2}} + \frac{(\vec{c_{i}}\cdot\vec{u})^{2}}{2 c_{s}^{4}}\right]
\label{eq:Equilibre_Maxwell_Boltzmann}
\end{equation}
where the factor $w_{0} = 1/3$, $w_{(1, 2, 3, 10, 11, 12)} = 1/18$ and the rest of $w_{i} = 1/36$.  $w_{i}$ depend on 
the scheme with the requirement of rotational invariance \cite{Satoh:2011aa}. The LBM sound speed 
is then given by $c_{s}=\sum_{i}w_{i}c_{i}^{2} = 1/\sqrt 3$. 

The velocities evolve according to the Boltzmann equation. In its discretized form, it requires an explicit expression 
of the collision term. We used the Bhatnagar-Gross-Krook (BGK) model in which the collision term for each 
direction $i$ is simply proportional to the distance from the Maxwell distribution \cite{Bathnagar:1954aa}:
\begin{equation}
\frac{\partial f_i}{\partial t}_{coll} = \frac{1}{\tau} \left(   f_{i}^{eq}(\vec{r},t) - f_{i}(\vec{r},t) \right)
\end{equation}
where $\tau$ is a characteristic time. Hence, for the D3Q19 scheme, we have a system of $18$ discrete 
equations governing the distribution functions: 
\begin{equation}
f_{i}(\vec{r}+\vec{c_{i}}\Delta t,t+\Delta t) = f_{i}(\vec{r},t)+\mbox{\ensuremath{\frac{1}{\tau}}}\left(f_{i}^{eq}(\vec{x},t)-f_{i}(\vec{r},t)\right)
\end{equation}
These equations are solved in two steps. In the collision step, the variations of the distribution functions 
are calculated from the collisions: 
\begin{equation}
\tilde{f}_{i}(\vec{r},t+\Delta t)=f_{i}(\vec{r},t)+\mbox{\ensuremath{\frac{1}{\tau}}}\left(f_{i}^{eq}(\vec{r},t)-f_{i}(\vec{r},t)\right)
\label{eq:Collision_step}
\end{equation}
where the functions $\tilde{f}_{i}$ designate the post-collision functions. In the streaming step, the new distributions 
are advected in the directions of their propagation velocities:
\begin{equation}
f_{i}(\vec{r}+\vec{c_{i}}\Delta t,t+\Delta t)=\tilde{f}_{i}(\vec{r},t+\Delta t).
\end{equation}

The above equations imply an equation of state \cite{Chapman:1970aa,Chen1998,He1997}:
\begin{equation}
P(\rho)=\rho c_{s}^{2}.
\label{eq:Equation_d_Etat}
\end{equation}
The kinematic viscosity is then given by \cite{Guo2000,He1997}
\begin{equation}
\eta=c_{s}^{2}\left[\tau-\frac{1}{2}\right]
\label{eq:ViscositeLBM}
\end{equation}
with the requirement $\tau>1/2$. 

As discussed in \cite{Chapman:1970aa}, the Lattice Boltzmann method holds only when the pressure wave lengths are small  compared 
to the lattice spacing unit. This imposes a limitation on Mach number $Ma = u/c_s \ll 1$ and therefore fluid speeds higher than the sound speed
cannot be simulated. 

In nature, for a given fluid we have: sound speed $c * c_s$, density $\rho_{_f}$ and viscosity $\eta_{_f}$. From Eq.~\ref{eq:Collision_step}, 
we need the relaxation time $\tau$. This is related to $c$, $\eta_{_f}$ and $\Delta x$ by:
\begin{equation}
\tau = 0.5 + \frac{\eta_{_f}}{c_s^2} \frac{1}{c \Delta x}
\label{eq:tauPhysLBM}
\end{equation}
Equation.~\ref{eq:tauPhysLBM} shows that since $c$ and $\eta_{_f}$ are fixed from  fluid properties, only $\Delta x$ can  be used to ensure the stability 
of LBM, which becomes unstable  when $\tau \rightarrow 1/2$. Numerically, there is a limitation in computer capability regarding the smallest value of $\Delta x$. 
To handle this, a \textit{sub-grid turbulent model} based on LES with a Smagorinsky turbulence model is used 
\cite{Yu2005,kraichnan1976,Smaqorinsky1963,Moin1982}. The viscosity formulation is:
\begin{equation}
\eta^* = \eta + \eta_{_t} = c_s^2\left[\tau^* - \frac12\right] = c_s^2\left[\tau +  \tau_{_t} - \frac12\right]
\end{equation}
where $\eta_{_{t}} = c_s^2\tau_{_t}$ is the sub-grid LES viscosity and $\tau_{_t}$ is the sub-grid LES LBM relaxation time. 
The LES viscosity is calculated from the filtered strain rate tensor $S_{\alpha, \beta} = \frac 12 \left(\partial_{\alpha} u_{\beta} + \partial_{\beta} u_{\alpha} \right)$ 
and a filter length scale $l_x$ through the relation $\eta_{_t} = \left(C l_x\right)^2 S$ where $C$ is the Smagorinsky constant. In LBM, $S$ is 
obtained from the second momentum \cite{hou1996} $\Pi_{\alpha, \beta} = \sum_{i\neq0}c_{i\alpha}c_{i\beta}(f_i - f_i^{eq})$ as 
$S = \frac{\Pi}{2\rho c^2_s \tau^*}$, where $\Pi = \sqrt{2 \Pi_{\alpha, \beta} \Pi_{\alpha, \beta}}$. From $S$ and $\Pi$, $\tau_{_t}$ is expressed as:

\begin{equation}
\tau_{_t} = \frac 12 \left[\sqrt{\tau^2 + 2 \sqrt{2} (C l_x)^2 \left(\rho c^4_s \delta t\right)^{-1} \Pi } -\tau \right]
\end{equation}
where the filter length scale $l_x = \Delta x$ is the spatial lattice discretization.

\subsection{LBM-DEM coupling}
\label{subsec:lbmdemcoupling}
\begin{figure}[!ht]
	\centerline{
	\includegraphics[width=.7\textwidth]{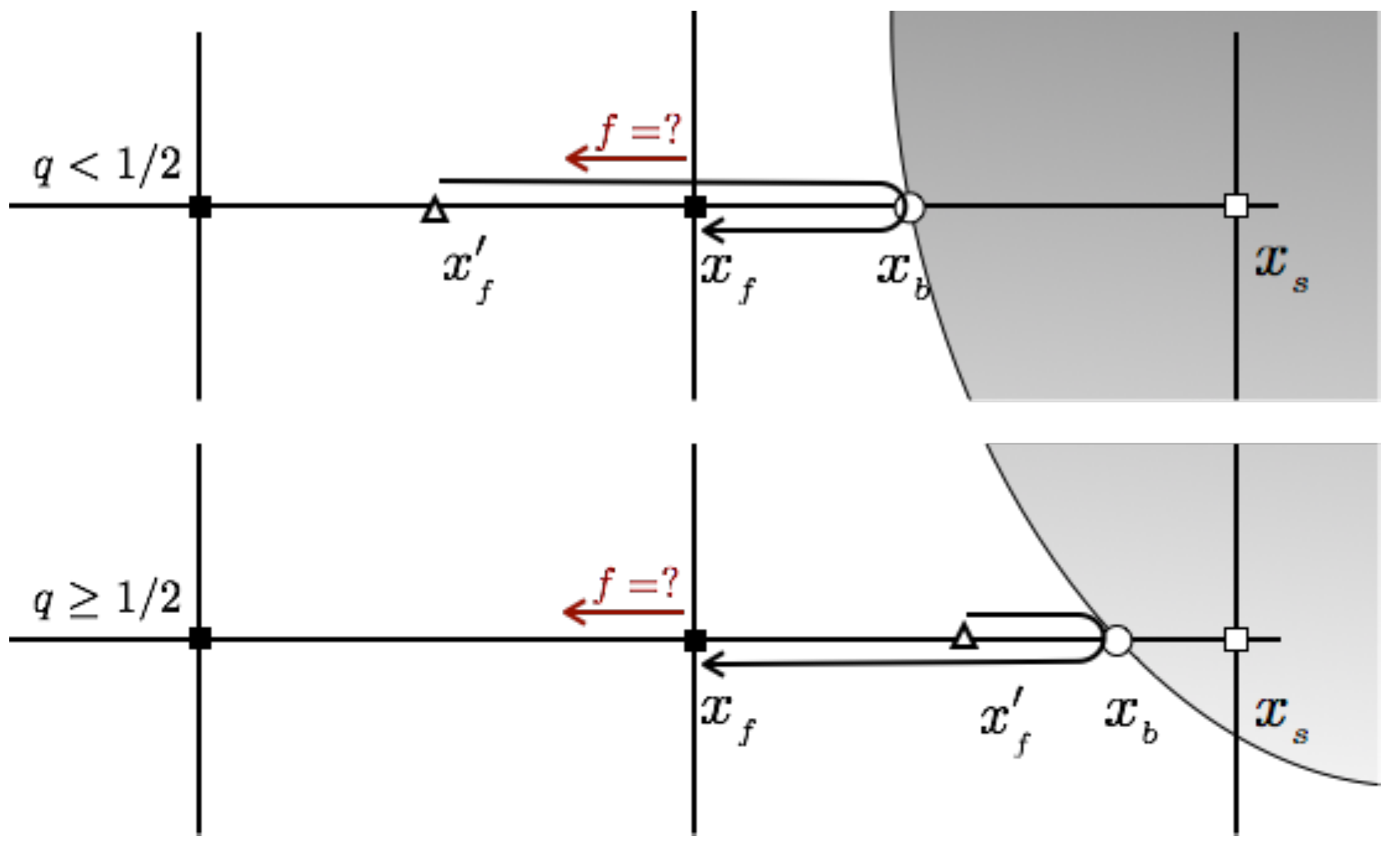}
	}
	\caption{A 2D illustration of the fluid-structure interaction scheme. Black squares are fluid nodes, triangles are 
	fictitious nodes, empty squares are solid nodes and circles are boundary nodes.
	}
\label{fig:fluideparticule_CB}
\end{figure}

There exist different techniques to model fluid-structure interaction. The most used in CFD is stress integration, 
however, in LBM the preferred approach is based on momentum exchange 
\cite{Bouzidi2001,Feng2007,Iglberger2008,Yu2003,Li2004,Peskin_1972,Wan_2006,Wan_2007}. 
This approach is simple in LBM, since in LBM each node already contains information about the derivatives of the hydrodynamic variables 
in each distribution function $f_i$ \cite{Chen1998}. Due to the presence of a solid boundary, after the collision step but before the streaming step, 
we know the population of $f_i$ except those which reflect off the solid wall as shown in Fig.~\ref{fig:fluideparticule_CB} ($f_i = ?$). In our simulations, 
we use the method developed by Bouzidi \cite{Bouzidi2001}, which mimics a macroscopic no-slip condition. For clarity, we describe the interaction 
rules using lattice units. This means that the time step is $one$, the space discretization is $one$, etc. Out of the characteristic scales, 
we will denote, around the fluid-solid boundary, $x$ as position and the subscripts $f$, $s$, and $b$ respectively will indicate either the \textit{fluid} 
or \textit{solid} domain and the \textit{fluid-solid boundary}.  We present the method in simplified 1D form. For more clarity in the way momenta 
are exchanged between fluid and solid domains, let us introduce $q=|x_b-x_f|/\Delta x$. According to the LBM scheme (collide and stream 
Sect.~\ref{subsec:lbm}) in the presence of a solid wall, we have two scenarios depending on the wall position $x_b$ (see Fig.~\ref{fig:fluideparticule_CB}): 
\begin{enumerate}
\item $q < \frac 12$ where fluid leaves  node $x_f$, reflects off the wall, and reaches $x_f$ again in time less than $\Delta t$.
\item $q \ge \frac 12$ where fluid leaves  node $x_f$, reflects off the wall, and reaches $x_f$ again in time greater than $\Delta t$.
\end{enumerate}
To handle these scenarios, we introduce a fictitious node $x_f'$(see Fig.~\ref{fig:fluideparticule_CB}) such that 
after streaming, a fluid particle leaving $x_f'$ arrives at $x_f$ exactly in time $\Delta t$.

As shown in Fig.~\ref{fig:fluideparticule_CB}, if $x_f$ is the last fluid node before we reach the solid boundary, 
$x_s = x_f + c$ should be a solid node. Let $f_{i'}$ be the distribution function such that $f_{i'}$ is the opposite direction of $i$ where $i$ 
is the direction oriented from fluid node to solid boundary. By using a linear interpolation, $f_{i'}$ is expressed as follow:

\begin{equation}
   \begin{array}{lcr}
	f_{i'}(x_f, t + \Delta t) = 2qf_i^c(x_f , t) + (1 - 2q)f_i^c(x_f-\vec{c}_i, t) + \partial f_{i'}^w				& \ \mbox{  for} & q <    \frac{1}{2} \\
	f_{i'}(x_f, t + \Delta t) = \frac{1}{2q}f_i^c(x_f, t) + \frac{2q - 1}{2q}f_{i}^c(x_f+\vec{c}_i, t) + \partial f_{i'}^w & \ \mbox{  for} & q \ge \frac{1}{2}
   \end{array}
\end{equation}
where $f_i^c$ corresponds to the distribution function of $f_i$ after the collision step but before streaming and $f_{i}^c(x_f+\vec{c}_i, t) = f_{i'}^c(x_f, t)$. 
The term $\partial f_{i'}^w$ is calculated from the boundary velocity and is zero if the boundary is stationary.
$\partial f_{i'}^w$ is calculated by considering that the fluid velocity $u$ evolves linearly between $x_{f}$ 
and $x_{s}$. If $u_0$ is the boundary velocity, $u$ is then defined by

\begin{equation}
   u = u_0+(x_f-q)\frac{\partial u}{\partial x}
\end{equation}
at first order in $u$. The equilibrium value of $f_i$ is given by $f_i = f_i^0+3\omega_i u \cdot c_i$ where $f_i^0$ is constant and depends on the lattice 
discretization scheme \cite{Bouzidi2001,Ladd2001}. Using a linear interpolation, $\partial f_{i'}^w$ is given by:
\begin{equation}
   \begin{array}{lcr}
	\partial f_{i'}^w = 6\omega_i\ u_0\ c_i			& \ \mbox{         for} & q <    \frac{1}{2} \\
	\partial f_{i'}^w = \frac{3}{q}\omega_i\ u_0\ c_i	& \ \mbox{         for} & q \ge \frac{1}{2}
   \end{array}
\label{eq:moving_boundary_velocity}
\end{equation}
Hydrodynamic forces acting on the structure are calculated by using momentum exchange 
\cite{Bouzidi2001,Iglberger2008,Yu2003,Lallemand2003}. 
Let $\vec{Q}_f$ and $\vec{Q}_s$ be fluid and solid momentum calculated near 
the boundary. The exchanged momentum is given by:
\begin{equation}
\Delta \vec{Q} = \vec{Q}_s - \vec{Q}_f.
\end{equation}
$\vec{Q}_f$ and $\vec{Q}_s$ are calculated as follows:
\begin{equation}
	\vec{Q}_f = \Delta x^D \sum_{all} \sum_i f_i^f \vec{c}_i
\end{equation}
\begin{equation}
	\vec{Q}_s = \Delta x^D  \sum_{all} \sum_i f_{i'}^s \vec{c}_{i'}
\end{equation}
where $D$ is the space dimension, and $f_{i}^f$ and $f_{i'}^s$ are respectively the fluid and the solid distribution functions. To be clear, $f_{i'}^s$ is constructed at a lattice point occupied by solid by taking the solid velocity $\vec{u}_s$ and density $\rho_s$ and assigning a Maxwell equilibrium distribution, per Eq.~\ref{eq:Equilibre_Maxwell_Boltzmann}.
The hydrodynamic force $\vec{F}$ and torque $\vec{T}$ are then given by:
\begin{equation}
\vec{F}=\frac{\Delta \vec{Q}_{fs}}{\Delta t}= \frac{\Delta x^D}{\Delta t}\sum_{all}\sum_i(f_i^f+
f_{i'}^s)\vec{c}_{i'}
\end{equation}
\begin{equation}
\vec{T}= \frac{\Delta x^D}{\Delta t}\sum_{all}\sum_i l\times(f_i^f+f_{i'}^s)\vec{c}_{i'}
\end{equation}
where $l$ is the distance between the center-of-mass of the solid domain and the boundary node $x_b$.

\subsection{LBM-DEM-Rubber coupling}
\label{subsec:lbmdemrubbercoupling}

Unlike the coupling between LBM and DEM, the LBM-Rubber and DEM-Rubber coupling is indirect. We focus our explanation below on the case of a rubber ring 
component of a valve, but the idea can be generalized to other cases.  

\begin{figure}[!ht]
\centerline{
\includegraphics[width=1.\textwidth]{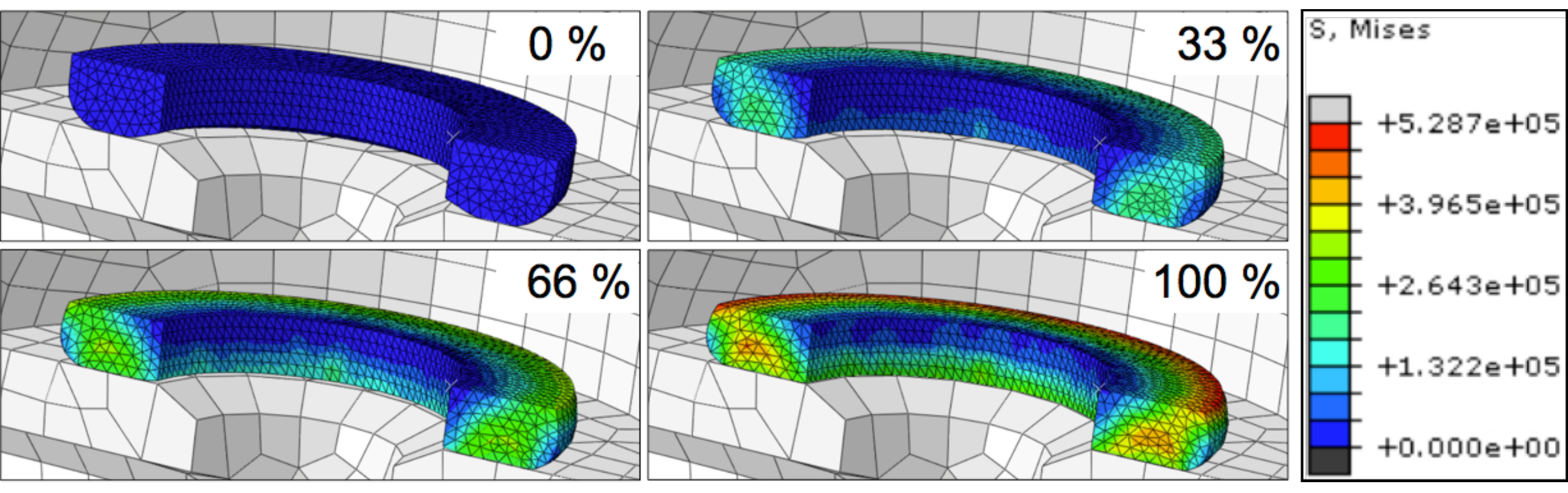}
}
\caption{Snapshots of rubber deformation during the Abaqus simulation at 0 \%, 33 \%, 66 \% and 100 \% of the simulation 
duration where the color map shows the Mises stress in Pa.}
\label{fig:abq_deform_snapshot}
\end{figure}

\begin{figure}[!ht]
\centerline{
\includegraphics[width=.75\textwidth]{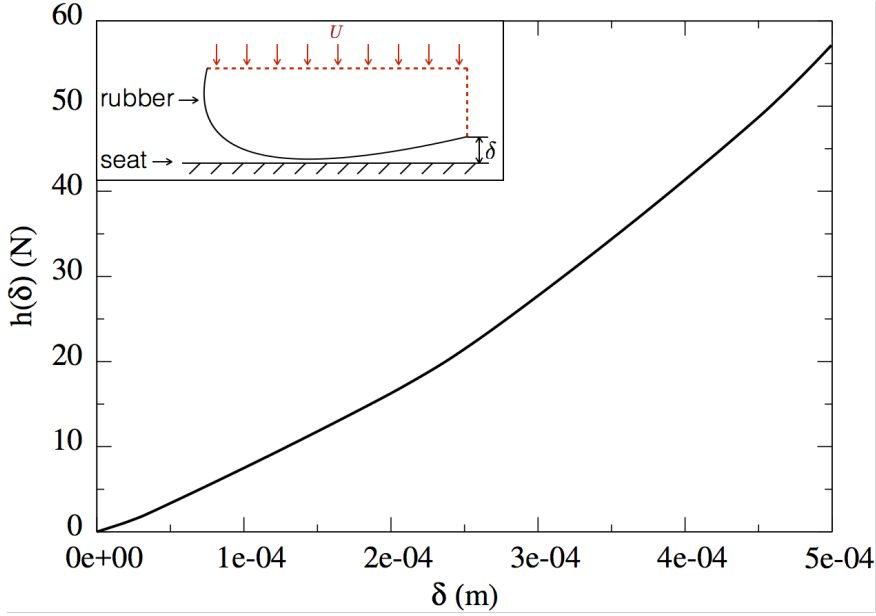}
}
\caption{Net-force of valve seat on rubber as function of displacement. The inset shows the configuration of Abaqus simulation 
where the dot line represent the imposed velocity boundary $U$.}
\label{fig:force_vs_displacement}
\end{figure}

A first 
simulation is performed using Abaqus from which the deformed rubber shape and reaction force of the valve seat on the rubber are 
saved for many states of rubber compression.  This simulation uses no fluid or particles. The Abaqus simulation consists of compressing the rubber ring geometry against the bare valve seat (see inset of 
Fig.~\ref{fig:force_vs_displacement}).  The rubber is simulated as a nearly incompressible neo-Hookean elastic solid with a strain energy $\Psi$  (per unit reference volume) given by $
    \Psi = (G_r/2)\left(\overline{I}_1 - 3\right) + (K_r/2)\left(J - 1\right)^2$
where  $\overline{I}_1$ is the first 
deviatoric strain invariant defined as $
   \overline{I}_1 = \overline{\lambda}_1^2 + \overline{\lambda}_2^2 +\overline{\lambda}_3^2$,
 the deviatoric stretches are given by $\overline{\lambda}_i = J^{\frac{1}{3}}\lambda_i$, $J$ is the total volume ratio, and $\lambda_i$ are the principal stretches. The (small-strain) shear modulus is $G_r$ and bulk modulus is
$K_r$.
A frictionless contact between rubber and valve seat is used for simplicity. Figure~\ref{fig:abq_deform_snapshot} shows several snapshots during 
the Abaqus simulation and Fig.~\ref{fig:force_vs_displacement} gives the seat net-force on rubber as a function of (downward) displacement 
$\delta$ of the rubber ring, $h(\delta)$.  We index each deformed rubber configuration by the value of $\delta$ it corresponds to. 
The Abaqus tests are performed under quasi-static conditions but we also assume damping can exist such that upward the force on the rubber satisfies the relation
\begin{equation}\label{constit}
F=h(\delta)+\nu\dot{\delta}.
\end{equation}


Then, the data from the Abaqus simulation is used in a weak coupling routine to describe rubber configurations when the ring takes part in a slurry simulation. 
In short, the method determines which of the deformed rubber configurations from the stand-alone Abaqus tests is the best representation 
of the actual deformed rubber state at that moment in the slurry problem.   Hence, the utility of this method lies in the fact that the rubber deformation that occurs in 
the actual slurry problem largely resembles the modes of deformation seen in a purely solid compression experiment.  Situations where the rubber surface becomes heavily locally deformed could be problematic for this approach.

From the LBM-DEM point of view, the rubber is composed of tetrahedra; this allows us to 
compute contact forces for DEM and the exchanged momentum for LBM as if it were a simple collection of polyhedral objects. 
Since the Abaqus simulation is performed without fluid or particles, to use its solutions we need to deduce an effective upward force 
from LBM-DEM acting on the bottom rubber surface, which can then be referenced against the Abaqus data to infer a deformed rubber shape. 
The effective force is needed because the rubber in the Abaqus simulations has contact only with the valve seat, whereas in the slurry case, 
there can be additional forces from fluid and particles extending to the lateral surfaces of the rubber.  

\begin{figure}[!ht]
\centerline{
\includegraphics[width=1.\textwidth]{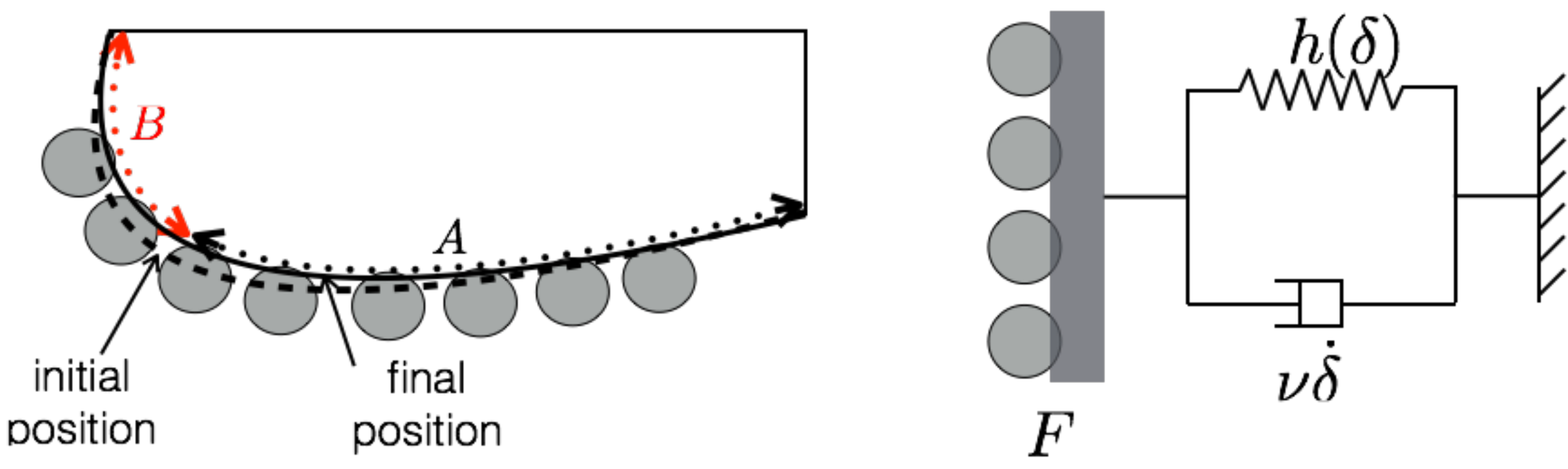}
}
\caption{Configuration of rubber-LBM-DEM coupling. The left figure gives the geometrical definitions of area $A$ and $B$, and the initial 
and final shape of the rubber as it proceeds through the iteration loop to solve Eq.~\ref{constit}, with a representation of particles in 
contact shown. The right figure gives a 1D analog of the model to compute the rubber deformation.}
\label{fig:abq_coupling_config}
\end{figure}

Key to our routine is to identify two subsets of the exposed rubber surface, 
denoted surface A and surface B.  Surface A is the part that makes contact with the valve seat and surface B remains free 
in the Abaqus simulations (see left Fig~.\ref{fig:abq_coupling_config}). In particular, surface $A$ and $B$ are geometrically defined 
using the the last frame of the Abaqus simulation where the rubber is fully compressed. In the slurry case, 
we add uniform hydrostatic stress to the observed rubber loading distribution until the mean normal stress acting on surface B vanishes. 
This leaves us with a loading state that resembles one from Abaqus. Because the rubber is essentially incompressible, changing 
the hydrostatic stress uniformly along the surface does not affect the deformed rubber configuration.  To be specific, we compute 
the normal stress on surface $A$ using $P_{_A}  = \frac 1 A \int_A \vec{n}\cdot \vec{f} dS$ and on surface 
$B$ using $P_{_B} = \frac 1 B \int_B \vec{n}\cdot \vec{f} dS$ where $\vec{f}$ is the stress from hydrodynamic, particle, 
and valve-seat forces and $\vec{n}$ is the normal vector on section $dS$.  Since the rubber deformation is caused by the 
shear part of the stress, we uniformly subtract the traction $P_{_B}$ from all normal stresses acting on the rubber.  
This modified loading now resembles the Abaqus loading (inset Fig.~\ref{fig:force_vs_displacement}) in which only 
an upward force on surface $A$ exists. Therefore, we define the effective upward force on surface $A$ as ${F} = \text{Area}_A\cdot (P_{_A } - P_{_B}) $. 

The rubber shape is updated using a four step loop, which is performed after particle positions and fluid data are updated.  
The goal of the iteration routine is to implicitly solve Eq.~\ref{constit} for $\delta$ so that the effective force from particles, fluid, 
and the valve seat on the final rubber state matches the force from Eq.~\ref{constit}.
\begin{itemize}
\item Step 1: Compute effective upward force $F$ from fluid, particle, and valve-seat interactions on rubber given the current guess for $\delta$
\item Step 2: Use this force and Eq.~\ref{constit} to update to a new guess for $\delta$. 
\item Step 3: Check if new and old $\delta$ differ by less than a tolerance. If so break, if not update the rubber shape based on the new $\delta$.
\item Step 4: Update applied force on surfaces $A$ and $B$ according to the rubber shape given by the new guess for $\delta$.
\end{itemize}
We assume that fluid forces do not change throughout the iteration procedure. This is true by assuming a small incremental rubber shape 
change between iterates, so only particle and valve-seat forces on the rubber are updated during Step 4.

 In Step 2 we utilize the following update rule
 \begin{equation}
\delta \leftarrow \frac {\left[ F - h(\delta) + \delta_0 \frac{\nu}{\Delta t} \right] \frac{\Delta t}{\eta} + \delta}{1 + \frac{\nu}{\eta}}
\label{eqn:rubber_eqn_2}
\end{equation}
where $\delta_0$ is the actual displacement of the rubber at the beginning of the time-step.  The coefficient $\eta$ is numerically 
selected to aid convergence. Note that if the updated value of $\delta$ matches the value inputted, then Eq.~\ref{constit} is solved.  
The update rule above attempts to move $\delta$ toward such a solution with each pass through the iteration loop. We check convergence 
of $\delta$ (Step 3) using a tolerance that is scaled by the error of the first iterate, $\delta_1-\delta_0$, where $\delta_1$ is 
the value obtained for $\delta$ after the first pass of the loop. 

We use simple linear interpolation to compute values of $h(\delta)$ when $\delta$ is between two neighboring values from the Abaqus output.

The third step consists of updating the rubber shape using the new guess for $\delta$ obtained in Step 2. Again, a linear interpolation is applied as 
is done for $h(\delta)$.  For the shape, 
we export the Abaqus mesh, which allows the interpolation.  For example, if the new $\delta$ lies directly between two frames of the Abaqus data, 
the rubber shape is updated by moving the nodes of the rubber to positions halfway between those of the neighboring frames. 

The fourth step consists of recomputing the contact force between the tetrahedra representing the rubber, and the particles/valve-seat. 
The contact force is computed using Eq.~\ref{eq:normal_force} for the normal part and Eq.~\ref{eq:friction_force} for the tangential part. One 
specificity here is that we do not update $\vec{\delta_t}$ during the coupling loop routine unless when $|\vec{f\ }^t| > \mu_s \mbox{f}^n$ where 
$\vec{\delta_t} = \frac{1}{k_t} \vec{f\ }^t$.  A schematic model of one update through the iteration loop is presented in Fig.~\ref{fig:abq_coupling_config} 
(left) and a 1D mechanical model of the treatment of the rubber interaction is visualized in Fig.~\ref{fig:abq_coupling_config} (right).

\subsection{Numerical algorithm}
\label{subsec:NumericalAlgorithm}

\begin{figure}[!ht]
\centerline{
\includegraphics[width=1.\textwidth]{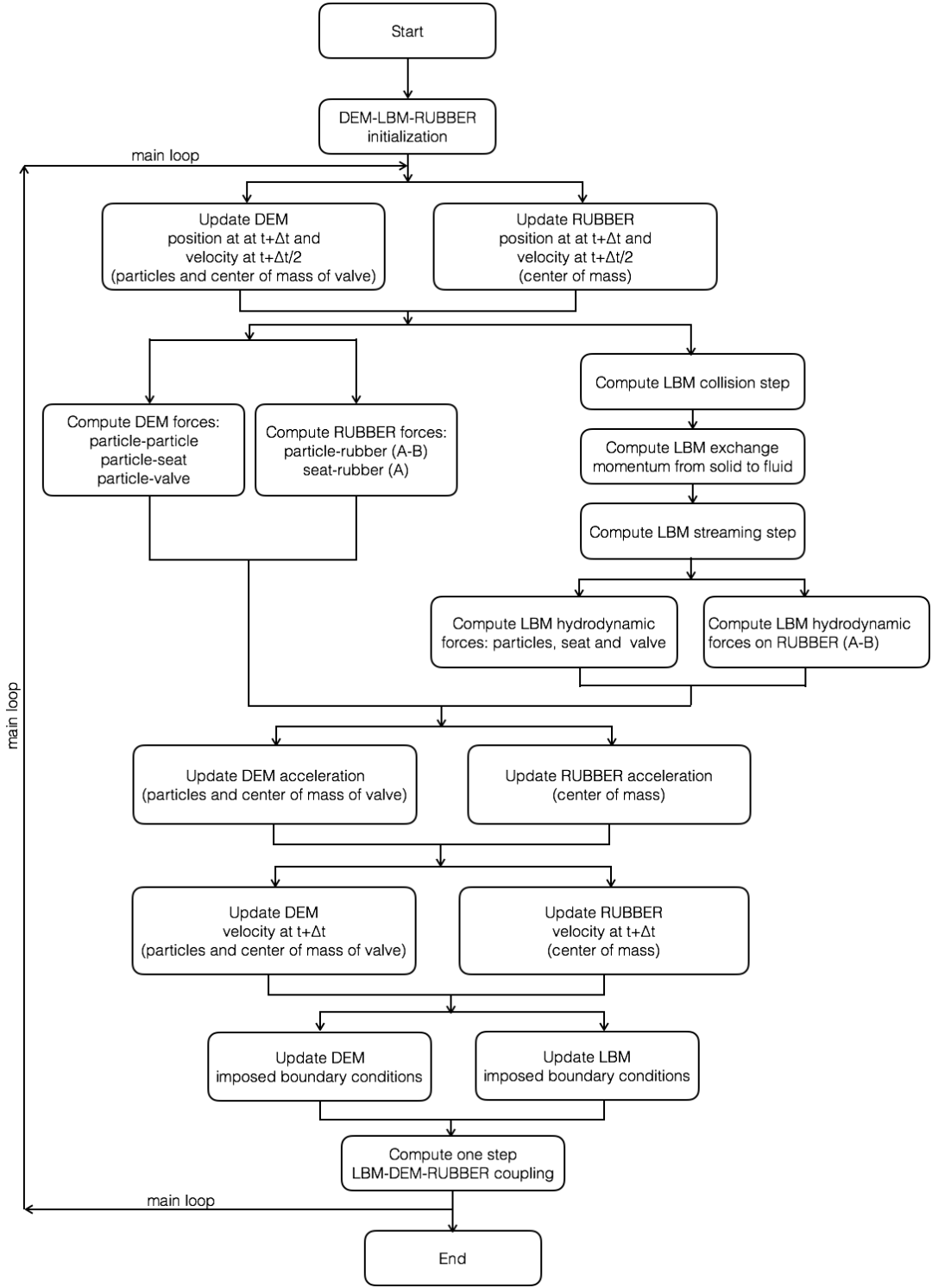}
}
\caption{LBM-DEM-RUBBER implementation algorithm.}
\label{fig:algo}
\end{figure}

The numerical method described in previous sections is implemented using the algorithm displayed in Fig.~\ref{fig:algo}. Note that we compute the valve acceleration by enjoining the applied force and mass of the rubber and valve components together in the Verlet update; rubber deformations are not included in the calculation of net acceleration of the valve/rubber composite as they are small compared to the movement of the valve overall. As shown in Fig.~\ref{fig:algo}, the 
LBM step is computed in parallel with DEM force calculation  and Rubber force calculation.


\subsection{Zoom-in with Effective Boundaries (ZIEB)}
\label{subsec:zoominlinktech}

The \textit{Zoom-In with Effective Boundaries} (ZIEB) technique replaces a fluid 
reservoir/domain with an analytical solution that interacts dynamically with the remainder of the domain. 
The challenge is to determine the correct effective dynamics at the fictitious interface, and to transfer the analytical result to
LBM distribution functions. In this study we model valves, which can be positioned far from the pump
that leads the slurry to the valve. The goal is to avoid having 
to calculate flow in the expansive pump region. From a computational point of view, one might assume a simple input velocity boundary condition should solve the problem, however, for a compressible fluid, the imposed flow and pressure may depend on the total removed volume and  feedback with the dynamics within the valve system. 
In this section, we first detail how to obtain the analytical solution then explain how to implement 
this solution as an LBM boundary condition.

\subsection*{ZIEB analytical solution}

\begin{figure}
    \centering
    \includegraphics[width=3.5in]{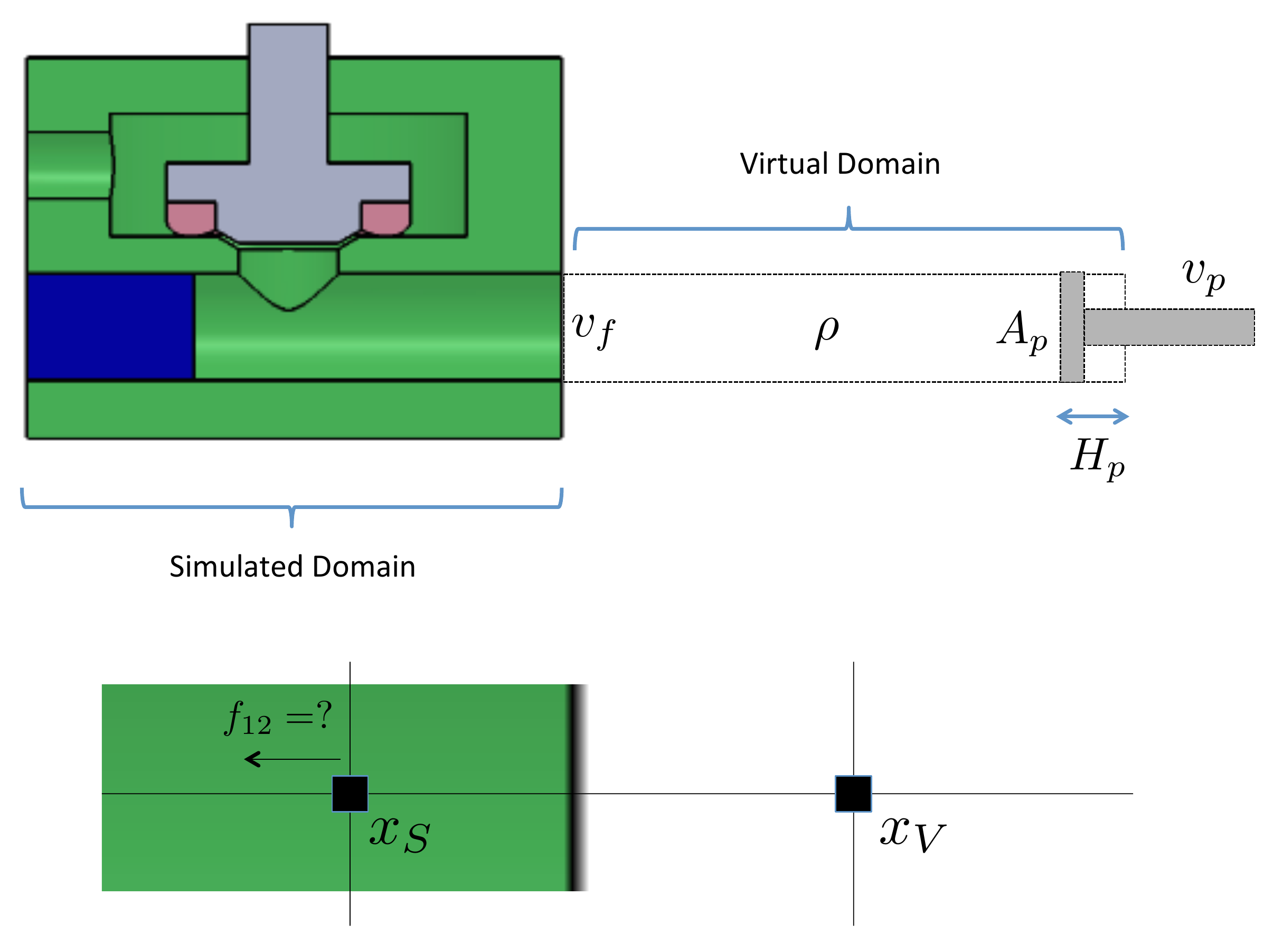}
    \caption{Top: Full geometry of problem to be solved --- valve region connected to a cylinder and piston region.  ZIEB technique allows removal of the piston/cylinder domain from the simulation, by replacing it with effective boundary conditions at the fictitious interface. Bottom:  Zoom-in near the fictitious interface showing the first simulated lattice point (at $x_S$) and the closest virtual lattice point (at $x_V$). }
    \label{virtual}
\end{figure}

Per Fig.~\ref{virtual}, we assume the virtual (i.e. removed) domain is a cylinder and piston.  The cylinder is initially full of fluid and has total volume $V_0$.  As the piston moves, fluid is pushed into the simulated domain.  Let the movement of the piston be given by some prescribed $H_p(t)$, where $H_p$ measures the piston displacement.  The cross-sectional area of the piston (and cylindrical domain) is $A_p$.  The piston velocity, $v_p(t)$, is simply defined from the time-derivative of $H_p$. Define as $v_f$ the mean in-flowing fluid velocity component on the interface between the domains.  Let $\rho$ be the average density of fluid in the virtual cylinder between the piston head and the interface. Further, we make the simplifying assumption that in the cylinder region the fluid density is in fact uniform, such that it is equal to $\rho$ throughout.

Conservation of fluid mass in the cylinder domain can be expressed by balancing the mass rate within the cylinder against the mass flux into the simulated domain:

\begin{equation}
    \frac{d}{dt}\rho\cdot(V_0-A_pH_p)=-\rho v_fA_p \ \ \Rightarrow \ \  \frac{d\rho}{dt}=\rho\left[\frac{v_pA_p-v_fA_p}{V_0-A_pH_p}\right]
    \label{eq:mass_conv}
\end{equation}
In a fully continuum framework, the above equation would need to be augmented with momentum balance in order to provide the in-flowing velocity, $v_f$, at the fictitious interface.  However, using the LBM description, we can update $v_f$ in another way, which is consistent with momentum balance on the small scale.

\subsection*{Implementation of ZIEB analytical solution in LBM}

At a given time $t^n$, we assume $\rho^n$ is given in the cylinder domain and is equal to the density at $x_S$, where, per Fig.~\ref{virtual}, $x_S$ is the lattice point in the simulated domain that is adjacent to the interface with the virtual domain.  We suppose the velocity at $x_S$ is the interfacial velocity $v_f^n$.  Both density and velocity at time $t^n$ at $x_S$ are defined by Eq.~\ref{dens} and \ref{momentum} through distribution functions $f_i^{n}$.  

The distribution functions are updated to $t^{n+1}$ under the following procedure, which is applied after the collision step but before streaming.  First we  update and store the density $\rho^{n+1}$ at $x_S$ using explicit integration of Eq.~\ref{eq:mass_conv}:

\begin{equation}
\rho^{n+1}=\rho^n\exp\left[ \frac{v_p^nA_p-v_f^{n}A_p}{V_0-A_pH^n_p}\Delta t\right]
\label{eq:ZIEBdesity8}
\end{equation}

Next, a partial LBM streaming step is performed at $x_S$ using the distributions at time $t^n$.  During this step $x_S$ streams to and from its neighboring `real' lattice points within the simulated domain. However, it only streams out of the interface with the virtual domain and does not receive any distributions from the virtual domain. Define $\rho^*$ as the density at $x_S$ after this partial streaming step.

The next step is to back-solve the needed distributions to be streamed in from the virtual domain in order to guarantee the final density at $x_S$ equals $\rho^{n+1}$.  For example, consider a setup as shown in Fig.~\ref{virtual} and suppose the fictitious interface is normal to the $\hat{z}$ direction (see Fig.~\ref{fig:SchemaLattice}).  After the partial streaming step, updated (though not finalized) distribution values exist for all the $f_i$ except for the values associated to
$i=7,\ 9,\ 12,\ 15,$ and $17$. These five distribution values are all unknown after the partial streaming step. To compute them, first we modify only the value of the $f_{12}$ distribution, which is the distribution that streams into the simulated domain normal to the fictitious boundary:
\begin{equation}
f_{12} = \rho^{n+1} - (\rho^* - f^{n}_{12})
\label{eq:ZIEBdesity9}
\end{equation}
This advects all the missing density at $x_s$ from a fictitious node $x_V$ (see bottom of  
Fig.~\ref{virtual}). With these distributions, the velocity at time $t^{n+1}$ is computed at $x_S$ 
according to $u=\frac{1}{\rho^{n+1}}\sum f_i$. Because the distributions in the $i=7,\ 9,\ 15,$ and 
$17$ directions are still unknown at this point, the Maxwell equilibrium function 
Eq.~\ref{eq:Equilibre_Maxwell_Boltzmann} is then used to redistribute all the distributions at $x_S$
to a more natural, equilibrium state.  This updates all the distributions to their final values, at 
$t^{n+1}$.

We notice here that the initial value of $\rho$ at the beginning of the simulation should be a 
normalized value (in lattice units) otherwise an additional step of normalizing by the physical 
fluid reference density will be necessary before using it in Eq.~\ref{eq:mass_conv} and Eq.~\ref{eq:ZIEBdesity9}.

\subsection{Tests and validations}
\label{subsec:testsnvalidations}

We first test some of the individual components of the routine.  In this section, we provide separate 
numerical validations of  the ZIEB technique, the rubber deformation model, and the LBM.

\begin{figure}[!ht]
\centerline{
\includegraphics[width=1.\textwidth]{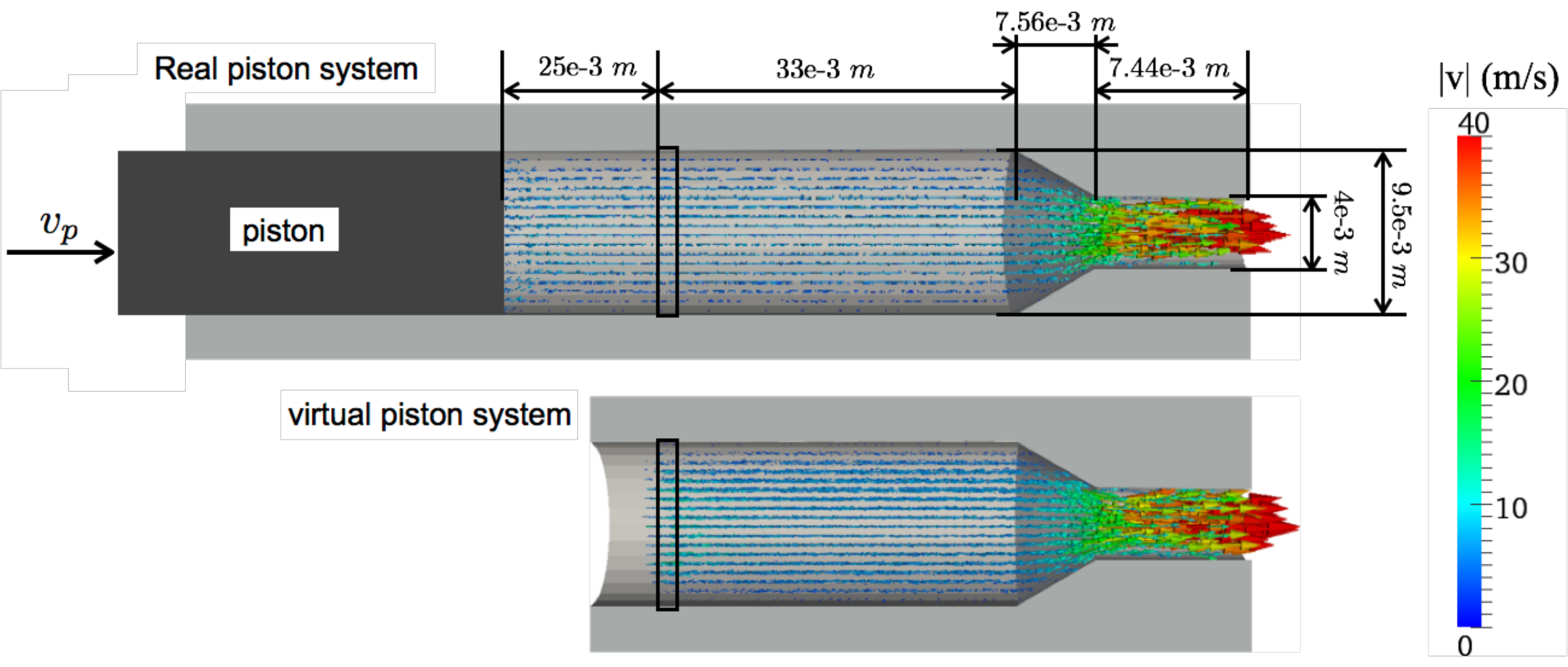}
}
\caption{Configuration of tests for the \textit{Zoom-in with Effective Boundaries} (ZIEB) technique. 
The real piston geometry is displayed in the top figure (mid-simulation) and the virtual piston 
geometry is displayed beneath, where ZIEB has been applied on the left boundary to mimic the removed piston domain.}
\label{fig:zoomnlink_config}
\end{figure}

\begin{figure*}[!ht]
\centerline{
\includegraphics[width=.4625\textwidth]{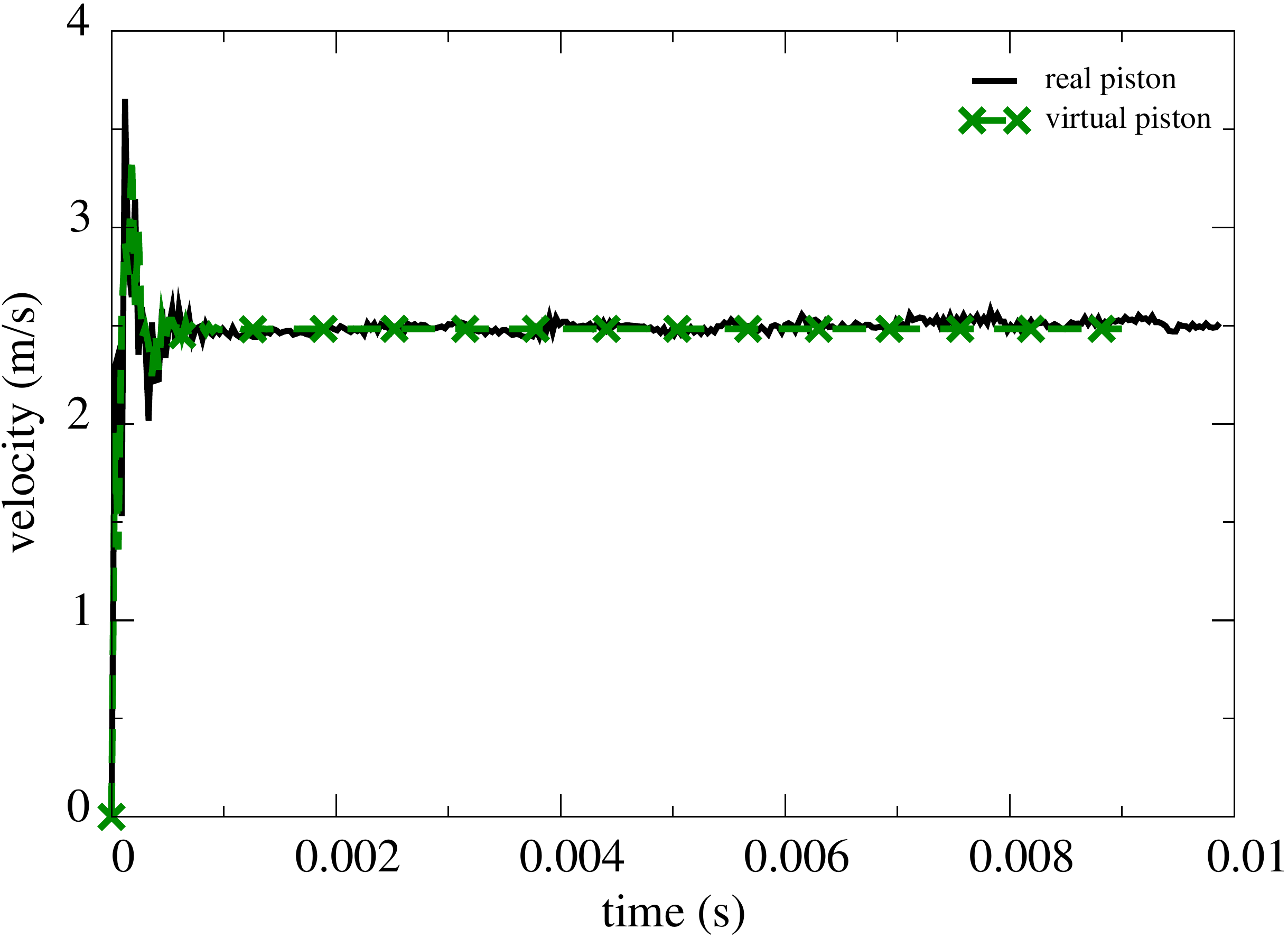}
\includegraphics[width=.5\textwidth]{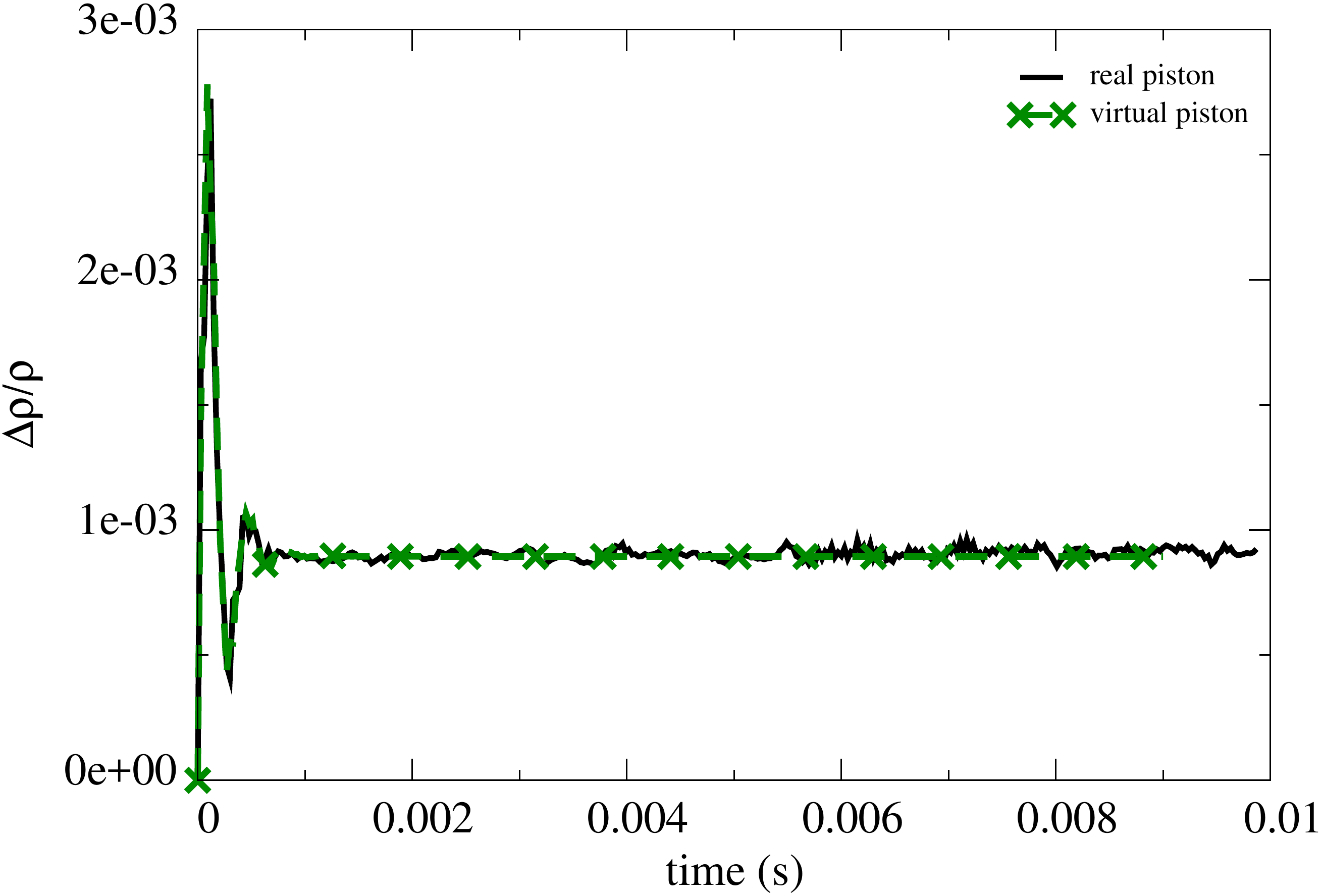}
}
\caption{Velocity and normalized input fluid density as functions of time for the real and virtual 
piston setups. The velocity and density are calculated within the boxed subdomain in 
Fig.~\ref{fig:zoomnlink_config}.}
\label{fig:velocity_density_opened}
\end{figure*}

To validate the ZIEB method, we performed an analysis of fluid flow in a geometry comprised of a 
piston driving fluid passing through a narrow restriction.  This flow field is then compared to that 
obtained using a ``virtual piston'' in which the domain containing the moving piston is removed and 
in its place an effective boundary condition from ZIEB is used, see Fig.~\ref{fig:zoomnlink_config}. 
The real piston begins positioned such that the fluid volume between the piston and input section is 
$V_{0}$; the same $V_0$ is used in Eq.~\ref{eq:ZIEBdesity8} for the virtual piston. We use 
$V_{0} = 1.77\mbox{e-}06\ m^3$ and $A_p = 7.09\ m^2$. As input parameters, we use a (pressure-free) 
fluid density $\rho_{_f} = 1000\ kg/m^3$, dynamic viscosity $\eta = 0.001\ Pa.s$ and a Smagorinsky 
constant of $C = 0.4$ for the sub-grid turbulence model \cite{Feng2007,hou1994,germano1991}. Figure~\ref{fig:velocity_density_opened} 
shows the comparison between the two simulations regarding fluid velocity and the normalized input 
fluid density computed in the same domain (see Fig.~\ref{fig:zoomnlink_config}). The agreement 
is strong, even the time-dependent fluctuations, confirming the correctness of the ZIEB method.

\begin{figure*}[!ht]
\centerline{
\includegraphics[width=.5\textwidth]{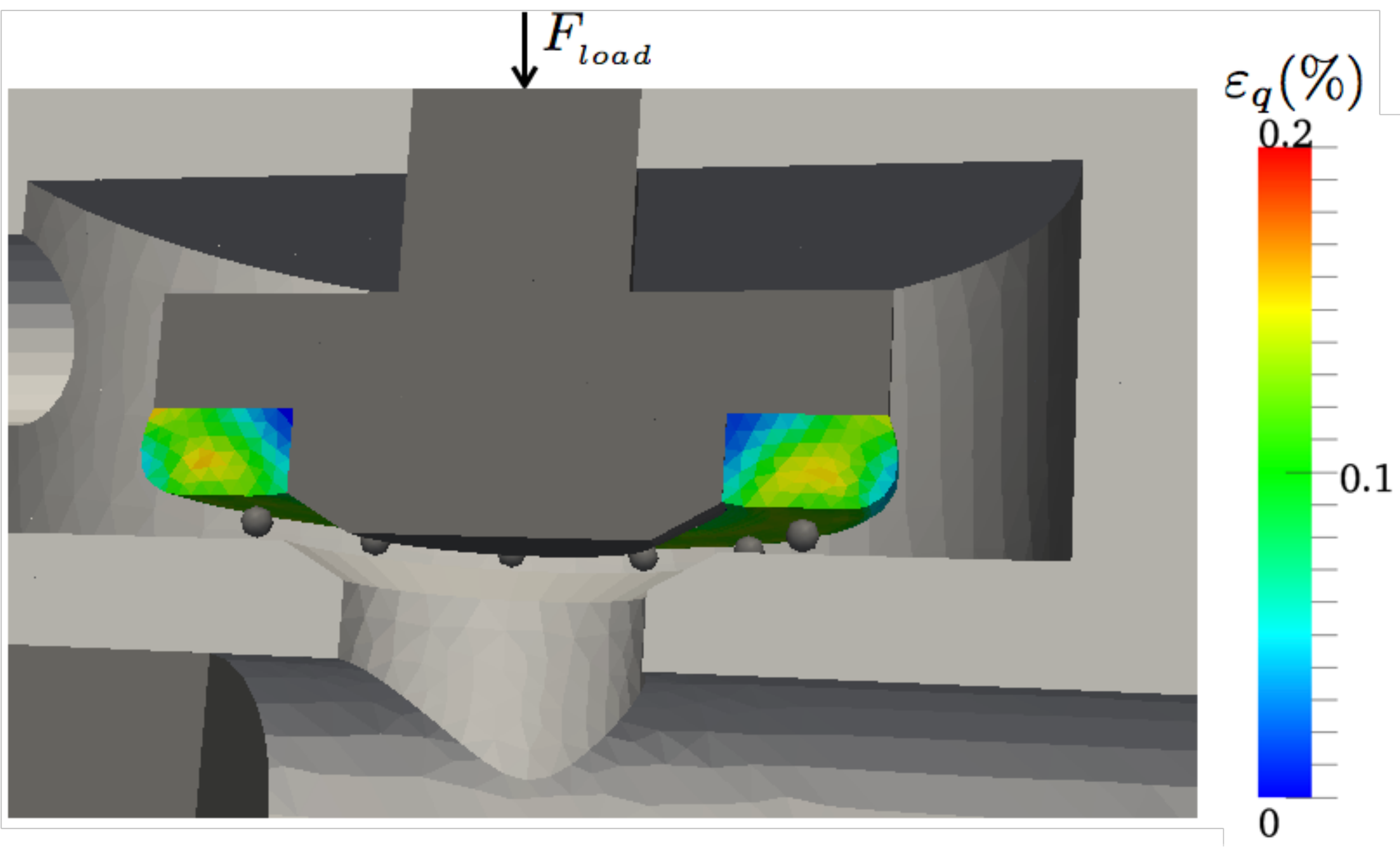}
\includegraphics[width=.5\textwidth]{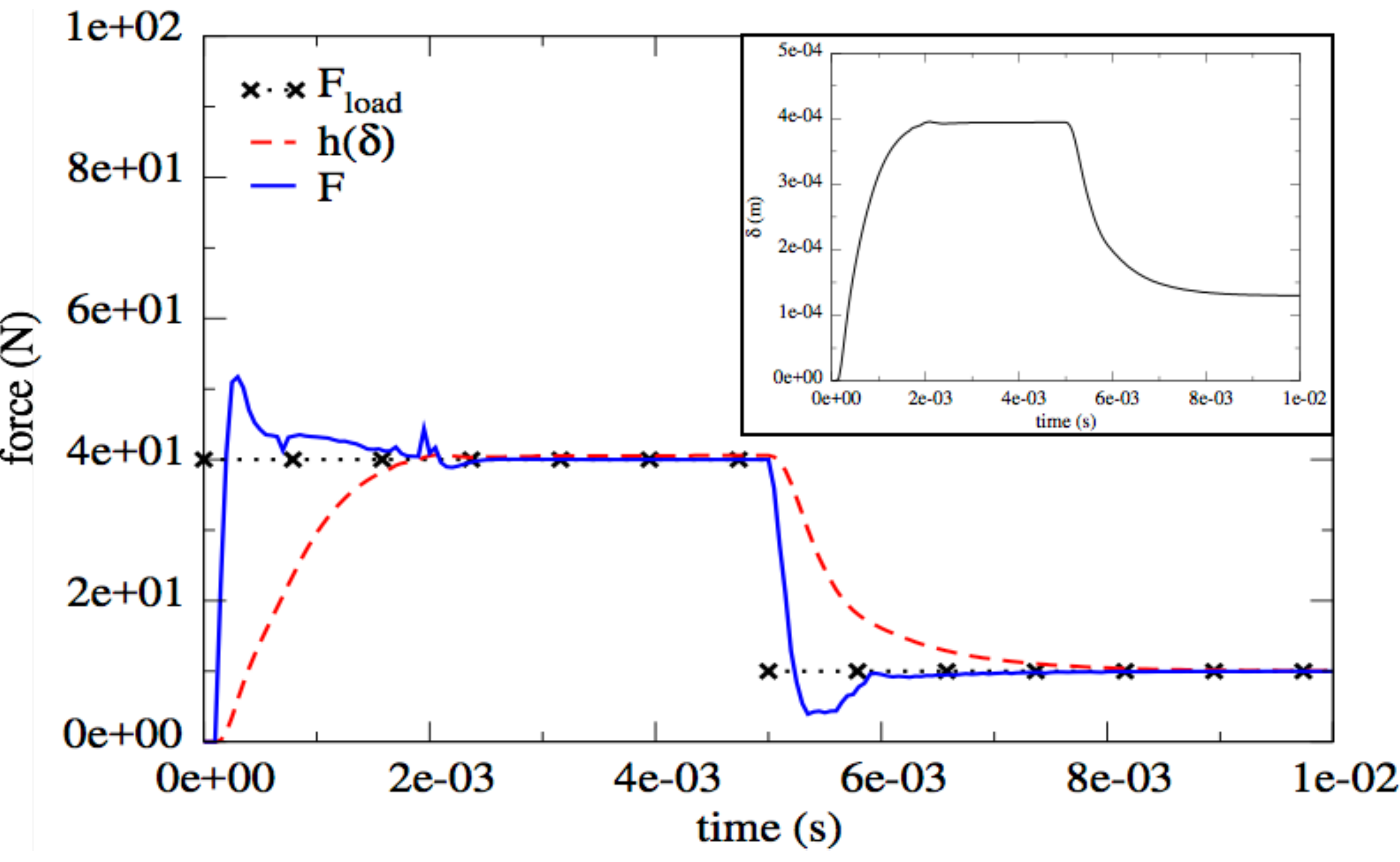}
}
\caption{A dry numerical test of the simplified DEM-Rubber coupling. The left figure shows the configuration 
of the test where the loading/unloading force $F_{load}$ is applied directly to the valve. Color corresponds 
to equivalent shear strain magnitude $\varepsilon_q$ in the rubber. Spheres are held fixed on the valve seat. 
The right figure shows loading/unloading force $F_{load}$, $h(\delta)$ and the net reaction force $F$ 
of spheres on rubber. The inset shows the rubber deformation $\delta$ as a function of time.}
\label{fig:RubberLoadUnloadTest}
\end{figure*}

The test of the rubber coupling and rubber deformation is performed running a loading/unloading test 
without fluid. A force (loading/unloading) $F_{load}$ is directly applied on the valve which presses the rubber
into contact with twelve frozen spheres (see Fig.~\ref{fig:RubberLoadUnloadTest}). Two phases are considered: a loading 
phase with $F_{load} = 40\ N$ then an unloading phase with $F_{load} = 10\ N$. We use a frictionless contact type 
between the rubber and spheres where normal stiffness is set to $1\mbox{e+}05\ N/m$ and no normal damping is used to 
insure that all dissipation comes from internal rubber damping. The rubber coupling parameters 
(Eq.~\ref{eqn:rubber_eqn_2}) are set to $\nu = 80\ N\cdot s/m$ and  $\eta = 10\ N\cdot s/m$. The valve density is set 
to 7850  $kg/m^3$ and the rubber density to $1200\ kg/m^3$. The time step is set to $1\mbox{e-}07\ s$. 
Fig.~\ref{fig:RubberLoadUnloadTest} shows the loading/unloading force, the reaction force $F$ of spheres on 
the rubber, the displacement $\delta$ (right inset) and the corresponding force $h(\delta)$.  The agreement of $F_{load}$, $h$, and $F$ after a relaxation time verifies the coupling.

\begin{figure}[!ht]
\centerline{
\includegraphics[width=1.\textwidth]{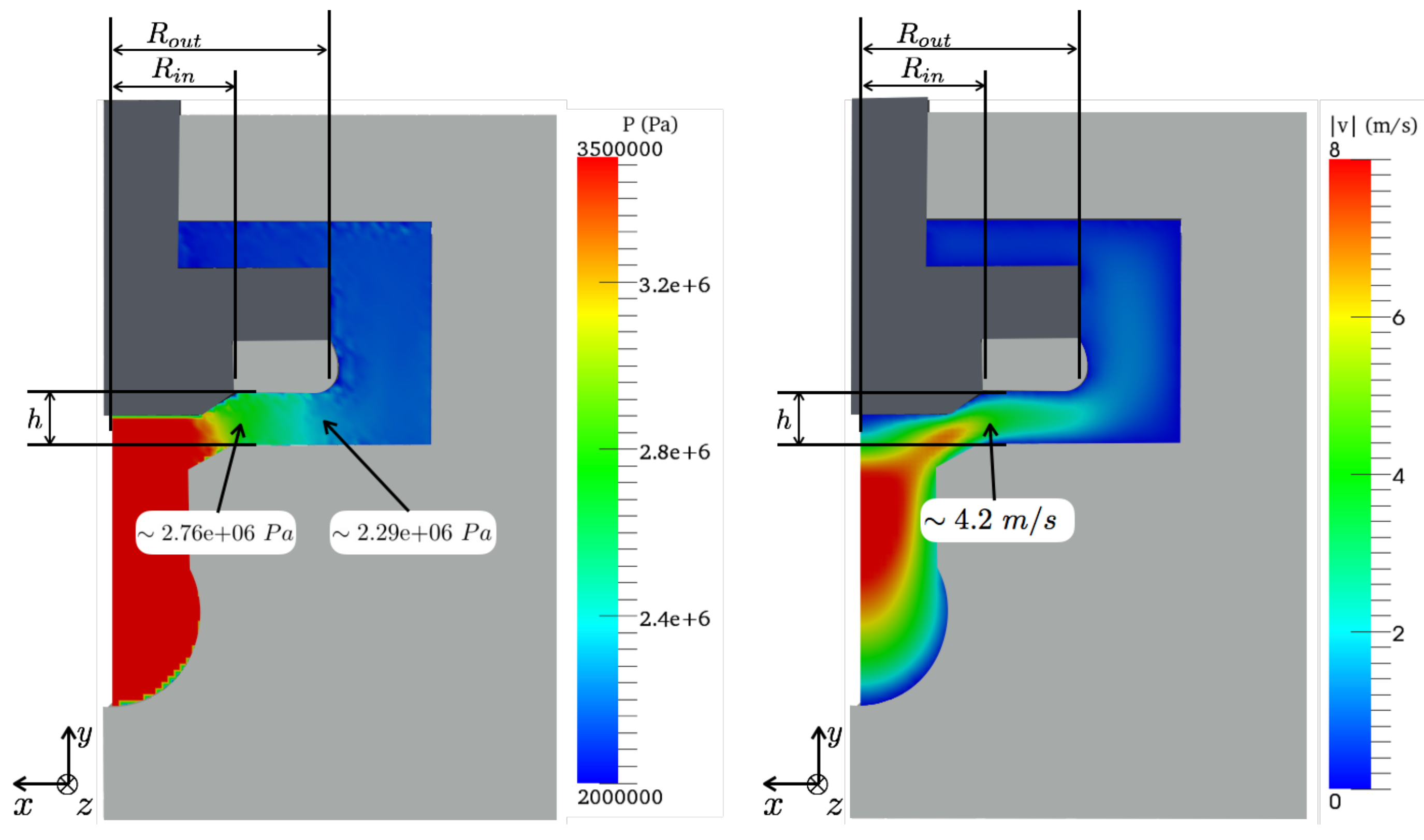}
}
\caption{Fluid flow and pressure in the rubber channel. On the left we show the fluid pressure and on the right, the velocity magnitude.}
\label{fig:channel_velocity_pressure}
\end{figure}

\begin{figure*}[!ht]
\centerline{
\includegraphics[width=.5\textwidth]{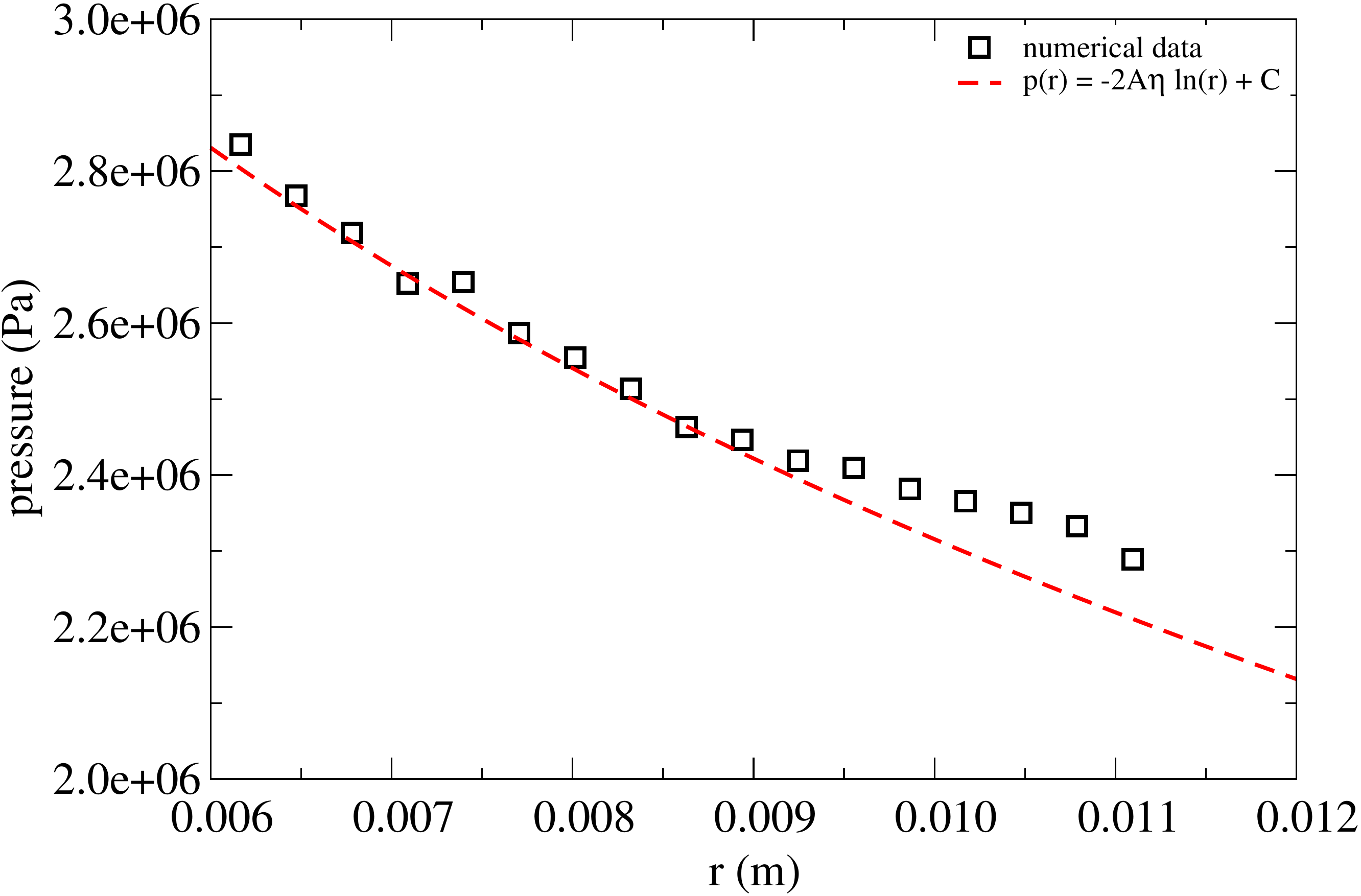}
\includegraphics[width=.45\textwidth]{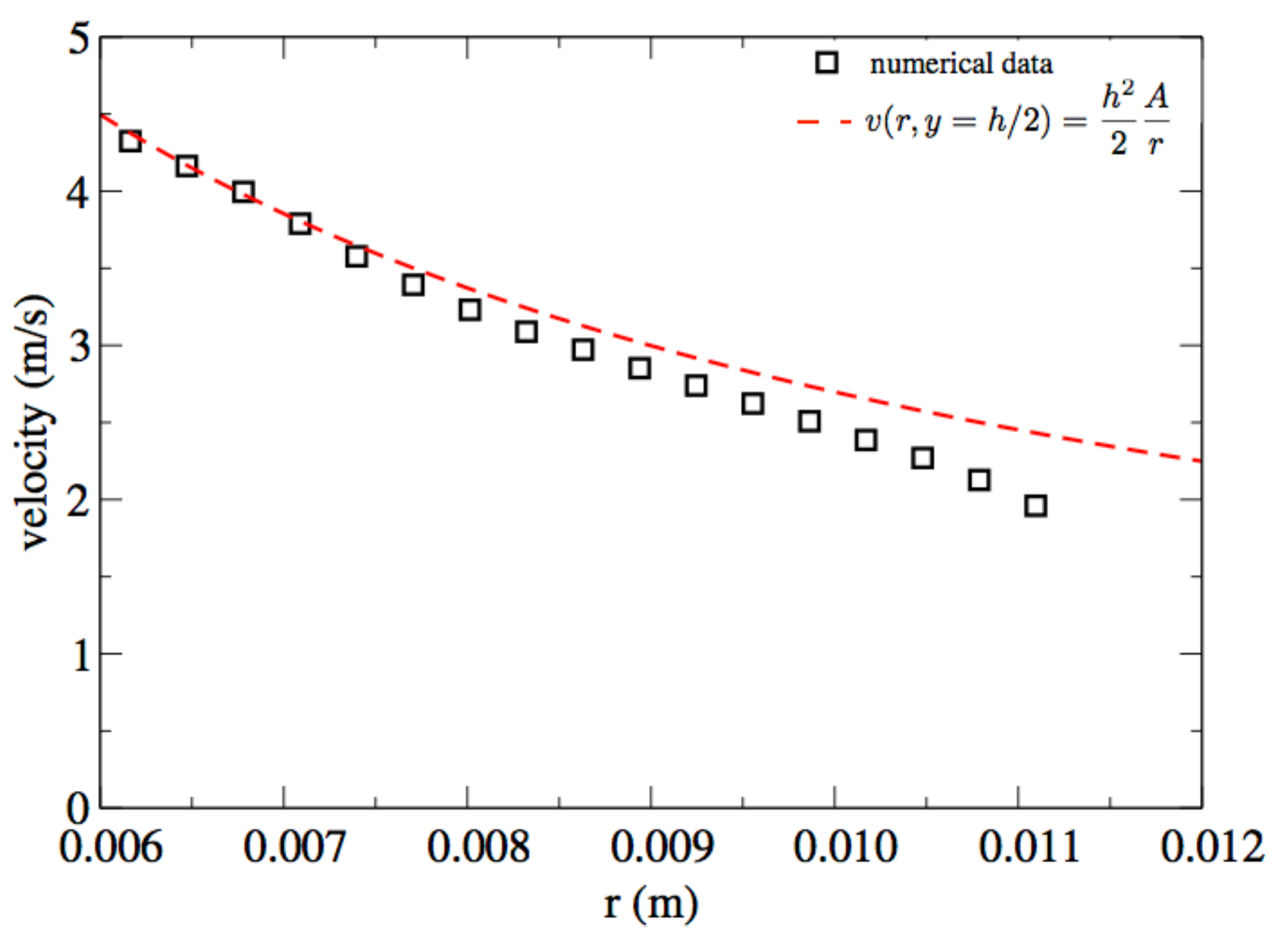}
}
\caption{Numerical versus theoretical comparison for pressure (left) and velocity (right) in the rubber channel. 
$A\sim 1.59\mbox{e+}04\ s^{-1}$, $y=h/2\simeq0.0013\ m$, $C\sim -2.33\mbox{e+}06\ Pa$.}
\label{fig:channel_velocity_pressure_vs_r}
\end{figure*}

The last test is focused on verifying the fluid LBM simulation by comparing flow of fluid in the rubber channel of 
a pressure-valve assembly against an analytical flow solution (recall Fig.~\ref{fig:opned_closed}). We can run a simulation where the fluid viscosity is large, such that the flow in the channel will be in the Stokes limit.  To aid in calculating an analytical solution, we treat the flow as radially directed and assert the lubrication approximation.  In view of Fig.~\ref{fig:opned_closed} for the definition of the $y$ direction, we obtain the following system of equations, which includes momentum and mass balance under the lubrication limit:


\begin{equation}
   \begin{array}{l}
	\frac{1}{r}\frac{\partial}{\partial r}\left(r v_r \right) = 0, \ \ \  \frac{\partial p}{\partial y}=0, \\
	\eta_{_f}  \frac{\partial^2 v_r}{\partial y^2} = \frac{\partial p}{\partial r  }.
   \end{array}
   \label{eq:channelgoveq1} 
\end{equation}
This is solved by $v(r, y) = \frac{A}{r} y(h-y)$ 
where $A$ is an undetermined constant and  $p(r) = -2A\eta_{_f} \ln(r) + C$ where $C$ is a constant and $A$ is the same as in the velocity field 
equation. From the $p(r)$ equation, the pressure difference $\Delta p$ between $r = R_{in}$ and outer at 
$r = R_{out}$ (see Fig.~\ref{fig:channel_velocity_pressure}) is given by: $\Delta p = -2A\eta_{_f}\ln (R_{out}/R_{in})$. Using $y=h/2 = 0.0026\ m$, 
$R_{in} = 0.0065\ m$ and $R_{out} = 0.0110\ m$, we find $v(R_{in}, h/2) \simeq 4.16\ m/s$ (see Fig.~\ref{fig:channel_velocity_pressure} left) 
from our numerical data, giving $A \sim 1.59\mbox{e+}04\ s^{-1}$. Hence, the predicted pressure difference is $\Delta p \simeq -5.3\mbox{e+}05\ Pa$ 
which is quite close to the obtained pressure difference from the simulation 
($-5.7\mbox{e+}05\ Pa$, see Fig.~\ref{fig:channel_velocity_pressure} right). Fig.~\ref{fig:channel_velocity_pressure_vs_r} 
shows the analytical solution as a function of $r$ in comparison with the numerical data and agreement is found. In this test, the fluid density is $\rho_{_f} = 1000\ kg/m^3$ and dynamic viscosity is 
$\eta_{_f} = 3.16\ Pa\cdot s$. The valve density is set to $7850\ kg/m^3$ and the rubber density to $1200\ kg/m^3$.  We use ZIEB on the input section (see Fig.~\ref{fig:system_cofig}) 
with a virtual piston velocity of $v_p = 6\ m/s$, and we apply a constant 
pressure of $P_{out} = 1.04\mbox{e+}04\ Pa$ at the output section (see Fig.~\ref{fig:system_cofig}).

\section{Examples}
\label{sec:num_procedure}

\begin{figure}[!ht]
\centerline{
\includegraphics[width=1.\textwidth]{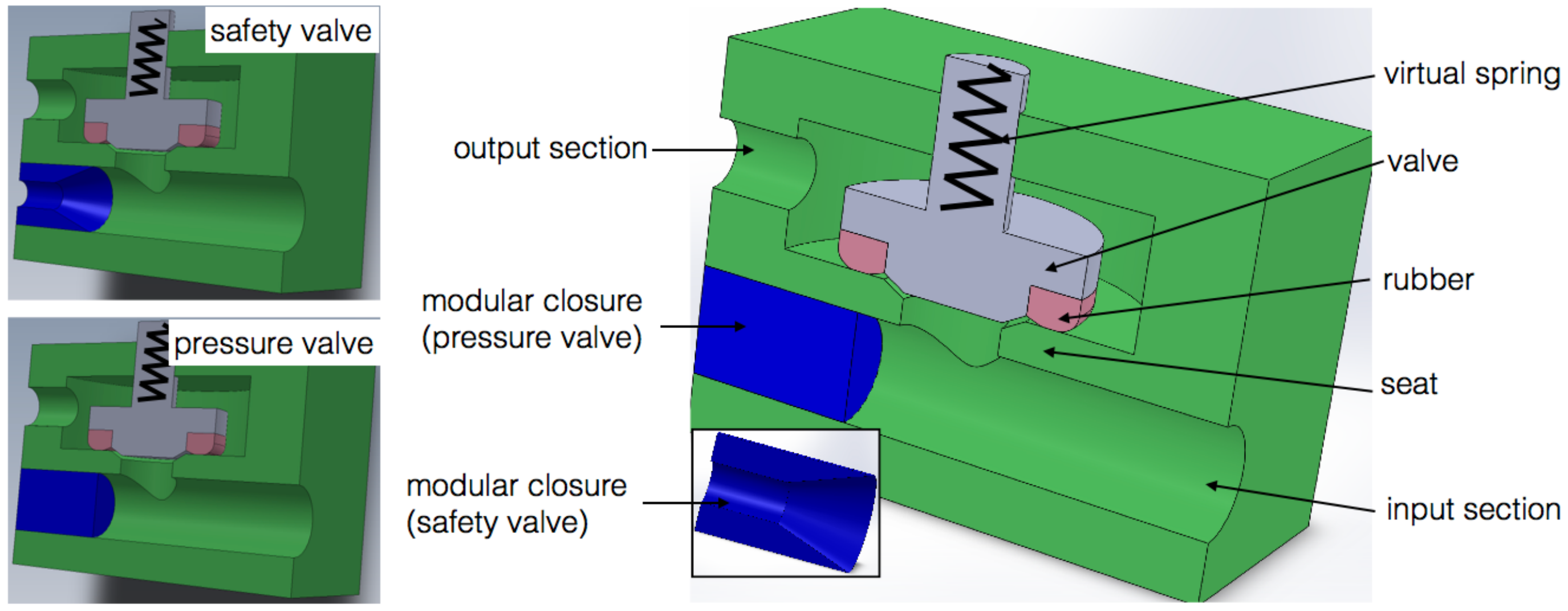}
}
\caption{Illustration of different parts of the safety and pressure valve.}
\label{fig:system_cofig}
\end{figure}

We present numerical examples utilizing the valve geometries presented in  Fig.~\ref{fig:system_cofig}. The two systems 
are used to mimic a safety and pressure valve. The different parts and their corresponding names are presented 
in Fig.~\ref{fig:system_cofig}. Their dimensions are given in Appendix~\ref{app:dimension}.

The valve is spring-loaded from above. Initially, the valve is closed due to force from the spring (and prescribed overpressure above in the pressure-valve geometry). The spring's force-displacement 
relation is chosen to be non-linear and is expressed as follows: 
$F_s =  k_v \delta_0 + k_v (a/\frac \pi 2)\tan(\frac {\pi}{2} \frac \delta a)$; $k_v$ is the stiffness, 
$\delta_0 = 0.092\ m$ is the preload spring displacement, $\delta$ is the load displacement and $a = 0.0045\ m$ 
is the spring maximum compression. The fluid region above the valve begins unpressurized in the safety-valve 
case and pressurized by the selected $P_{out}$ in the pressure-valve case. For each simulation, both for safety 
and pressure valve, we start by introducing fluid through the input section beneath valve domain using the 
ZIEB technique at constant virtual piston velocity $v_p$. When the beneath valve domain reaches a large enough 
pressure, it will overcome the spring-load (and possible top pressure) on the valve to open it. We continue 
displacing the virtual piston and then turn off $v_p$ when we assume that the flow has reached a steady state. 
We then wait until the valve is closed. We check 
the behavior of the valve systems with and without particles but the closure phenomena is only investigated 
for the case where we have particles. 

In the presence of particles, we start with a constant packing fraction in the domain beneath the valve corresponding 
to the imposed packing fraction $\phi$ at the input section. During the simulation, each particle that exits the 
simulated domain is removed and each new particle introduced at the input section starts with the fluid input velocity 
as its initial velocity. To control the input particle flow, we insert grains to ensure that the input particle packing fraction is 
constant within a width of $0.00425\ m$ from the interface with the virtual domain. We introduce/remove particles in the simulated system every 50 steps.

Physical parameters involved in the valve problem are displayed in Tab.~\ref{tab:summarizedparamvalvebh}), which include the geometry of the valve system (e.g. safety-valve or pressure-valve), each of which has fixed size dimensions (see Appendix).  For all tests, we fix the solid density of particles $\rho_{_s}=2500\ kg/m^3$, (pressure-free) fluid density $\rho_f=1000\ kg/m^3$
(small-strain) rubber shear modulus $G_{r}=3.0\mbox{e+}05\ Pa$ and bulk modulus $K_{_r}=8.0\mbox{e+}06\ Pa$, rubber damping $\nu=80\ N\cdot s/m$, rubber$+$valve 
mass $M_{_{vr}}=9.2\mbox{e-}03\ kg$, and valve spring stiffness $k_{_v}=625 N/m$.

Since mono-disperse particles may induce a crystallisation phenomena at higher packing, a random size distribution 
is used uniformly between $d_{min}$ and $d_{max}$. The distribution may be described by a mean particle size $d=(d_{min}+d_{max})/2$ and polydispersity $\Delta d = d_{max}/d_{min}$.

\begin{table}[!ht]
\caption{Parameters.}
\label{tab:summarizedparamvalvebh}

\begin{tabular}{llllllll}
\hline\noalign{\smallskip}
Particles & & Valve + Rubber & & Fluid & & System \\
\noalign{\smallskip}\hline\noalign{\smallskip}
mean diameter		& $[d]$		& mass				& $[M_{_{vr}}]$	& dynamic viscosity	& $[\eta_{_f}]$ &	system geometry &  $[\text{geo}]$\\
solid density			&$[\rho_{_s}]$	& rubber shear modulus	& $[G_{r}]$	& pressure-free density			& $[\rho_{_f}]$ & system size & $[r]$	\\
polydispersity		&$[\Delta d]$	& rubber bulk modulus	& $[K_{_r}]$	& output pressure	& $[P_{out}]$ & piston speed & $[v_p]$		\\
input packing fraction		&$[\phi]$		& valve spring stiffness	& $[k_{_v}]$	&		&					\\
& & rubber damping & $[\nu]$ & & & &\\
\noalign{\smallskip}\hline
\end{tabular}
\end{table}

To generalize the valve dynamics and flow behavior, we choose the natural units of our system to be the input section radius 
$[L] = r$ for length, time $[T] = \sqrt{\rho_{_f} r^3/k_{_v}}$ and mass $[M] = \rho_{_f} r^3$. From these units, a dimensionless parametric space is represented by: \\ 
$$\left\{\text{geo},  \frac{P_{out}}{k_{_v}}, v_p\sqrt{\frac{k_{_v}}{\rho_{_f}r}}, \frac d r, \Delta d, \phi, \frac{\rho_{_s}}{\rho_{_f}}, 
 \frac{\eta_{_f}}{k_{_v}\rho_{_f}r}, \frac{G_{r} r}{k_{_v}}, \frac{K_{r} r}{k_{_v}}, \frac{M_{_{vr}}}{\rho_{_f} r^3} \right\}$$ \\
where $\text{geo}$ is the system geometry. Taking into account the fixed parameters, the dimensionless parametric 
space we will explore is described by the following groups:\\
$$\left\{\text{geo}, \frac{P_{out}}{k_{_v}}, v_p\sqrt{\frac{k_{_v}}{\rho_{_f}r}}, \frac d r, \Delta d, \phi, \frac{\eta_{_f}}{k_{_v}\rho_{_f}r}\right\}$$ \\
The second group is only relevant to pressure valves and the latter five can be independently controlled 
through the selection of $v_p, d, \Delta d, \phi$, and $\eta_f$.

The parameters for all tests are summarized in Tab.~\ref{tab:spacerangeopenclosednp} and 
Tab.~\ref{tab:spacerangeopenclosedyp}.  As indicated in the Tables, the tests are conducted in order to observe 
the dependence of the valve behavior on each of $\eta_f, v_p, d, \Delta d,$ and $\phi$ independently;  for each 
variable, a sequence of tests is performed where it is varied over a range while the others are held fixed.

\begin{table}[!ht]
\caption{Range of parameters investigated for safety and pressure valve simulations without particles. 
All units are in $[\text{kg}] , [\text{m}], [\text{s}]$.}
\label{tab:spacerangeopenclosednp}

\begin{tabular}{c|c}
	\hline\noalign{\smallskip}
	geo$=$safety valve & geo$=$pressure valve	\\
	\begin{tabular}{lc|c}
		\hline\noalign{\smallskip}
		 & range & fixed		\\
		\noalign{\smallskip}\hline\noalign{\smallskip}
		$\eta_{_f}$		& 1e-03 to 3.16e+01	&	
		\begin{tabular}{ll}
			name & value 			\\
			\noalign{\smallskip}\hline\noalign{\smallskip}

			$P_{out}$		& 0		\\
			$v_p$		& 12.5	\\
			$d$			& 0		\\
			$\Delta d$		& 0		\\
			$\phi$		& 0		\\
		\end{tabular}				\\
		\noalign{\smallskip}\hline

		$v_p$		& 1 to 12.5	&	
		\begin{tabular}{ll}

			$P_{out}$		& 0		\\
			$d$			& 0		\\
			$\Delta d$		& 0		\\
			$\phi$		& 0		\\
			$\eta_{_f}$	& 1e-03	\\
		\end{tabular}				\\
	\end{tabular} &

	\begin{tabular}{lc|c}
		\hline\noalign{\smallskip}
		 & range & fixed 		\\
		\noalign{\smallskip}\hline\noalign{\smallskip}
		$\eta_{_f}$		& 1e-03 to 3.16e+01	&	
		\begin{tabular}{ll}
			name & value 			\\
			\noalign{\smallskip}\hline\noalign{\smallskip}

			$P_{out}$		& 10416	\\
			$v_p$		& 6		\\
			$d$			& 0		\\
			$\Delta d$		& 0		\\
			$\phi$		& 0		\\
		\end{tabular}				\\
		\noalign{\smallskip}\hline

		$v_p$		& 1 to 12	&	
		\begin{tabular}{ll}

			$P_{out}$		& 10416	\\
			$d$			& 0		\\
			$\Delta d$		& 0		\\
			$\phi$		& 0		\\
			$\eta_{_f}$	& 1e-03	\\
		\end{tabular}				\\
	\end{tabular}					\\
	\noalign{\smallskip}\hline
\end{tabular}
\end{table}
\begin{table}[!ht]
\caption{Range of parameters investigated for safety and pressure valve for simulations with particles. 
All units are in $[\text{kg}] , [\text{m}], [\text{s}]$.}

\label{tab:spacerangeopenclosedyp}
\begin{tabular}{c|c}
	\hline\noalign{\smallskip}
	geo$=$safety valve & geo$=$pressure valve	\\
	\begin{tabular}{lc|c}
		\hline\noalign{\smallskip}
		 & range & fixed		\\
		\noalign{\smallskip}\hline\noalign{\smallskip}
		$v_p$		& 1 to 12.5	&	
		\begin{tabular}{ll}

			$P_{out}$		& 0		\\
			$d$			& 0.8e-03	\\
			$\Delta d$		& 1.2		\\
			$\phi$		& 0.084	\\
			$\eta_{_f}$	& 1e-03	\\
		\end{tabular}				\\
		\noalign{\smallskip}\hline

		$d$			& 0.8e-03 to 1.5e-03	&	
		\begin{tabular}{ll}

			$P_{out}$		& 0		\\
			$v_p$		& 12.5	\\
			$\Delta d$		& 1.2		\\
			$\phi$		& 0.053	\\
			$\eta_{_f}$	& 1e-03	\\
		\end{tabular}				\\
		\noalign{\smallskip}\hline

		$\Delta d$		& 1.1 to 1.5	&	
		\begin{tabular}{ll}

			$P_{out}$		& 0		\\
			$v_p$		& 12.5	\\
			$d$			& 0.8e-03	\\
			$\phi$		& 0.053	\\
			$\eta_{_f}$	& 1e-03	\\
		\end{tabular}				\\
		\noalign{\smallskip}\hline

		$\phi$			& 0.026 to 0.128	&	
		\begin{tabular}{ll}

			$P_{out}$		& 0		\\
			$v_p$		& 12.5	\\
			$\Delta d$		& 1.2		\\
			$d$			& 0.8e-03	\\
			$\eta_{_f}$	& 1e-03	\\
		\end{tabular}				\\
	\end{tabular} &

	\begin{tabular}{lc|c}
		\hline\noalign{\smallskip}
		 & range & fixed 		\\
		\noalign{\smallskip}\hline\noalign{\smallskip}
		$v_p$		& 1 to 12	&	
		\begin{tabular}{ll}

			$P_{out}$		& 10416	\\
			$d$			& 0.8e-03	\\
			$\Delta d$		& 1.2		\\
			$\phi$		& 0.067	\\
			$\eta_{_f}$	& 1e-03	\\
		\end{tabular}				\\
		\noalign{\smallskip}\hline

		$d$			& 0.8e-03 to 1.4e-03	&	
		\begin{tabular}{ll}

			$P_{out}$		& 10416	\\
			$v_p$		& 6		\\
			$\Delta d$		& 1.2		\\
			$\phi$		& 0.053	\\
			$\eta_{_f}$	& 1e-03	\\
		\end{tabular}				\\
		\noalign{\smallskip}\hline

		$\Delta d$		& 1.1 to 1.5	&	
		\begin{tabular}{ll}

			$P_{out}$		& 10416	\\
			$v_p$		& 6		\\
			$d$			& 0.8e-03	\\
			$\phi$		& 0.053	\\
			$\eta_{_f}$	& 1e-03	\\
		\end{tabular}				\\
		\noalign{\smallskip}\hline

		$\phi$			& 0.026 to 0.117	&	
		\begin{tabular}{ll}

			$P_{out}$		& 10416	\\
			$v_p$		& 6		\\
			$\Delta d$		& 1.2		\\
			$d$			& 0.8e-03	\\
			$\eta_{_f}$	& 1e-03	\\
		\end{tabular}				\\
	\end{tabular}					\\
	\noalign{\smallskip}\hline
\end{tabular}
\end{table}

The contact model (DEM solver and DEM-Rubber coupling), fluid turbulence model (LES) and numerical parameters are 
displayed in Tab.~\ref{tab:contactcouplingnumericalparameter}

\begin{table}[!ht]
\caption{The contact parameters (DEM solver and DEM-Rubber coupling), fluid turbulence model (LES) and numerical 
parameters. All units are in $[\text{kg}] , [\text{m}], [\text{s}]$.}
\label{tab:contactcouplingnumericalparameter}

\begin{multicols}{2}

\begin{tabular}{lll}
\noalign{\smallskip}\hline
stiffness & normal & tangantial \\
\noalign{\smallskip}\hline\noalign{\smallskip}
seat-valve			& 1e+11	&	0 			\\
seat-rubber		    & 1e+06	&	0 			\\
seat-particle		& 1e+07	&	0.8e+07 		\\
valve-particle		& 1e+07	&	0.8e+07 		\\
rubber-particle		& 1e+05	&	0.8e+05 		\\
particle-particle	& 1e+07	&	0.8e+07 		\\
\noalign{\smallskip}\hline
\end{tabular}
\begin{tabular}{ll}
\noalign{\smallskip}\hline
friction coefficient &  \\
\noalign{\smallskip}\hline\noalign{\smallskip}
seat-valve			& 0		\\
seat-rubber		    & 0		\\
seat-particle		& 0.4		\\
valve-particle		& 0.4		\\
rubber-particle		& 0.4		\\
particle-particle	& 0.4		\\
\noalign{\smallskip}\hline
\end{tabular}
\begin{tabular}{lll}
\noalign{\smallskip}\hline
damping & normal & tangantial \\
\noalign{\smallskip}\hline\noalign{\smallskip}
seat-valve			& 1e+03	&	0 		\\
seat-rubber		    & 0		&	0 		\\
seat-particle		& 3		&	0 		\\
valve-particle		& 3		&	0 		\\
rubber-particle		& 0		&	0 		\\
particle-particle	& 3		&	0 		\\
\noalign{\smallskip}\hline
\end{tabular}
\begin{tabular}{ll}
\noalign{\smallskip}\hline
rubber and numerical parameters \\
\noalign{\smallskip}\hline\noalign{\smallskip}
Smagorinsky constant		            & 0.4000  \\
lattice speed				            & 2.5e+03 \\
fluid space discretization		        & 3.0e-04 \\
DEM time step				            & 5.0e-08 \\
numerical rubber convergence ($\eta$)	& 1.0e+01 \\
                                    	&         \\
\noalign{\smallskip}\hline
\end{tabular}

\end{multicols}
\end{table}

\subsection{Pressure valve lift behavior}
\label{sec:num_data_discuss}

In this section, we discuss the effect of fluid viscosity, piston velocity, and input packing fraction 
on the opening, steady flow, and closure  behavior of the valve for a pressure valve configuration. 
As shown in Tab. ~\ref{tab:spacerangeopenclosednp} and  Tab.~\ref{tab:spacerangeopenclosedyp}, when varying 
a parameter of interest, we fix the others to a control set taken from
$v_p = 6\ m/s$, $\eta_{_f} = 0.001\ Pa\cdot s$, $\phi = 0.053$ and $\Delta d = 1.2$.  We will focus our analysis 
on the pressure valve and will give a brief analysis of the results for the safety valve in 
Sec.~\ref{sec:SafetyValveLift}.

\subsubsection*{Valve opening phase}

\begin{figure*}[!ht]
\centerline{
\includegraphics[width=.5\textwidth]{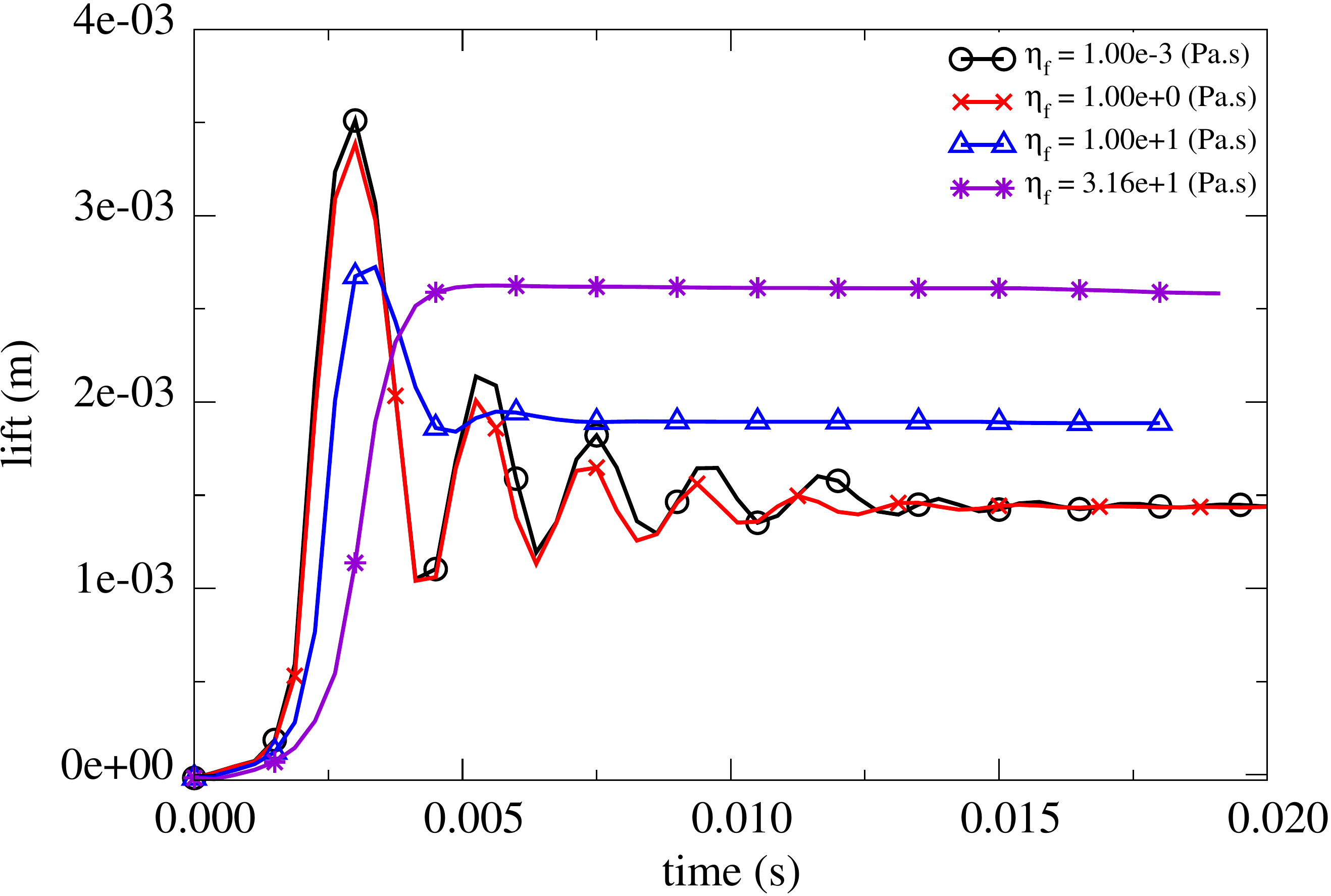}
\includegraphics[width=.5\textwidth]{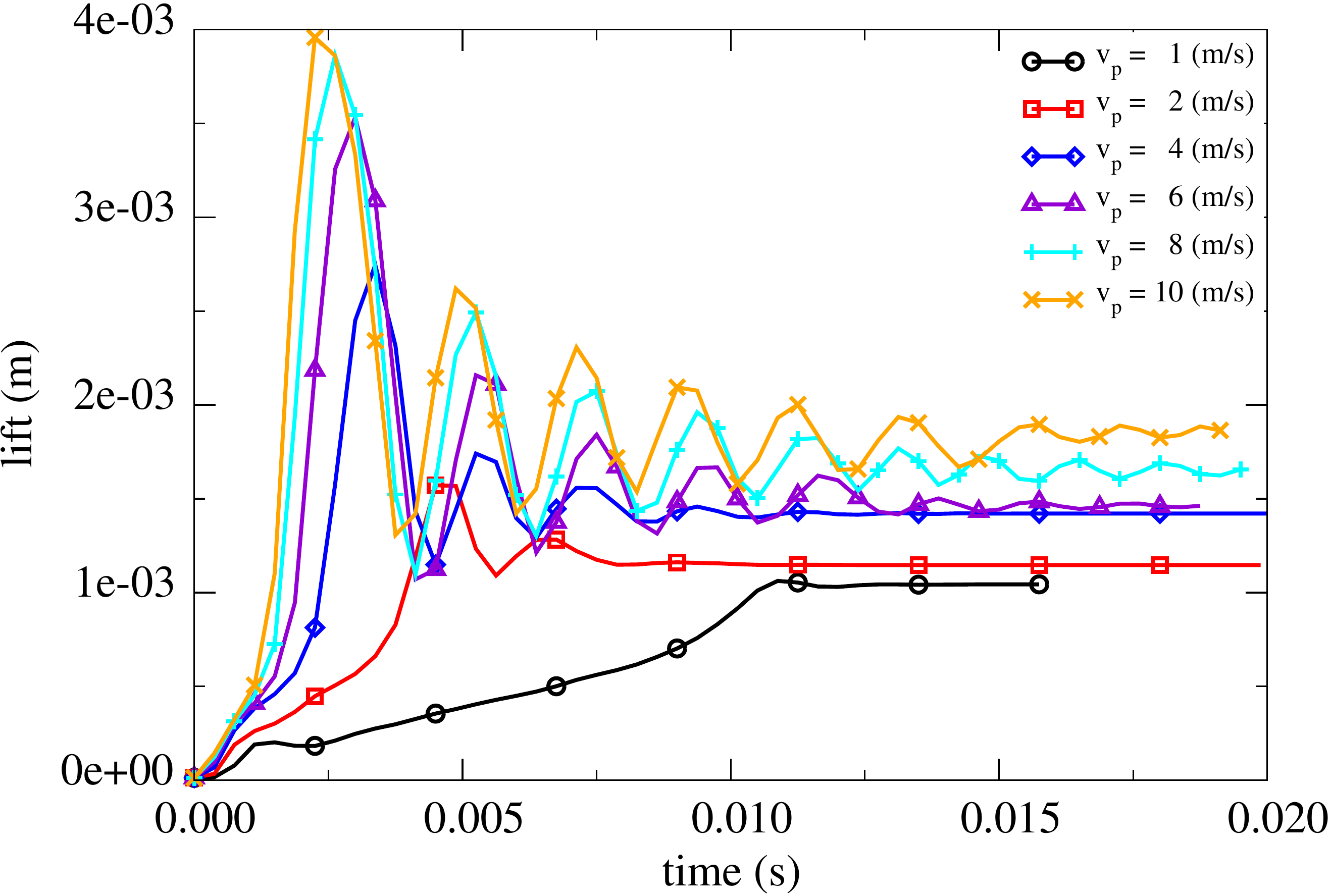}
}
\caption{Valve lift as function of time for different fluid viscosity (left) and different virtual piston velocity 
(right) (without particles).}
\label{fig:np_closed_lift_eta_v_xxoxxx}
\end{figure*}

\begin{figure}[!ht]
\centerline{
\includegraphics[width=.75\textwidth]{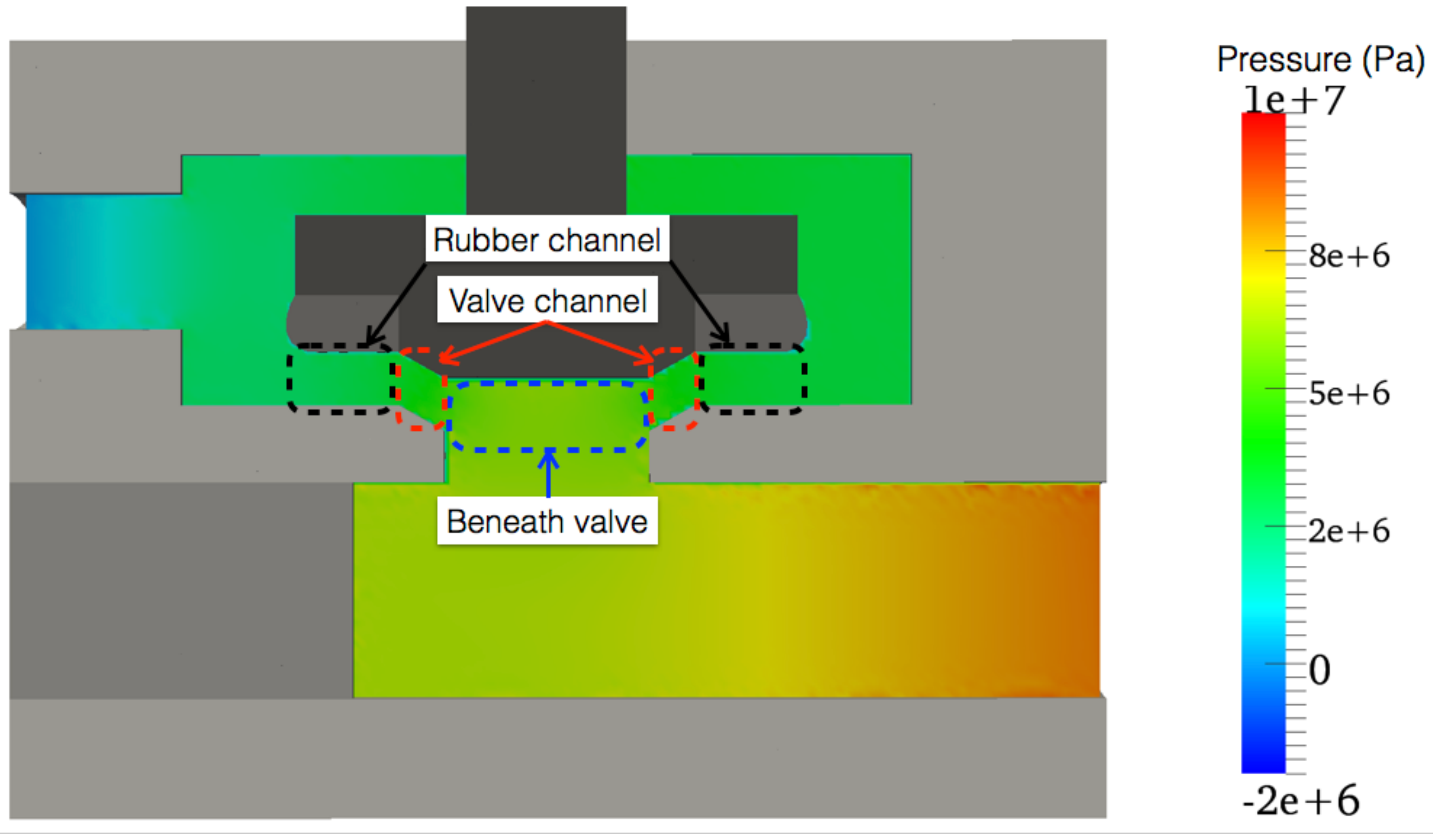}
}
\caption{Rubber channel, valve channel, and beneath valve domain configuration.}
\label{fig:np_closed_prb_conf}
\end{figure}

\begin{figure}[!ht]
\centerline{
\includegraphics[width=1.\textwidth]{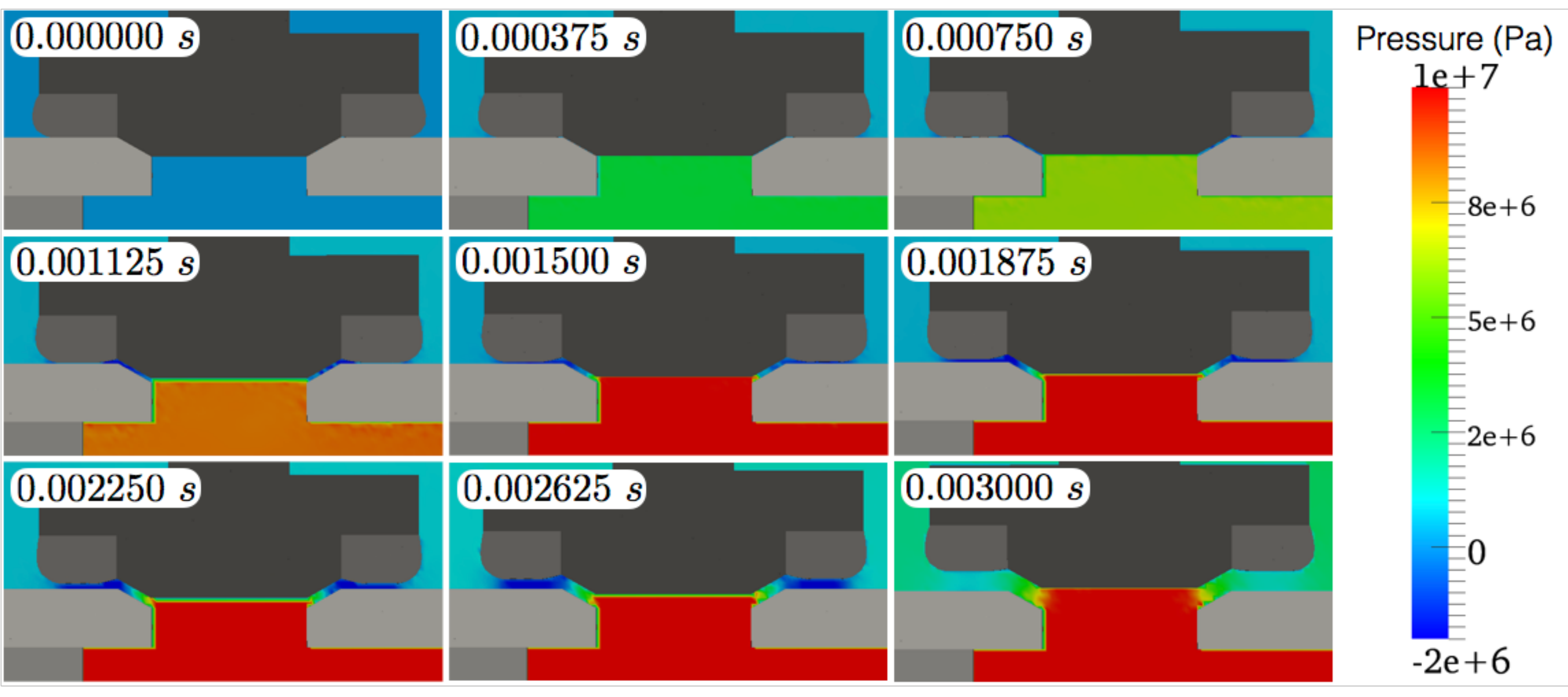}
}
\caption{Several snapshots showing valve  and rubber channel pressure evolution for $\eta_{_f} = 31.6\ Pa s$ and $v_p = 6\ m/s$. 
(without particles).}
\label{fig:np_closed_pressure_BC_BR_vs_t_zoom}
\end{figure}

\begin{figure*}[!ht]
\centerline{
\includegraphics[width=.5\textwidth]{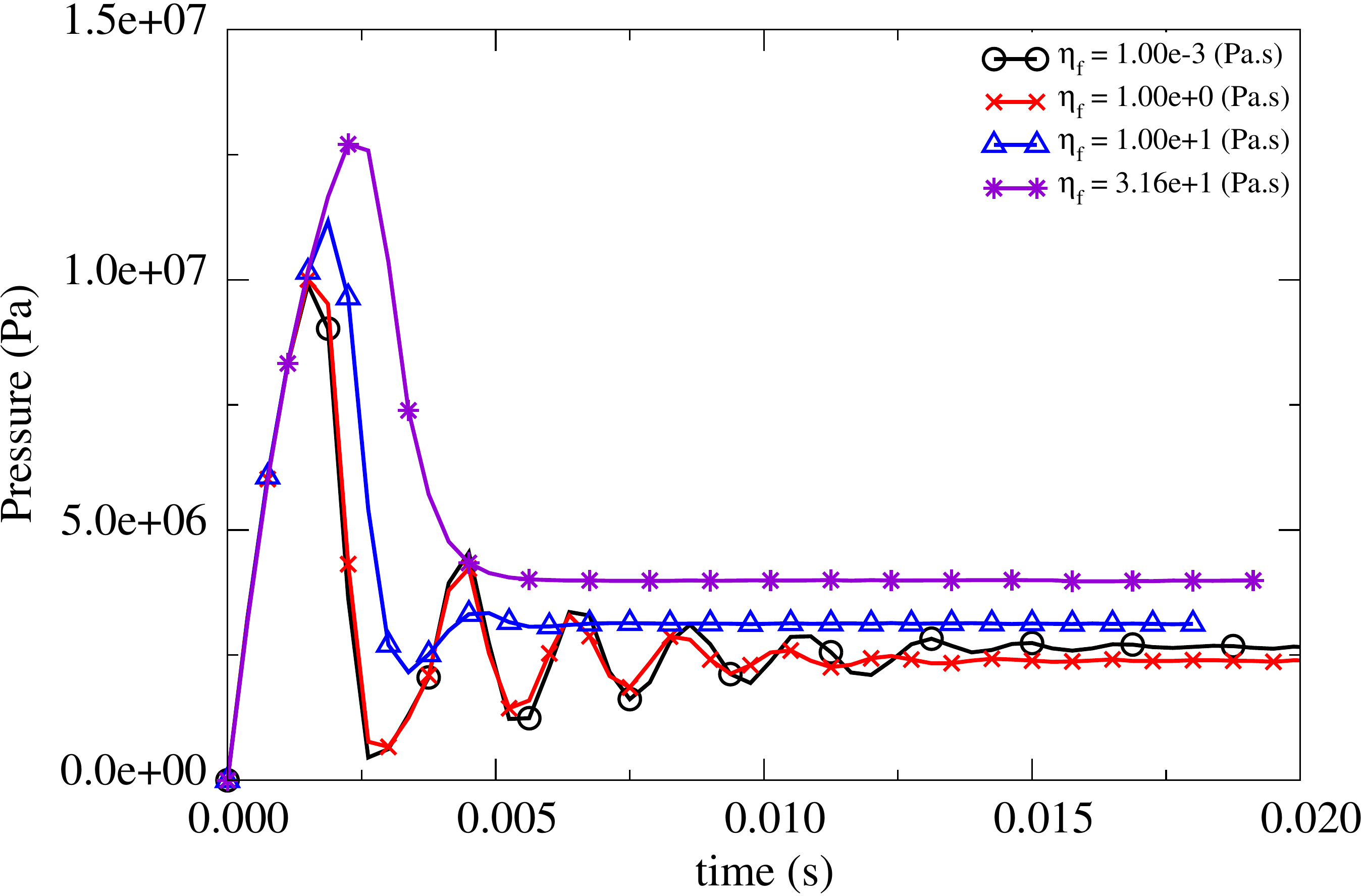}
\includegraphics[width=.5\textwidth]{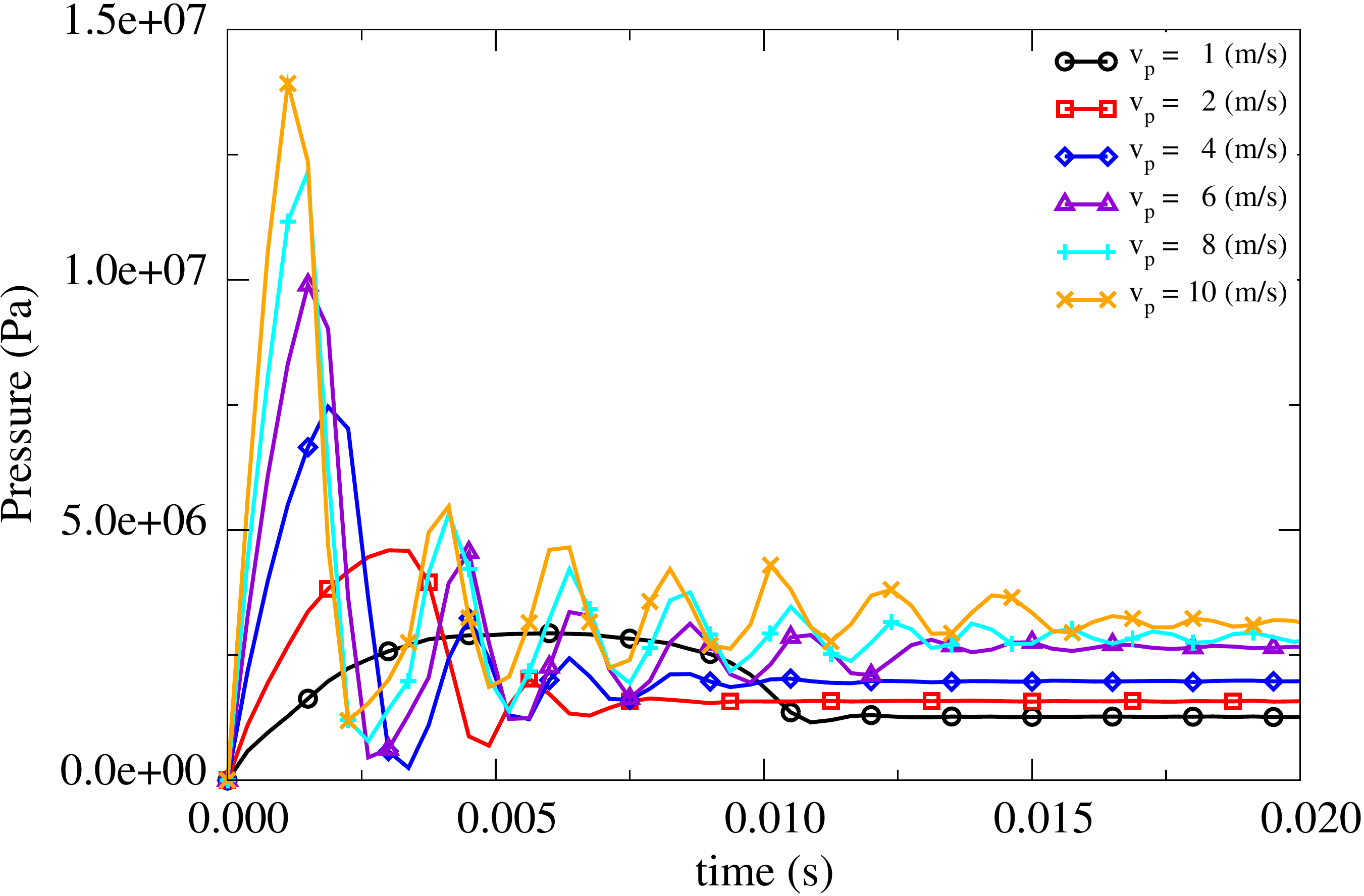}
}
\caption{Beneath valve pressure as a function of time for different fluid viscosity (left) and different piston velocity (right) 
(without particles).}
\label{fig:np_closed_pressure_eta_v_xxoxxx}
\end{figure*}

During the opening phase, we observe a delay between the initiation of the piston and the initiation of valve opening. The effect is not due to packing 
fraction Fig.~\ref{fig:closed_yp_lift_v_xxoxxx_sf_0o040_080_0o125_0o150_0o175} (left), polydispersity 
Fig.~\ref{fig:closed_np_lift_d_xxoxxx} (left) or mean particle diameter Fig.~\ref{fig:closed_np_lift_d_xxoxxx} (right), 
rather, the delay increases with fluid viscosity Fig.~\ref{fig:np_closed_lift_eta_v_xxoxxx} (left), 
and decreases when piston velocity increases Fig.~\ref{fig:np_closed_lift_eta_v_xxoxxx} (right) (simulation without 
particles) and Fig.~\ref{fig:closed_yp_lift_v_xxoxxx_sf_0o040_080_0o125_0o150_0o175} (right) (simulation with particles). 
The lack of dependence of the delay time on particle inputs is because the mean particle diameter is bigger than 
the initial valve lift so it does not modify fluid behavior in the valve-rubber channel (see schematic in 
Fig.~\ref{fig:np_closed_prb_conf}).  The more dominant effect is negative suction pressure, 
which develops in the valve-rubber channel as the valve initially displaces upward, as shown in
Fig.~\ref{fig:np_closed_pressure_BC_BR_vs_t_zoom}.  
The delay increases with increasing viscosity because this increases the suction force due to lubrication effects in the narrow valve-rubber channel. At the same time, in the beneath valve region where the fluid domain is not thin, the pressure is mostly independent 
of viscosity as we observe in Fig.~\ref{fig:np_closed_pressure_eta_v_xxoxxx} (left) where before the first peak 
of the valve lift (Fig.~\ref{fig:np_closed_lift_eta_v_xxoxxx} (left)) the pressure evolution is the same.

Increasing the piston velocity reduces the delay because the suction pressure in the valve-rubber channel is balanced by faster growth of the beneath valve pressure. Figure~\ref{fig:np_closed_pressure_eta_v_xxoxxx} (right) shows that during the pressurization phase, 
the pressure slope increases with piston velocity, as expected.

\subsubsection*{Quasi-steady open valve phase}

\begin{figure*}[!ht]
\centerline{
\includegraphics[width=.5\textwidth]{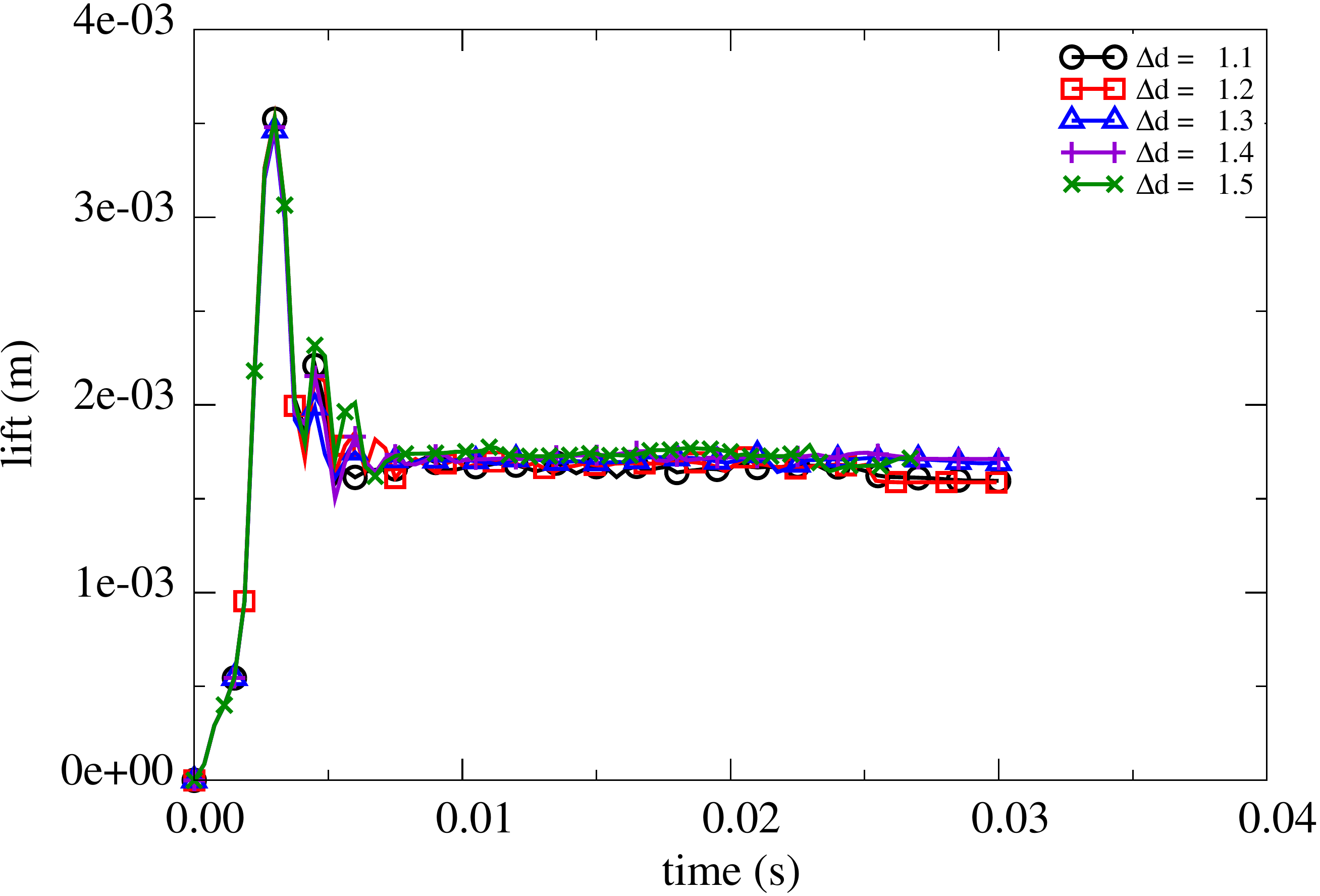}
\includegraphics[width=.5\textwidth]{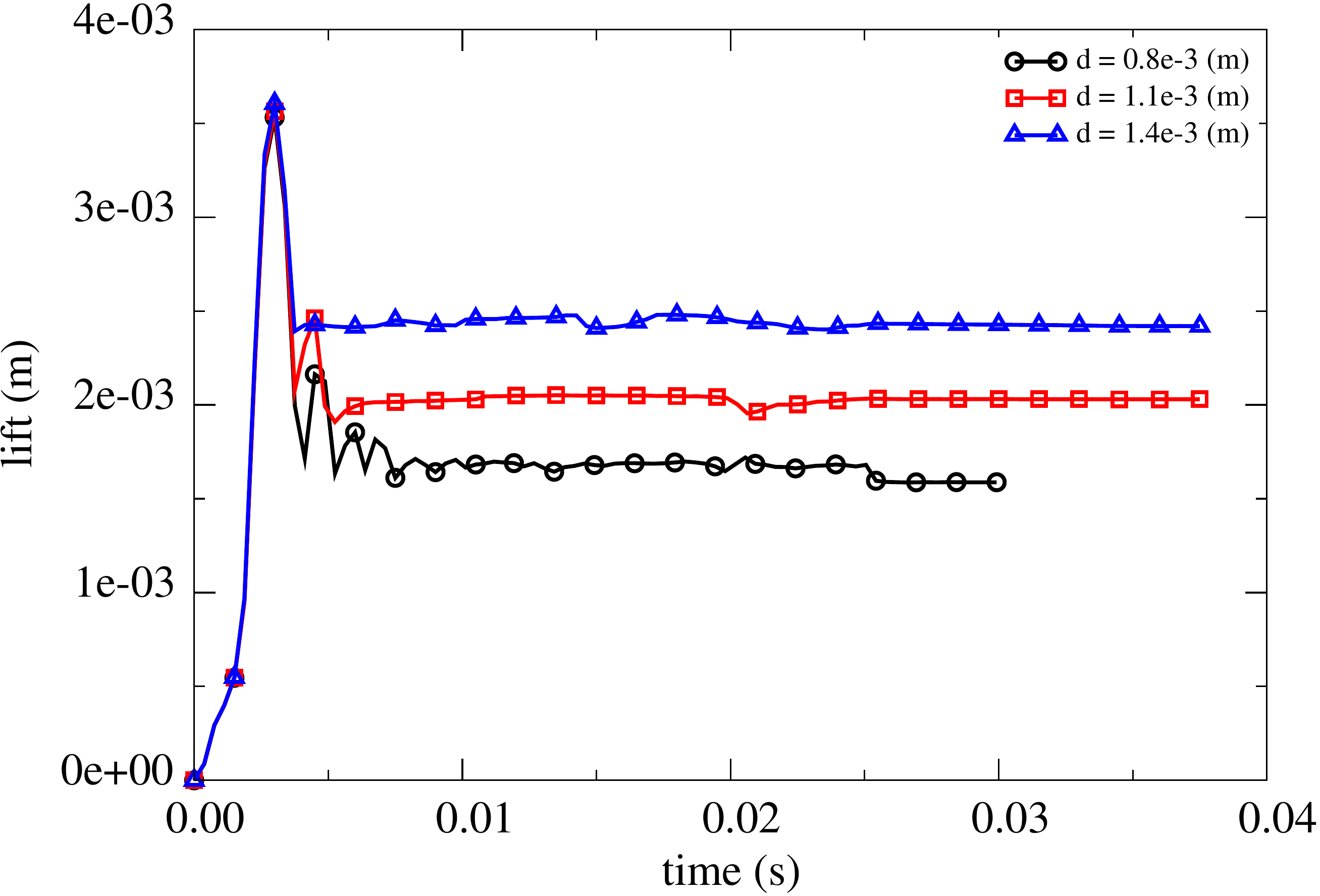}
}
\caption{Valve lift as function of time for different polydispersity (left) and different mean grain diameter (right).}
\label{fig:closed_np_lift_d_xxoxxx}
\end{figure*}

The valve displacement begins to approach a steady value (with or without oscillations) after the first peak in valve lift ($t \simeq 0.010\ s$) and 
ends when the piston motion is stopped at $t=0.025\ s$. As shown in Fig.~\ref{fig:np_closed_lift_eta_v_xxoxxx}, the steady 
lift position, for the simulations without particles, increases with fluid viscosity and piston velocity. 
For lower viscosity, the valve has an `underdamped' response characterized by decaying oscillations, whereas 
for larger viscosities, an `overdamped' response can be seen (see Fig. \ref{fig:np_closed_lift_eta_v_xxoxxx} (left)). 
The presence of particles can modify the valve lift behavior. Figure~\ref{fig:closed_np_lift_d_xxoxxx} (left) shows 
that virtually no effect on valve lift is observed for different polydispersity in the range we tested. Increasing the 
mean diameter of particles, Fig.~\ref{fig:closed_np_lift_d_xxoxxx} (right), increases the steady lift position but 
this appears to be primarily a particle size effect; after the initial upward motion of the valve, the valve descends 
downward and is stopped by a monolayer of mobile particles in the rubber channel, which holds the valve position 
at roughly $1d$ high. Further tests would be needed at higher fixed piston speeds to determine if the valve 
positioning depends more robustly on $d$.

\begin{figure*}[!ht]
\centerline{
\includegraphics[width=.5\textwidth]{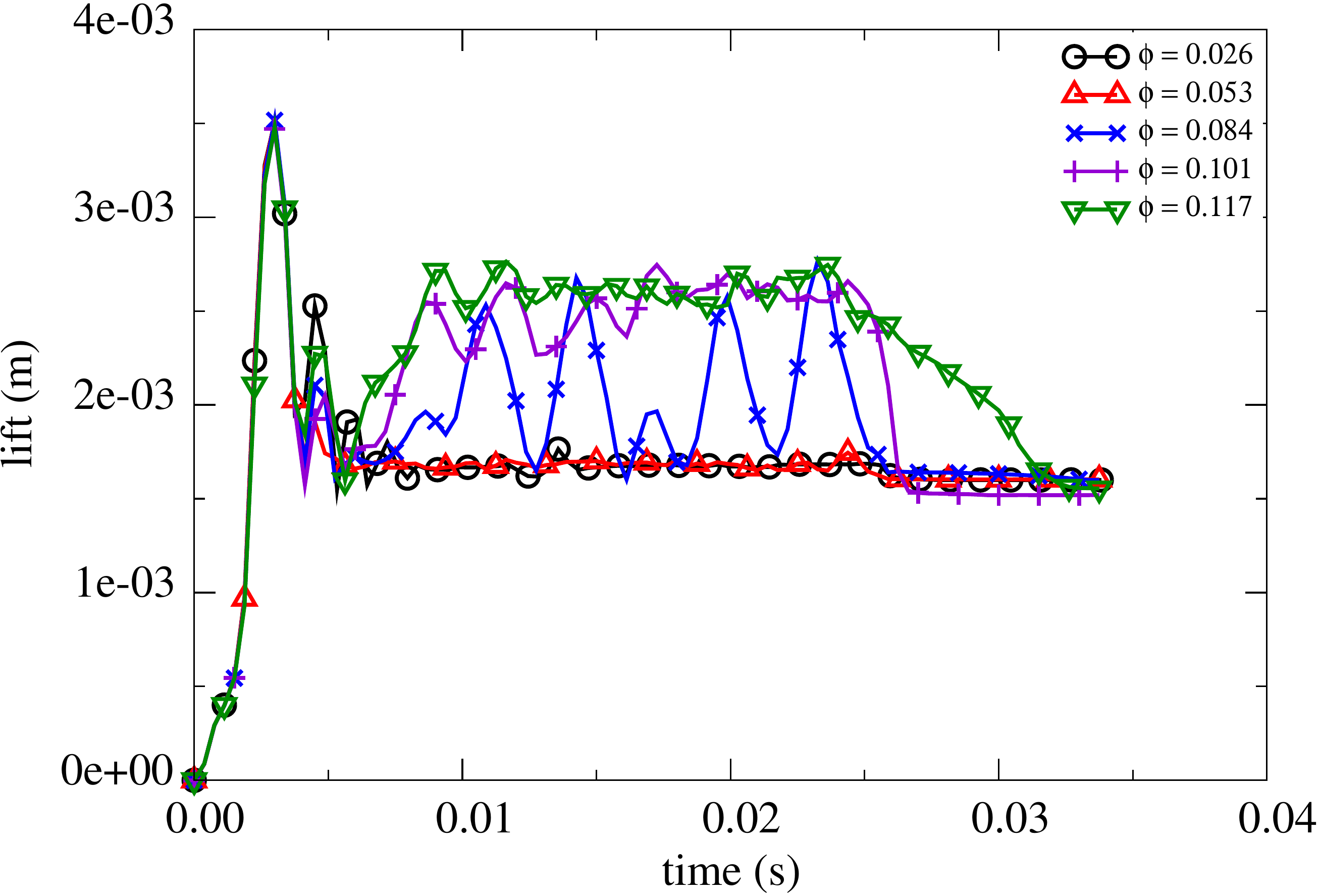}
\includegraphics[width=.5\textwidth]{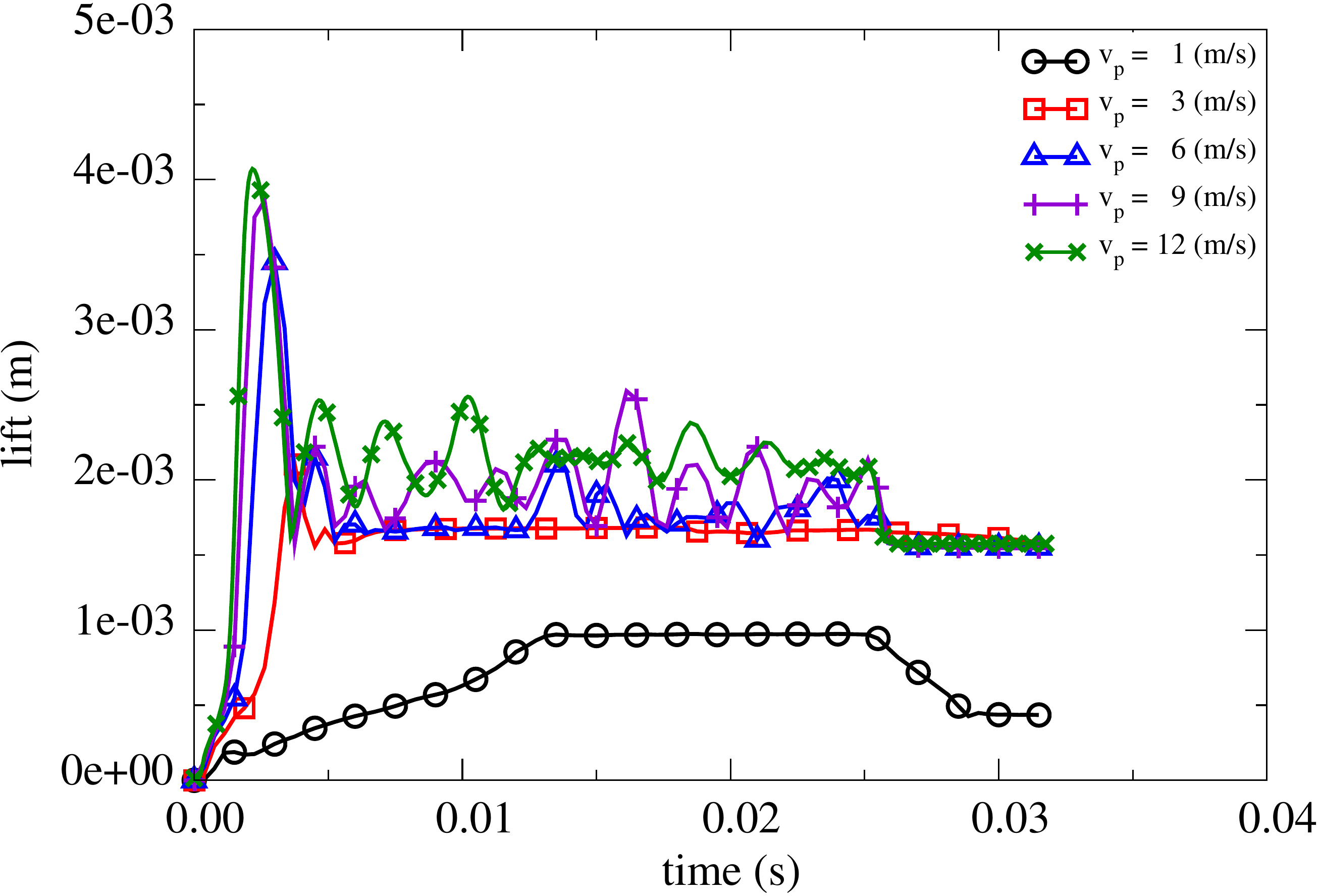}
}
\caption{Valve lift as function of time for different packing fraction (left) and different  piston velocity (right).}
\label{fig:closed_yp_lift_v_xxoxxx_sf_0o040_080_0o125_0o150_0o175}
\end{figure*}

Tests involving variation of the packing fraction or the piston velocity, 
Fig.~\ref{fig:closed_yp_lift_v_xxoxxx_sf_0o040_080_0o125_0o150_0o175}, show a non-trivial valve lift behavior 
in which three approximate regimes can be identified: 
\begin{enumerate}
\item A lower input particle flux behavior: $\varphi<\varphi _{_l}$.
\item A transition input particle flux behavior: $\varphi _{_l} \leq \varphi \leq \varphi _{_u}$.
\item A higher input particle flux behavior: $\varphi _{_u} < \varphi$.
\end{enumerate}
where the input flux is defined by $\varphi = \phi\ v_p$.

\begin{figure}[!ht]
\centerline{
\includegraphics[width=.75\textwidth]{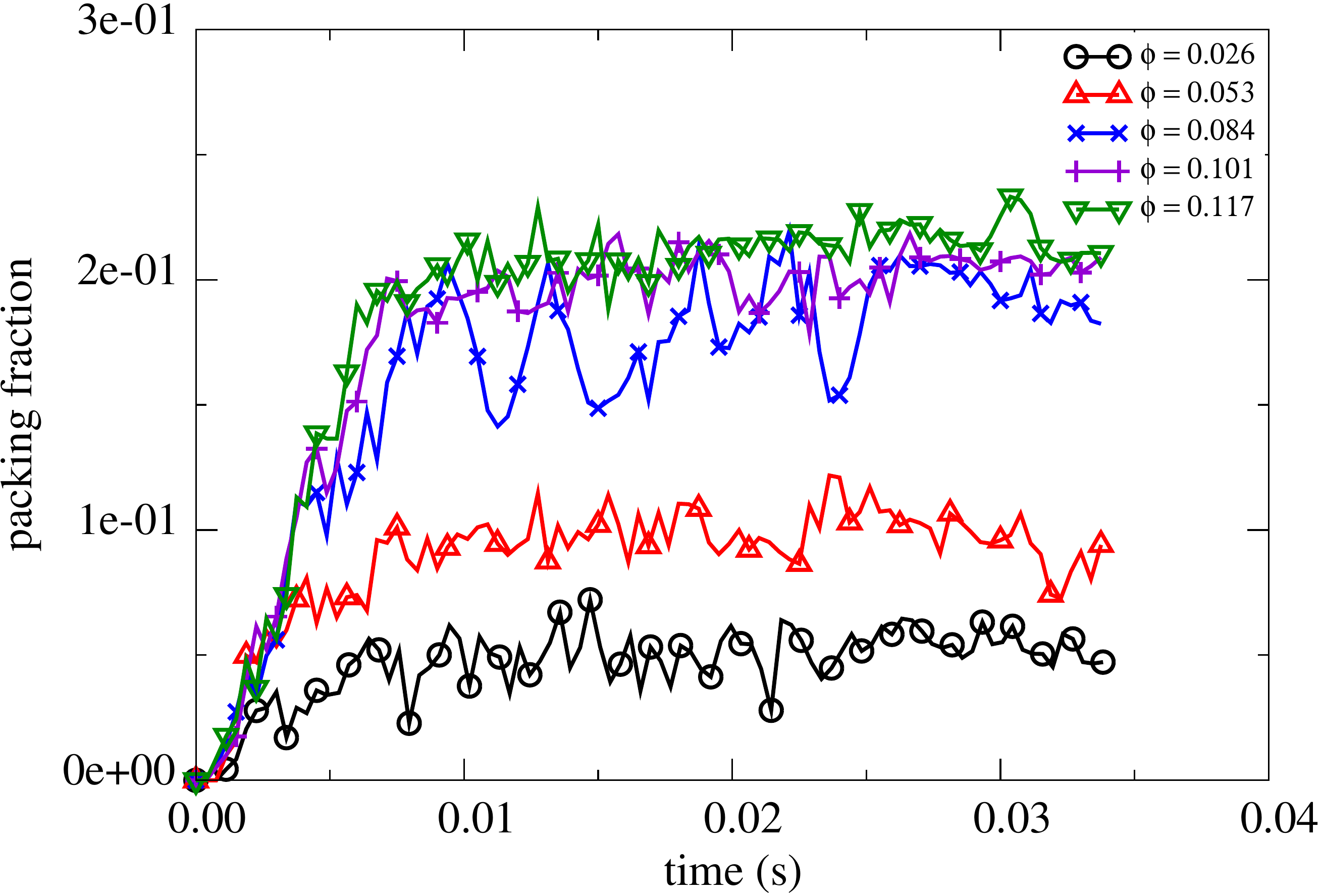}
}
\caption{Beneath valve packing fraction as function of time for different input packing fraction.}
\label{fig:closed_yp_BVPackingFraction_sf_xxoxxx_v_xxoxxx}
\end{figure}

In the first regime, we observe a simple particle suspension flow through the valve and rubber channel with a 
quasi-constant lift as shown in Fig.~\ref{fig:closed_yp_lift_v_xxoxxx_sf_0o040_080_0o125_0o150_0o175}. 
The beneath valve packing fraction shows also a quasi-constant value 
Fig.~\ref{fig:closed_yp_BVPackingFraction_sf_xxoxxx_v_xxoxxx}. This regime is observed 
for $\varphi<\varphi_{_l} \simeq 0.405$ (from Fig.~\ref{fig:closed_yp_lift_v_xxoxxx_sf_0o040_080_0o125_0o150_0o175}). 

\begin{figure*}[!ht]
\centerline{
\includegraphics[width=.5\textwidth]{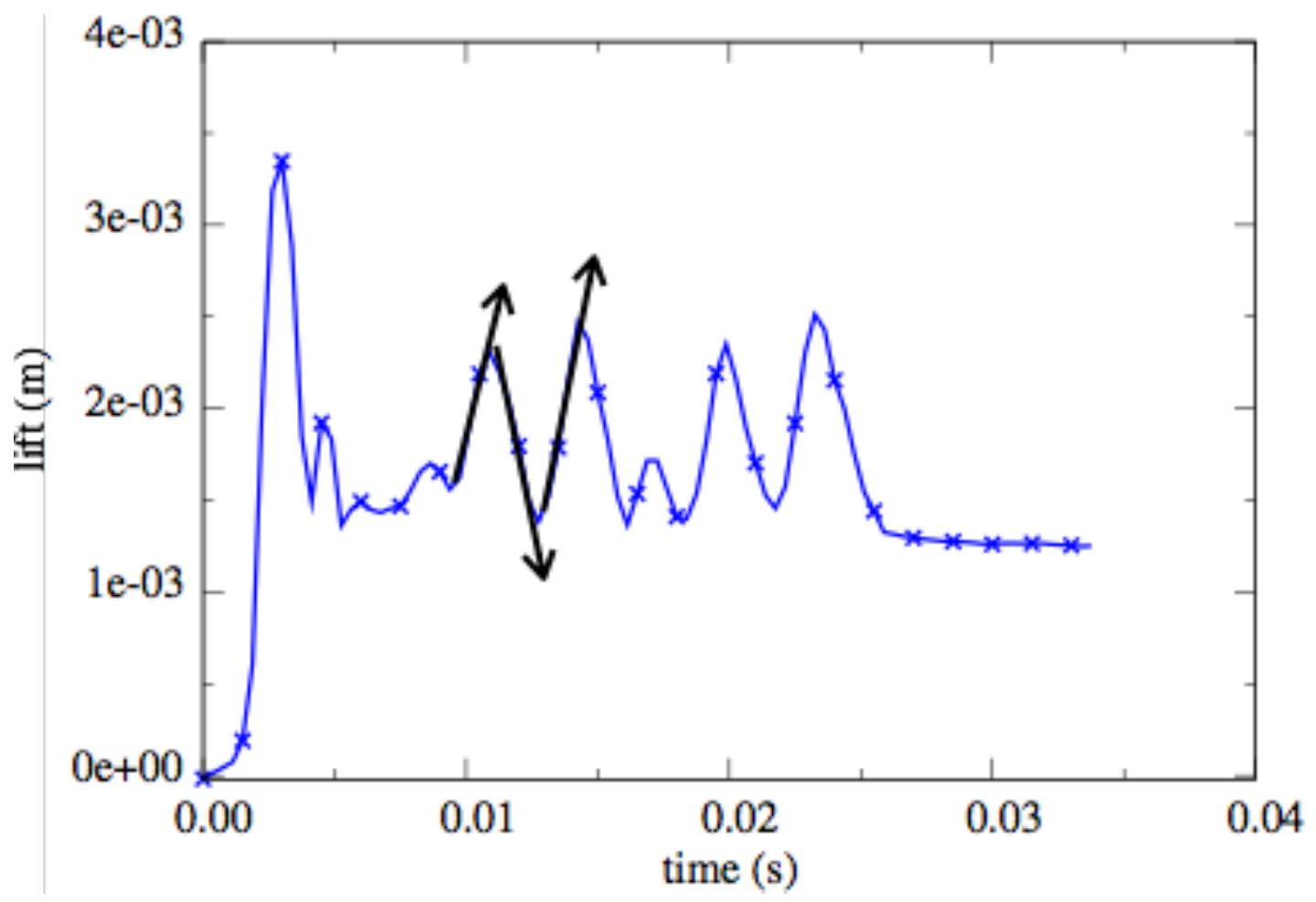}
\includegraphics[width=.5\textwidth]{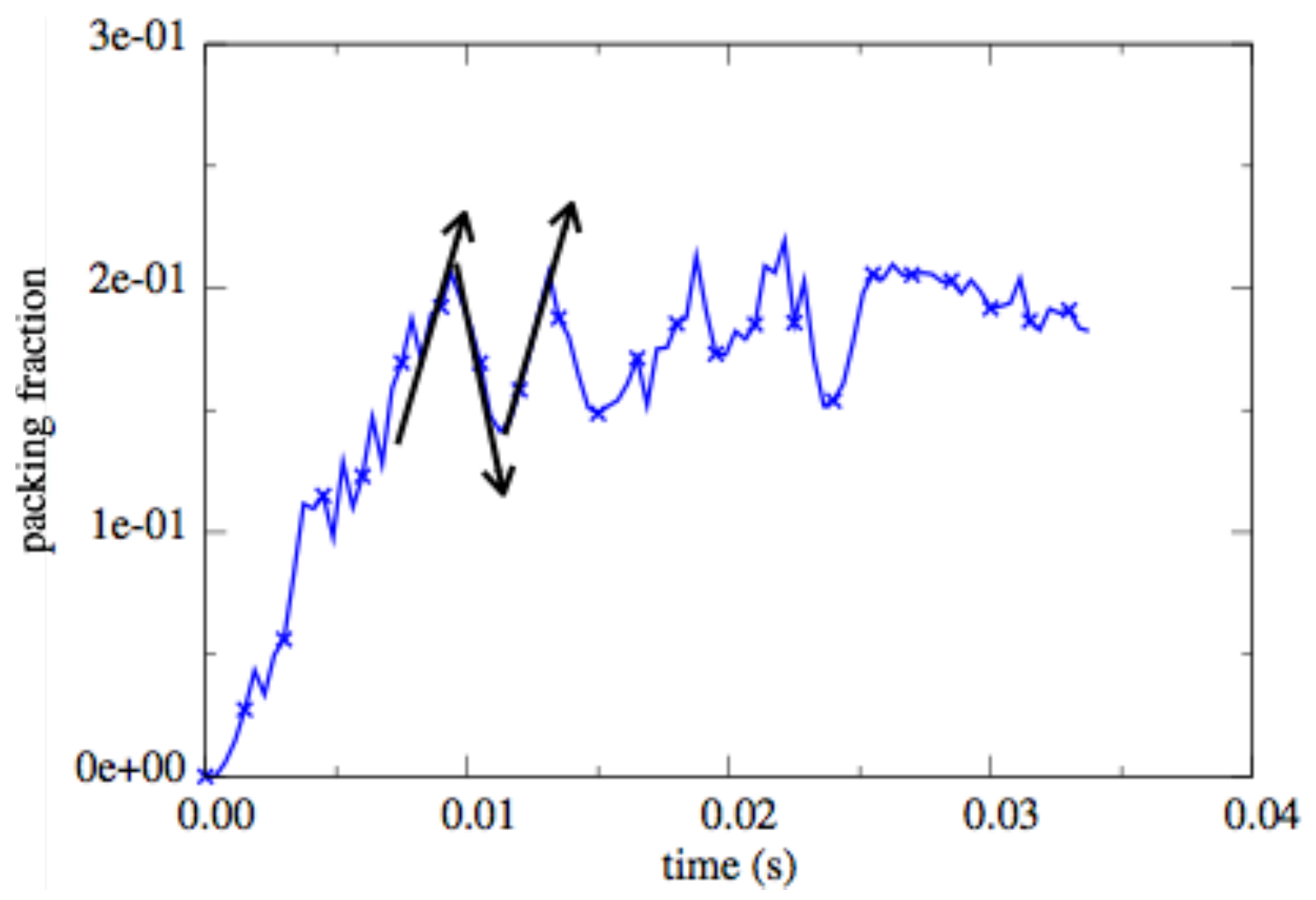}
}
\caption{Valve lift (left) and beneath valve packing fraction (right) for $\phi = 0.084$ as function of time.}
\label{fig:closed_yp_lift_packing_0o1250}
\end{figure*}

\begin{figure}[!ht]
\centerline{
\includegraphics[width=.75\textwidth]{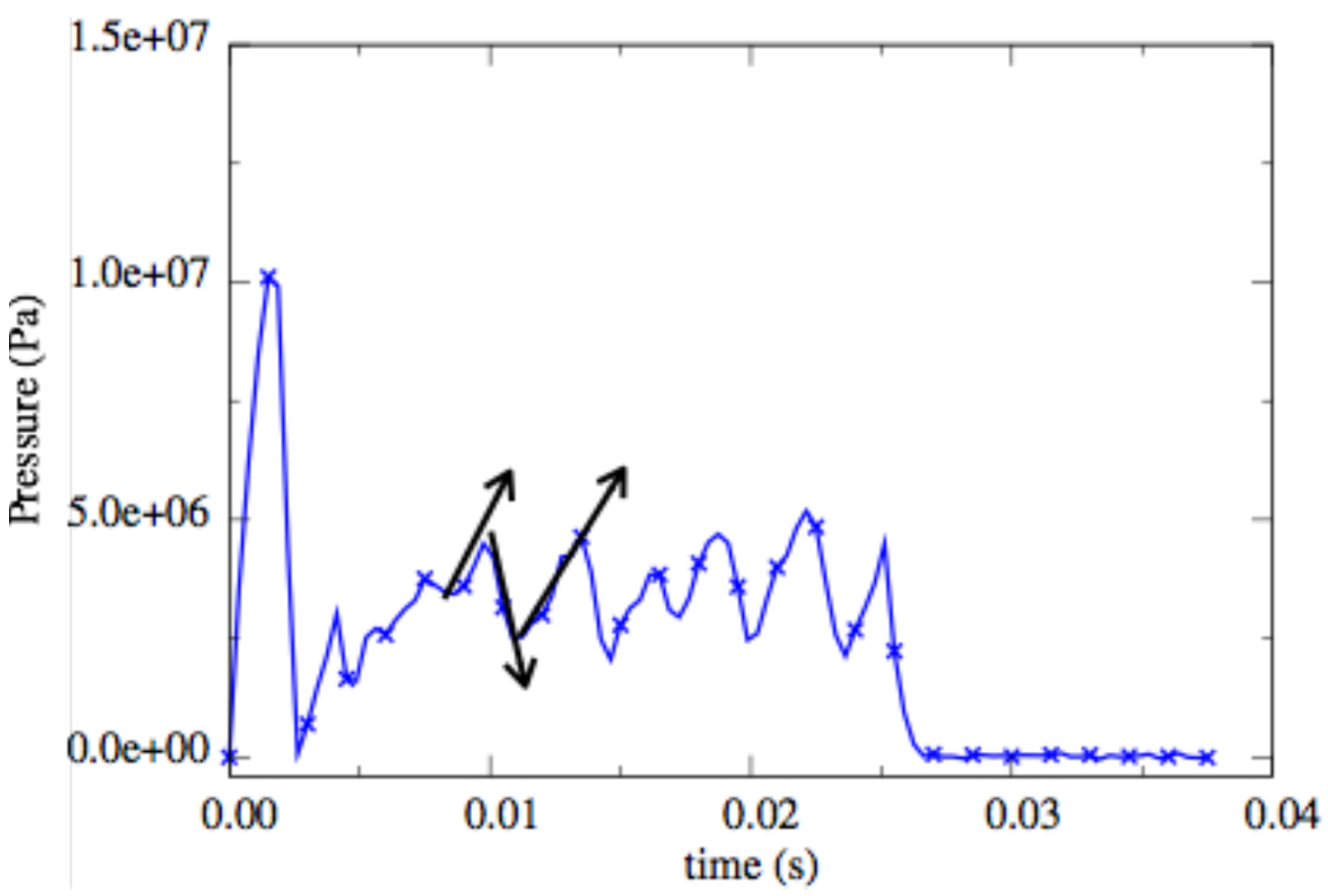}
}
\caption{Beneath valve pressure as function of time for $\phi = 0.084$.}
\label{fig:closed_yp_pressure_0o1250}
\end{figure}

The second regime is characterized by unsteady motion of the valve that oscillates between notably disparate 
high and low positions. This regime, for the range of parameters we tested, appears to be limited by 
$\varphi>\varphi_{_l} \simeq 0.405$  and $\varphi<\varphi_{_u} \simeq 0.611$. 
To better understand the valve lift behavior in this regime let us analyze the lift for $\phi = 0.084$. 
Figure~\ref{fig:closed_yp_lift_packing_0o1250} and Fig.~\ref{fig:closed_yp_pressure_0o1250} show the time 
dependence of the lift, beneath valve packing fraction, and beneath valve pressure. Notice that 
the peaks and valleys of the beneath valve pressure and packing fraction are relatively in sync. The peaks 
in the lift plot are delayed with respect to those of the pressure and packing fraction. This can be understood 
as follows: when the valve position is low, particles aggregate under the valve and as they do so, they 
form something of a `plug' that causes the pressure beneath the valve to build up.  When the pressure is 
sufficiently high, the valve will open up to release the pressure, which causes the backed-up grains 
beneath the valve to escape through the open valve. When this happens it causes the beneath valve 
packing fraction and the pressure to decrease, which immediately allows the valve to recover to its 
initial lift (Fig.~\ref{fig:closed_yp_lift_packing_0o1250}, Fig.~\ref{fig:closed_yp_pressure_0o1250}). 
This phenomena can be distinguished from the lower input particle flux regime, in that in the lower flux case, 
the flux of grains into the system is not enough to back-up sufficiently under the valve and induce a pressure 
build-up.
\begin{figure}[!ht]
\centerline{
\includegraphics[width=1.\textwidth]{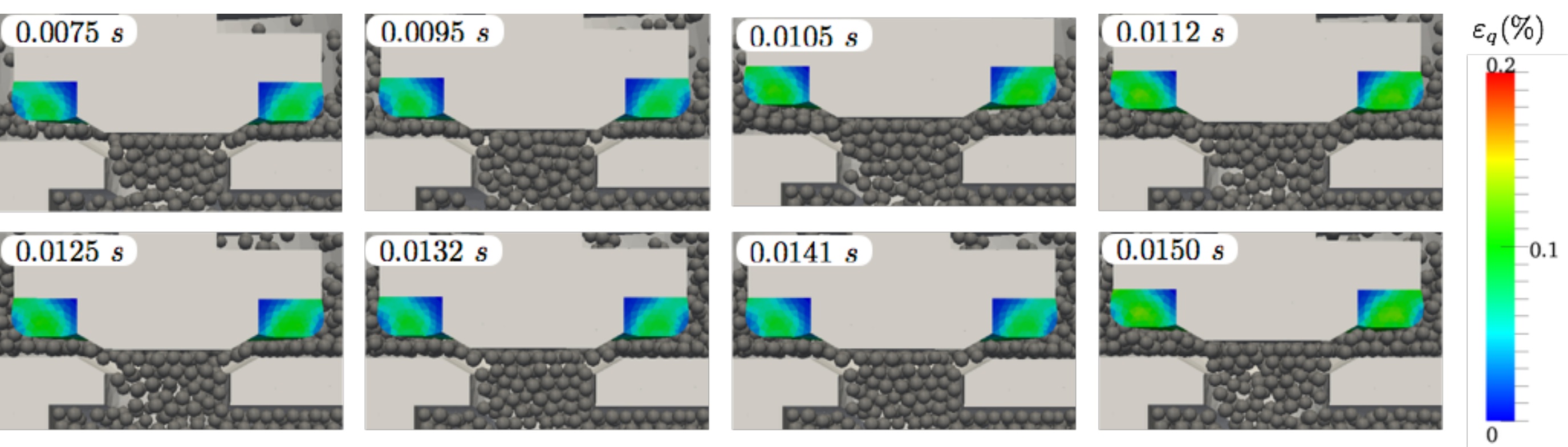}
}
\caption{Several snapshots showing particles beneath valve for $\phi = 0.084$. According to 
Fig.~\ref{fig:closed_yp_lift_packing_0o1250} (right), the first two packing fraction peaks correspond to 
$t = 0.0095\ s$ and $t = 0.0132\ s$, and the first two valley correspond to $t = 0.0112\ s$ and $t = 0.0145\ s$.}
\label{fig:closed_yp_packingsmapshot_0o1250}
\end{figure}

Figure~\ref{fig:closed_yp_packingsmapshot_0o1250} shows several snapshots of the beneath valve region for 
$\phi = 0.084$ (from Fig.~\ref{fig:closed_yp_lift_v_xxoxxx_sf_0o040_080_0o125_0o150_0o175} (left)). 
According to Fig.~\ref{fig:closed_yp_lift_packing_0o1250} (right), the first two packing fraction 
peaks correspond to $t = 0.0095\ s$ and $t = 0.0132\ s$, and the first two valleys to $t = 0.0112\ s$ and 
$t = 0.0145\ s$. The rest of the snapshots correspond to the time between the peaks and the valley. 
Comparing the first packing fraction peak and the first valve lift peak 
in Fig.~\ref{fig:closed_yp_packingsmapshot_0o1250}, we observe a delay; the first peak in
packing fraction occurs at $t = 0.0095\ s$, and the peak for valve lift occurs between $t = 0.0105\ s$ and $t = 0.0112\ s$. 
Using Fig.~\ref{fig:closed_yp_lift_packing_0o1250}, we find that the delay is $\sim 0.0014\ s$. The same delay 
is observed for all peaks and valleys. Contrary to the valve lift, between the pressure and packing fraction 
peak/valleys, no delay is observed. This is in agreement with the lift behavior 
Fig.~\ref{fig:closed_yp_lift_v_xxoxxx_sf_0o040_080_0o125_0o150_0o175} where the valve lift is a consequence 
of the packing fraction/pressure evolution. 

The third regime of valve behavior corresponds to a high particle flux such that the beneath valve slurry develops a sustainably high pressure able to push and hold the valve at a maximal lift. This is observed for $\varphi>\varphi_{_u} \simeq 0.611$ on 
Fig.~\ref{fig:closed_yp_BVPackingFraction_sf_xxoxxx_v_xxoxxx} (left) from $\phi \simeq  0.101$.

Out of the three phases, one outlier phenomenon is observed for $v_p = 1\ m/s$. In fact, Fig.~\ref{fig:closed_yp_lift_v_xxoxxx_sf_0o040_080_0o125_0o150_0o175} (right) 
shows that when the valve is opened, from $t = 0.013\ s$ to $t = 0.025\ s$, we have a constant but small lift. 
During this phase, it turns out that a portion of the particles are stuck at the entrance of the valve channel but 
without entering fully into the channel. The force from these stuck particles and fluid pressure is enough to hold open the 
valve at a constant lift.

\subsubsection*{Valve closing mechanisms}

In this section, we focus on the closure phase of the valve simulations with particles in order to investigate 
the effect particles have on the final lift of the valve and to study the degree to which particles become stuck 
in the valve-rubber channel, which could have detrimental effects on valve performance in practice. Since the rubber plays the role of a seal between valve and seat, preventing grain trapping during closure could be a key design goal in such systems.

\begin{figure}[!ht]
\centerline{
\includegraphics[width=1.\textwidth]{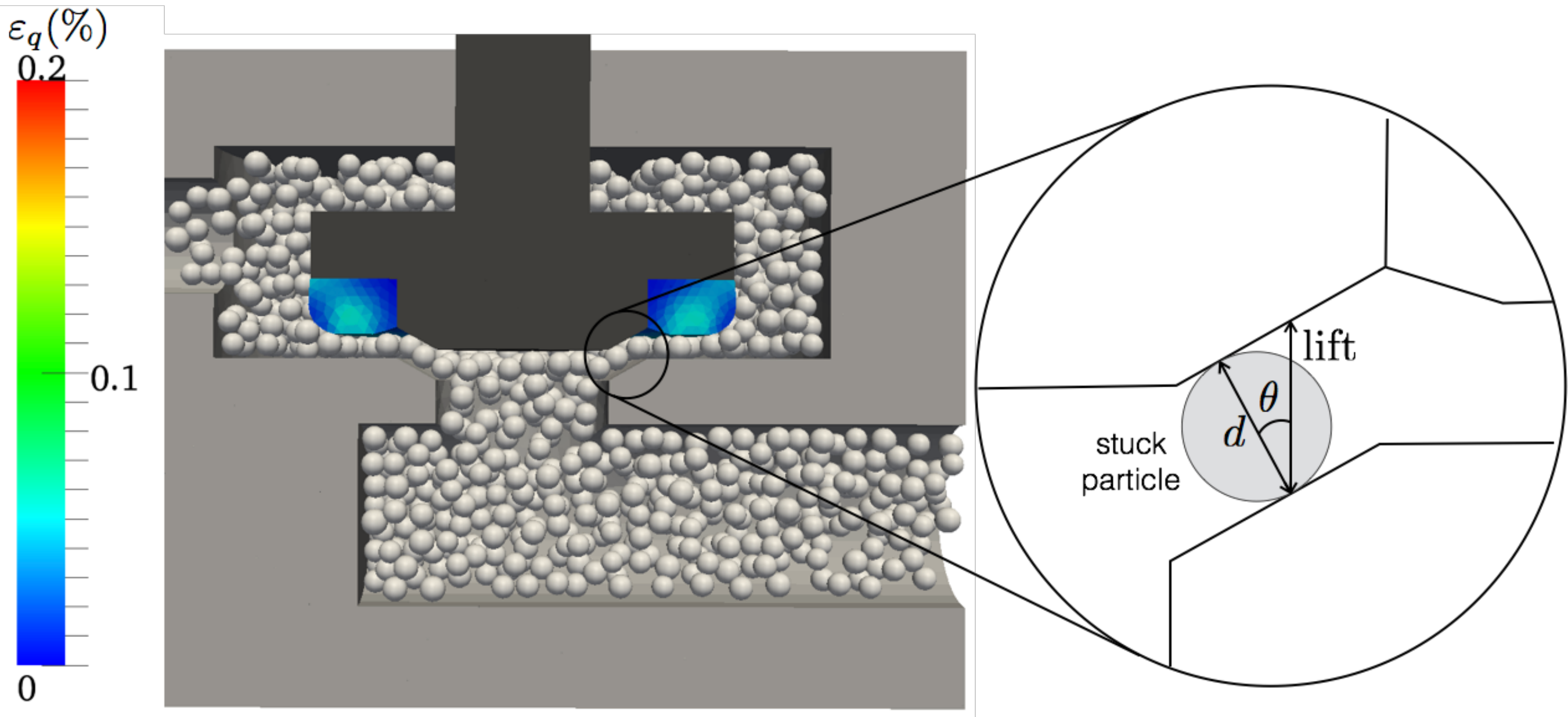}
}
\caption{Final lift configuration; relation to stuck particle size.}
\label{fig:closing_stuck_particles}
\end{figure}

\begin{figure*}[!ht]
\centerline{
\includegraphics[width=.5\textwidth]{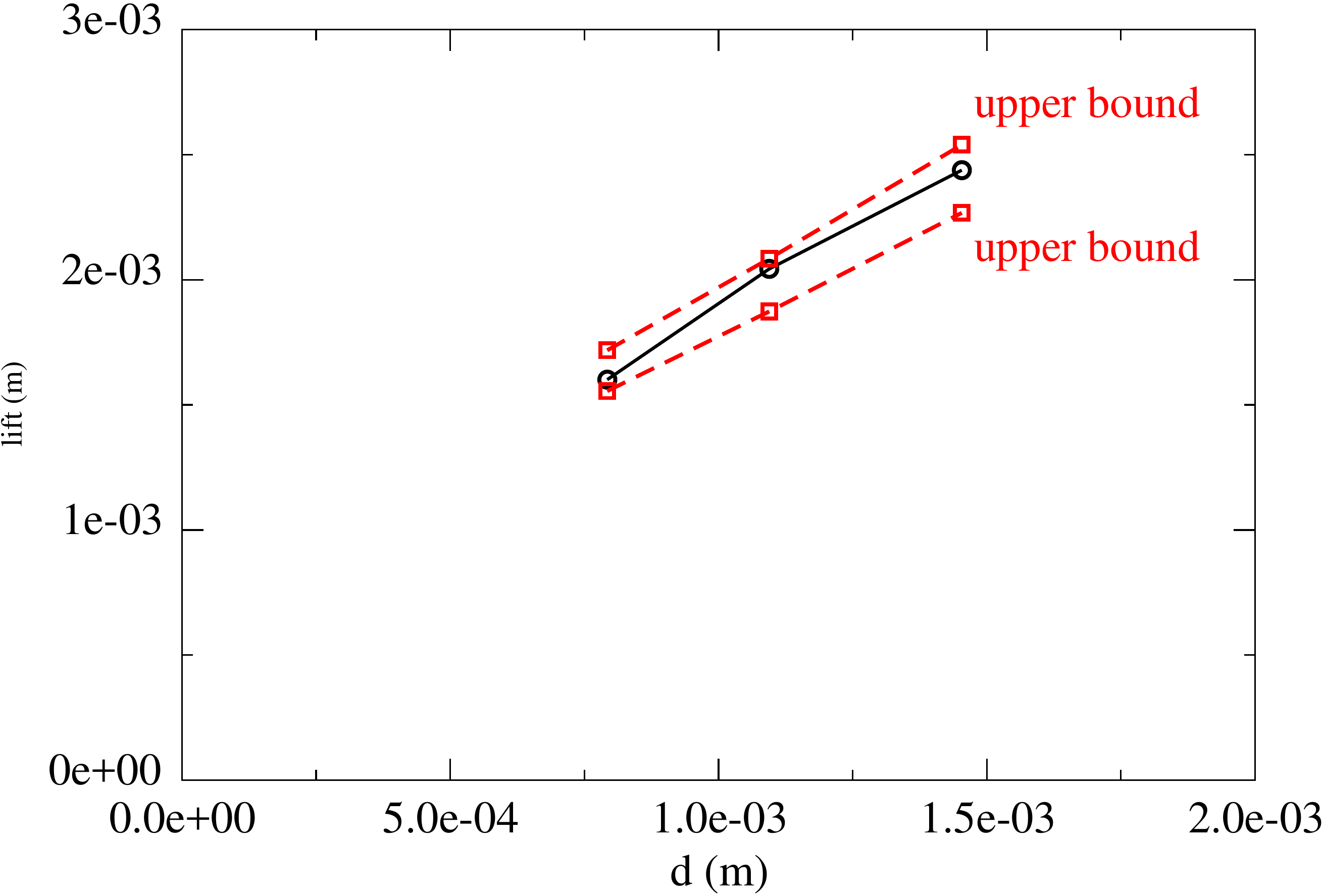}
\includegraphics[width=.5\textwidth]{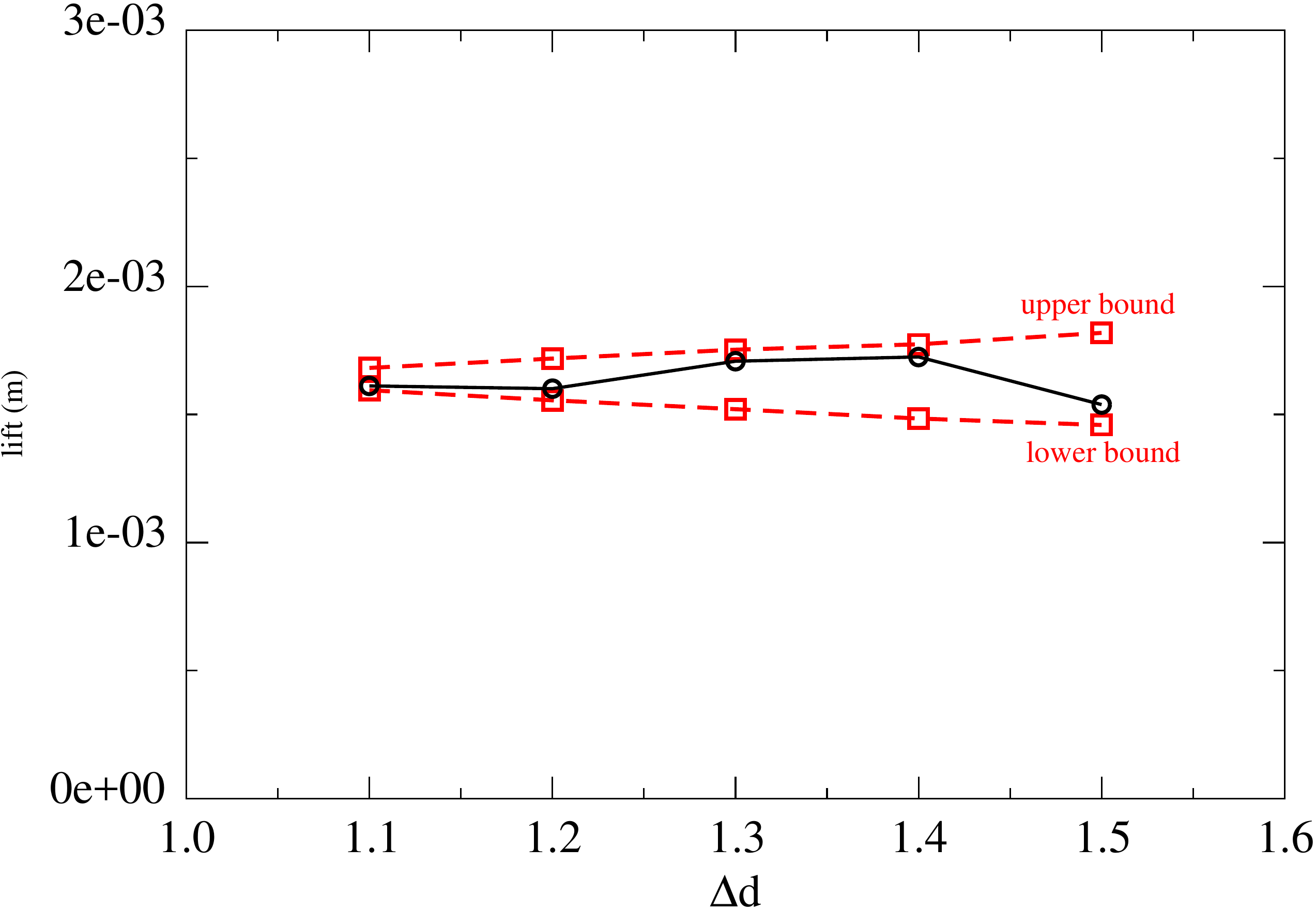}
}
\caption{Final lift for different $d$ (left) and $\Delta d$ (right).}
\label{fig:closed_yp_stuck_lift_dmean_poly}
\end{figure*}

\begin{figure*}[!ht]
\centerline{
\includegraphics[width=.5\textwidth]{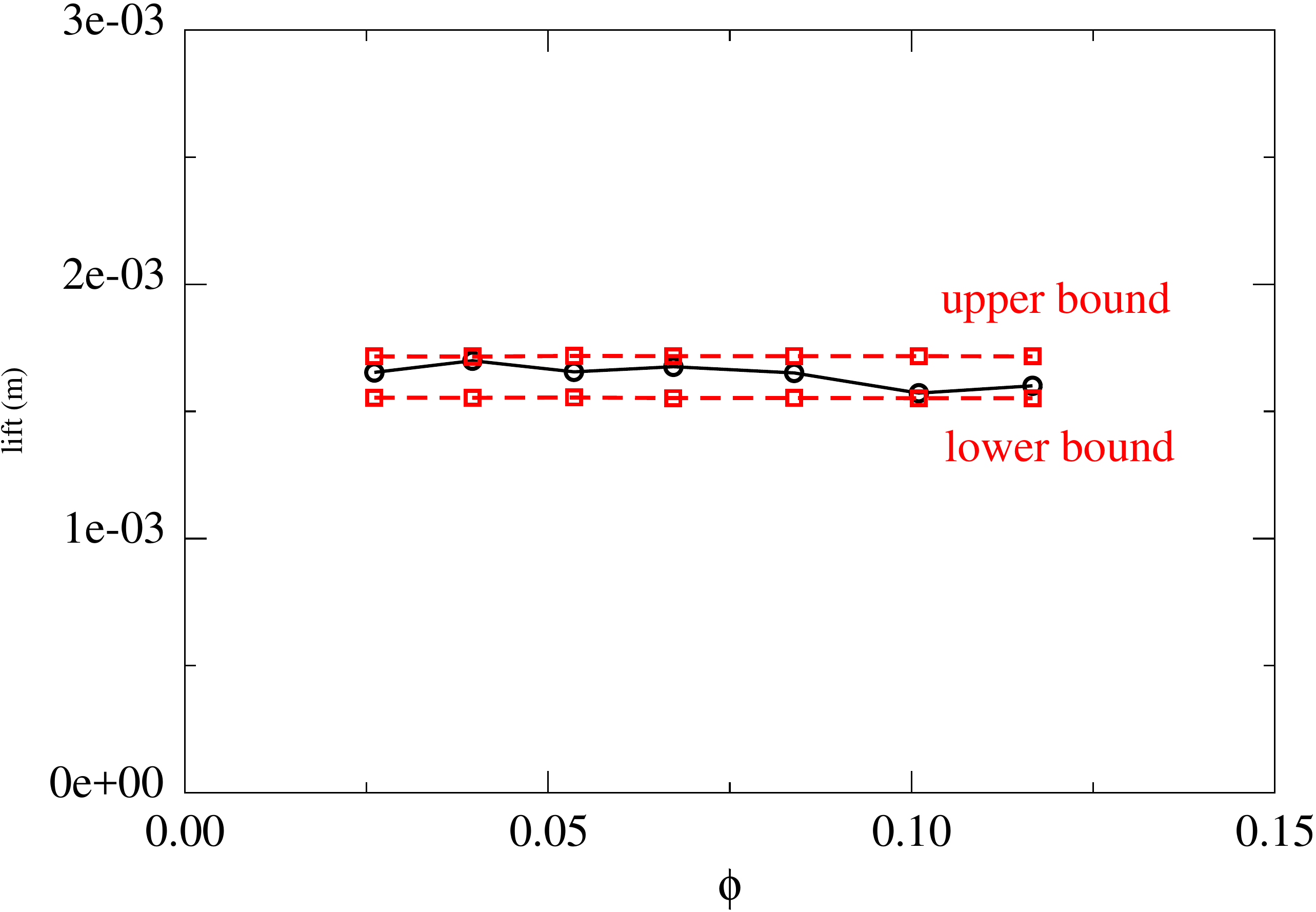}
\includegraphics[width=.5\textwidth]{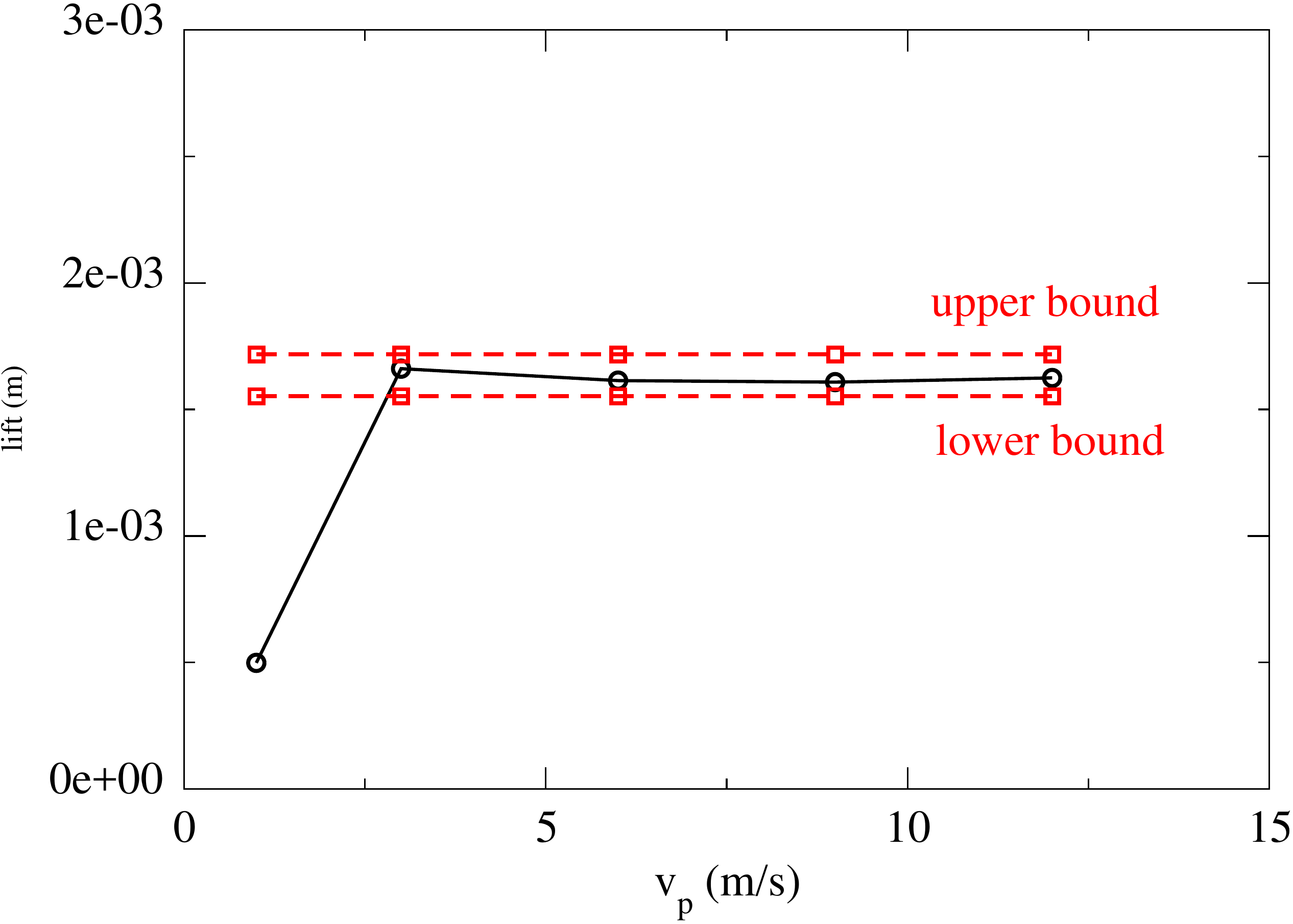}
}
\caption{Final lift for different $\phi$ (left) and $v_p$ (right).}
\label{fig:closed_yp_stuck_lift_packing_vp}
\end{figure*}

The closure phase starts when the piston velocity $v_p$ is turned off. In the range of the parametric study we investigated, the final lift is mostly affected by the mean particle diameter $d$ as shown in Fig.~\ref{fig:closed_yp_stuck_lift_dmean_poly} 
(for $d$ and $\Delta d$) and Fig.~\ref{fig:closed_yp_stuck_lift_packing_vp} (for $\phi$ and $v_p$). This behavior is simply explained by 
the fact that the closing valve squeezes out all but a monolayer of particles, which become stuck in the valve channel as illustrated in Fig.~\ref{fig:closing_stuck_particles} 
where a zoom-in on the valve channel shows how geometrically the lift depends on the stuck particle diameter $d$. Using the size of particles and 
the geometry of the valve channel Fig.~\ref{fig:closing_stuck_particles}, we calculated the envelope giving the maximum lift (upper bound) 
and minimum lift (lower bound) which should be obtained if only big particles (with $d_{_{max}}$) or small particles (with $d_{_{min}}$) 
were stuck. The two bounds are given by: 

\begin{equation}
\text{lift} = d^*/\cos(\theta)
\label{eq:lift}
\end{equation}
where $d^*$ is either $d_{_{max}}$ or either $d_{_{min}}$ and $\theta$ is the valve channel inclination as shown on 
Fig.~\ref{fig:closing_stuck_particles}. $\theta = 29.53^o$ is obtained from Fig.~\ref{fig:Seat} (DETAIL B). As expected, 
the final lift is always between these two limits. 

Figure~\ref{fig:closed_yp_stuck_lift_packing_vp} (right) for $v_p = 1\ m/s$ shows a lift of $\sim 0.00048 \pm 0.00001\ m$ which is less than 
the minimum lift ($\text{lift}_{_{min}} = 0.00155\pm0.000001\ m$) for the smallest particle diameter in the polydispersity to travel through the valve. 
In fact here, as discussed previously in Section \textit{Quasi-steady open valve phase} for the effect of $v_p$ on the lift, no particle flow is 
observed in the valve/rubber channel since the lift is less than one particle diameter. Therefore, when $v_p$ is turned off, the rubber descent is unimpeded by grains. Once the rubber makes 
contact with the valve seat, the fluid beneath the valve cannot escape and therefore a pressure and residual lift of $\sim 0.00048 \pm 0.00001\ m$ 
remains, which is the lift when the rubber is in contact with the seat with zero deformation. 

\begin{figure*}[!ht]
\centerline{
\includegraphics[width=.5\textwidth]{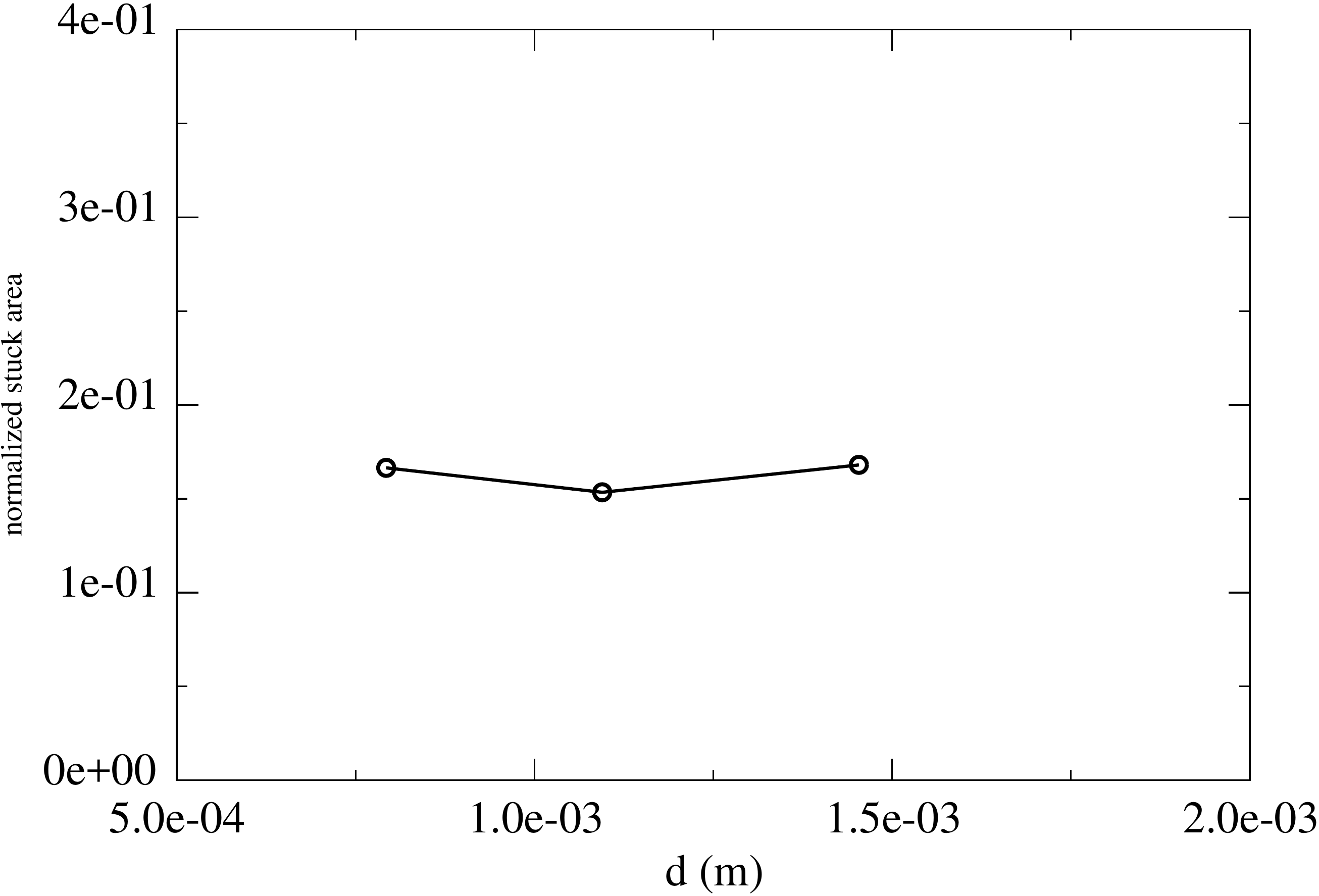}
\includegraphics[width=.5\textwidth]{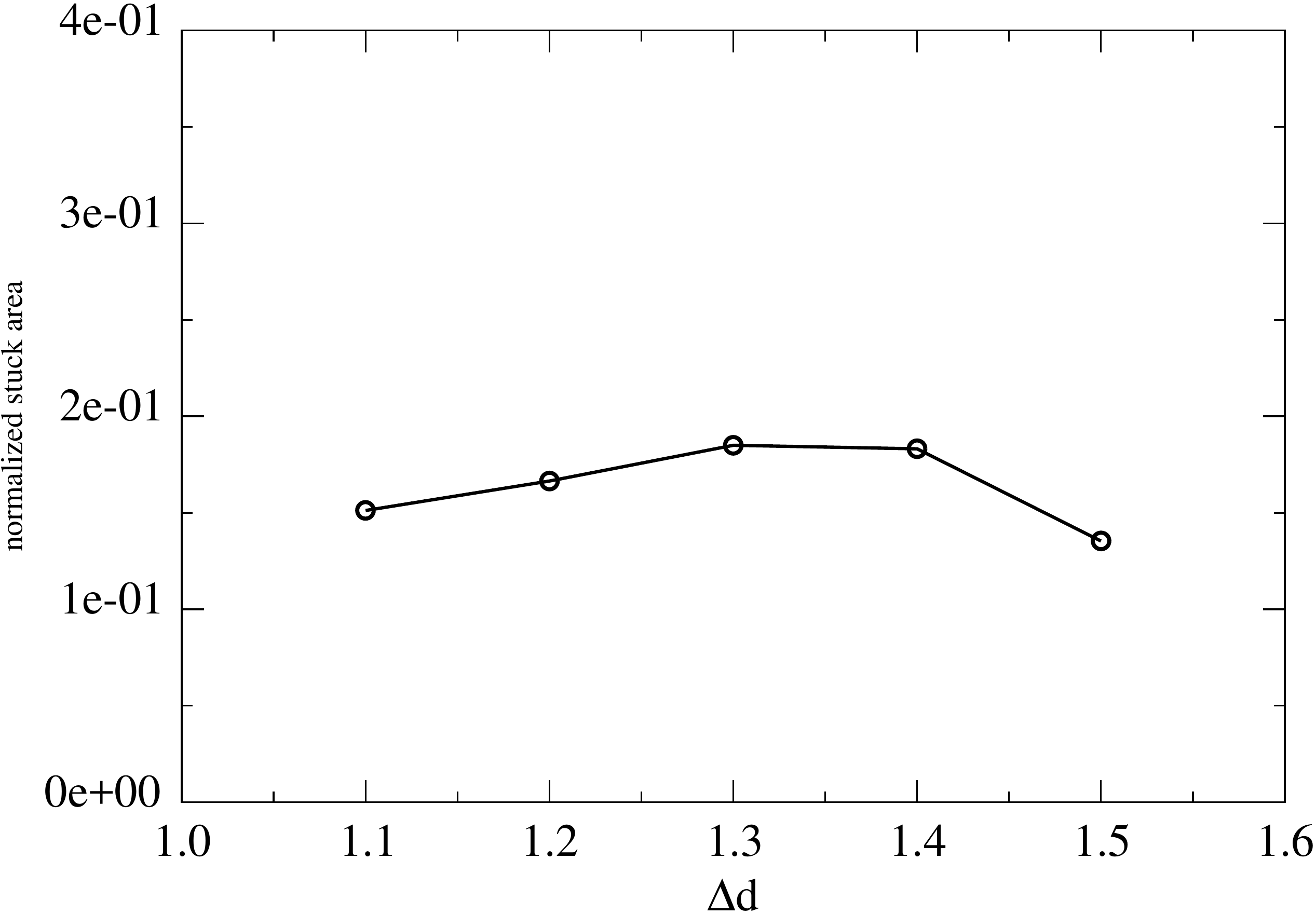}
}
\caption{NASP for different $d$ (left) and $\Delta d$ (right).}
\label{fig:closed_yp_stuck_norm_dmean_poly}
\end{figure*}

\begin{figure*}[!ht]
\centerline{
\includegraphics[width=.5\textwidth]{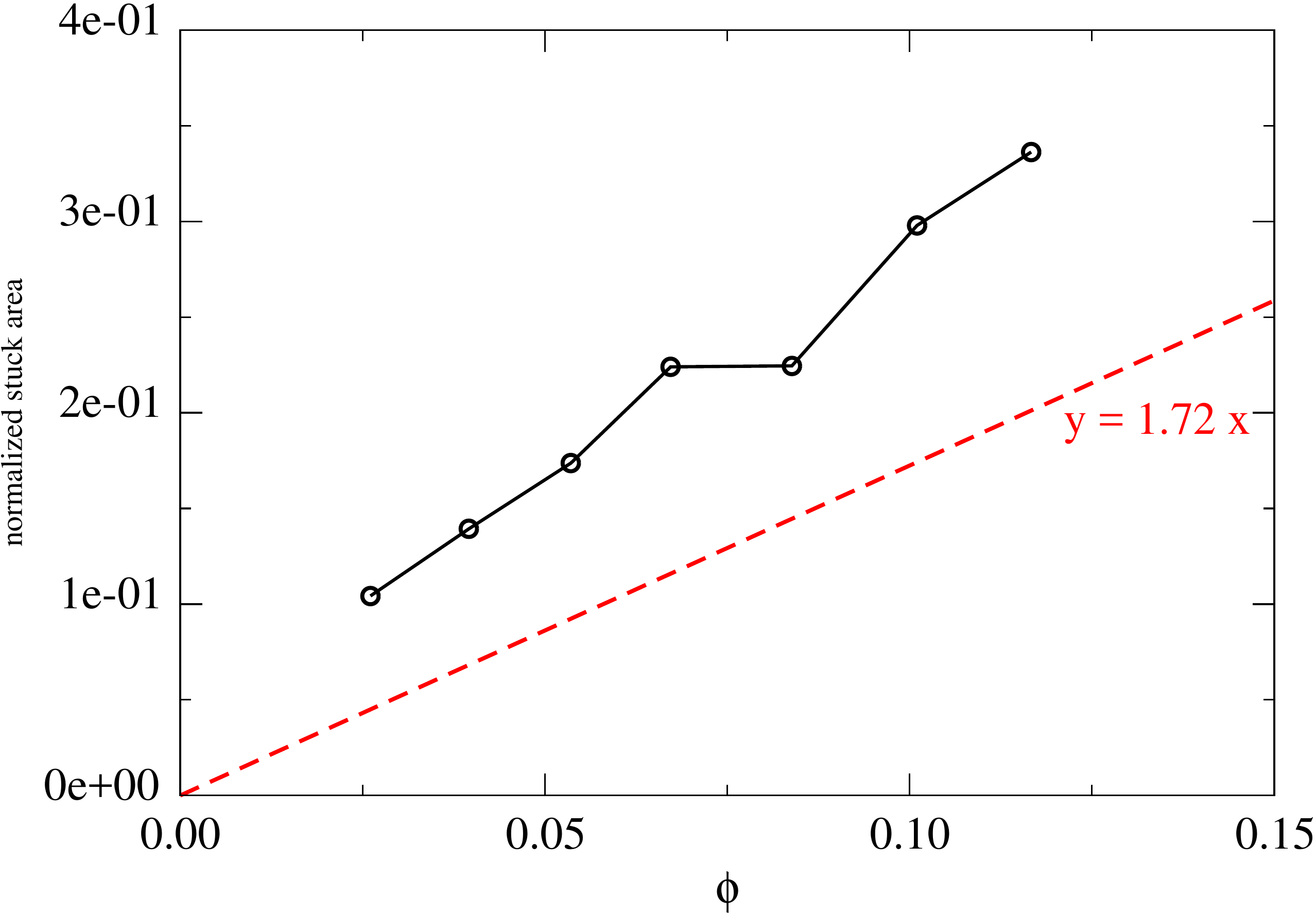}
\includegraphics[width=.5\textwidth]{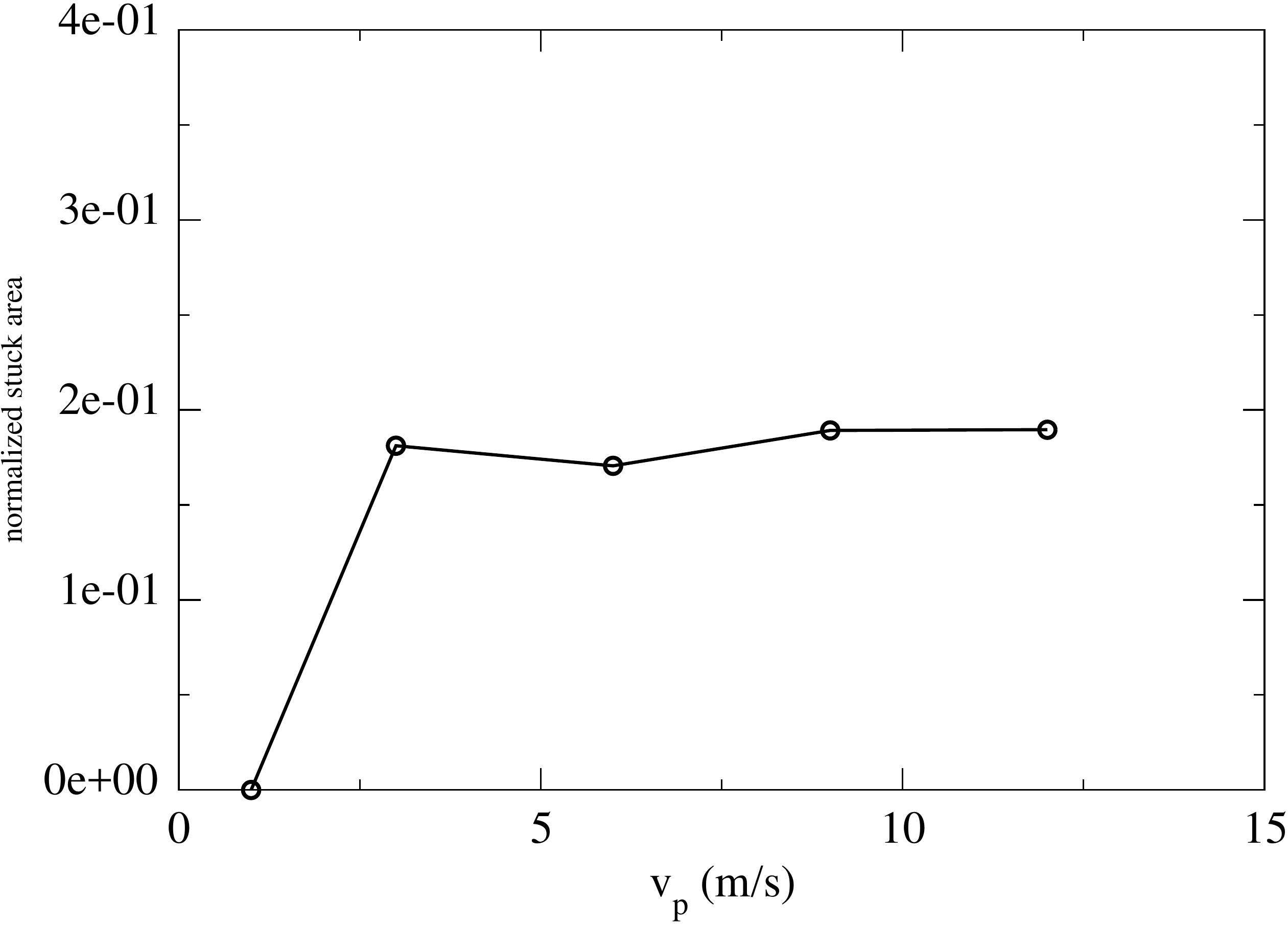}
}
\caption{NASP for different $\phi$ (left) and $v_p$ (right).}
\label{fig:closed_yp_stuck_norm_packing_vp}
\end{figure*}

We give here a first quantitative view on which variables --- among $d_{_{max}}, \Delta d, \phi$ and  $v_p$ --- matter 
most in affecting the quantity of particles that get stuck beneath the valve during closure. Figure~\ref{fig:closed_yp_stuck_norm_dmean_poly} 
and Fig.~\ref{fig:closed_yp_stuck_norm_packing_vp} show the projected area of particles on the seat (beneath valve-rubber area) normalized 
by the beneath valve-rubber area, i.e. the `normalized area of stuck particles' (NASP). On these four 
figures, we find that $\phi$ matters most whereas there is little variation due to changes in $d_{_{max}}$, $v_p$ and $\Delta d$. 

 We can assume 
the packing fraction in the open valve channel, during the quasi-steady valve phase, is bounded below by $\phi$. 
As the valve descends during closure, particles are squeezed out and, as a further lower bound, we can approximate 
that a single monolayer at the same packing fraction $\phi$ remains. This implies the total number of stuck particles 
in the rubber-valve channel is approximated by: $N_{stuck} \simeq \text{lift}\ A\ \phi\ \frac{6}{\pi\ d^3}$ 
where the final lift is given by $\text{lift} \simeq d/\cos(\theta)$, and $A$ is the projected rubber-valve area. 
The projected total particle area is: 
$S_{stuck} \simeq \frac{d}{\cos(\theta)}\ A\ \phi\ \frac{6}{\pi\ d^3}\ \frac{\pi\ d^2}{4} \simeq \frac{3}{2}\frac{A}{\cos(\theta)}\ \phi$ 
normalizing the $S_{stuck}$ by $A$, we obtain: 
\begin{equation}
\text{NASP} \simeq \frac{3}{2}\frac{1}{\cos(\theta)}\ \phi=1.72 \phi 
\label{eq:NASP}
\end{equation}

The above lower bound formula assumes that the final packing fraction of grains stuck in the valve is greater than the input value, $\phi$. 
In our tests we have observed that this is always true except for the one outlier case mentioned previously ($v_p = 1\ m/s$, 
Fig.~\ref{fig:closed_yp_lift_v_xxoxxx_sf_0o040_080_0o125_0o150_0o175} (right)) where no particles travel through the channel 
because the beneath valve fluid pressure is less than the necessary pressure to open the valve to a lift greater than 
$d_{min}/\cos(\theta)$. This case is observed in Fig.~\ref{fig:closed_yp_stuck_norm_packing_vp} (left) ($v_p = 1\ m/s$) 
where the normalized stuck area is zero.


If particles get stuck, they can also potentially break depending on the force of contact with the valve and seat, and grain properties such as the particle size and strength.
A loose approximation of the possible volume of debris created can be made by assuming stuck particles all break. 
This may be expressed in terms of the final lift and the NASP, and then normalized by $d^3$, giving
\begin{equation}
\mbox{Debris} = \mbox{lift}\cdot A\cdot \text{NASP}/d^3
\label{eq:Debris1}
\end{equation}
where $A$ is the valve-rubber projected area.

\begin{figure*}[!ht]
\centerline{
\includegraphics[width=.5\textwidth]{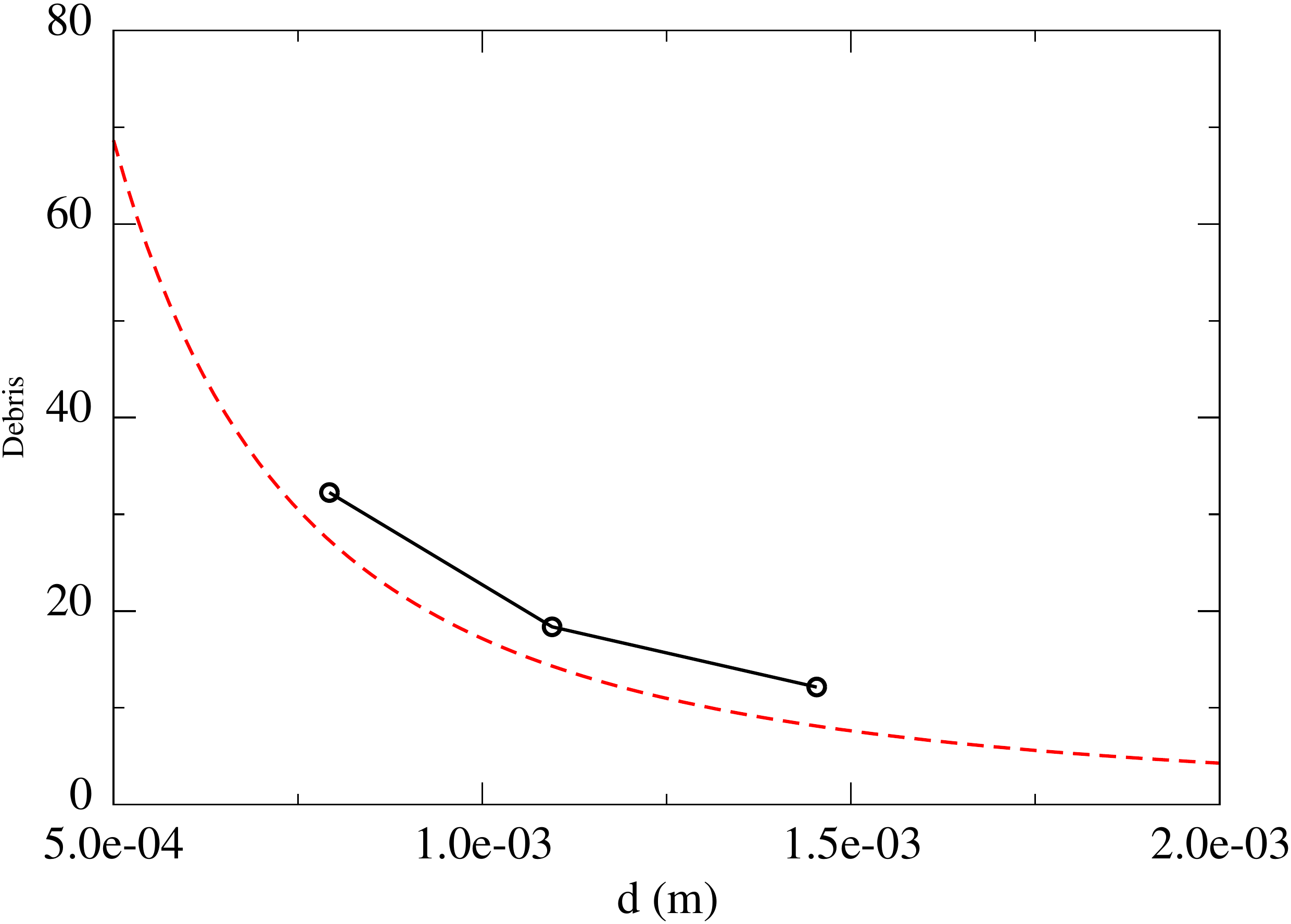}
\includegraphics[width=.5\textwidth]{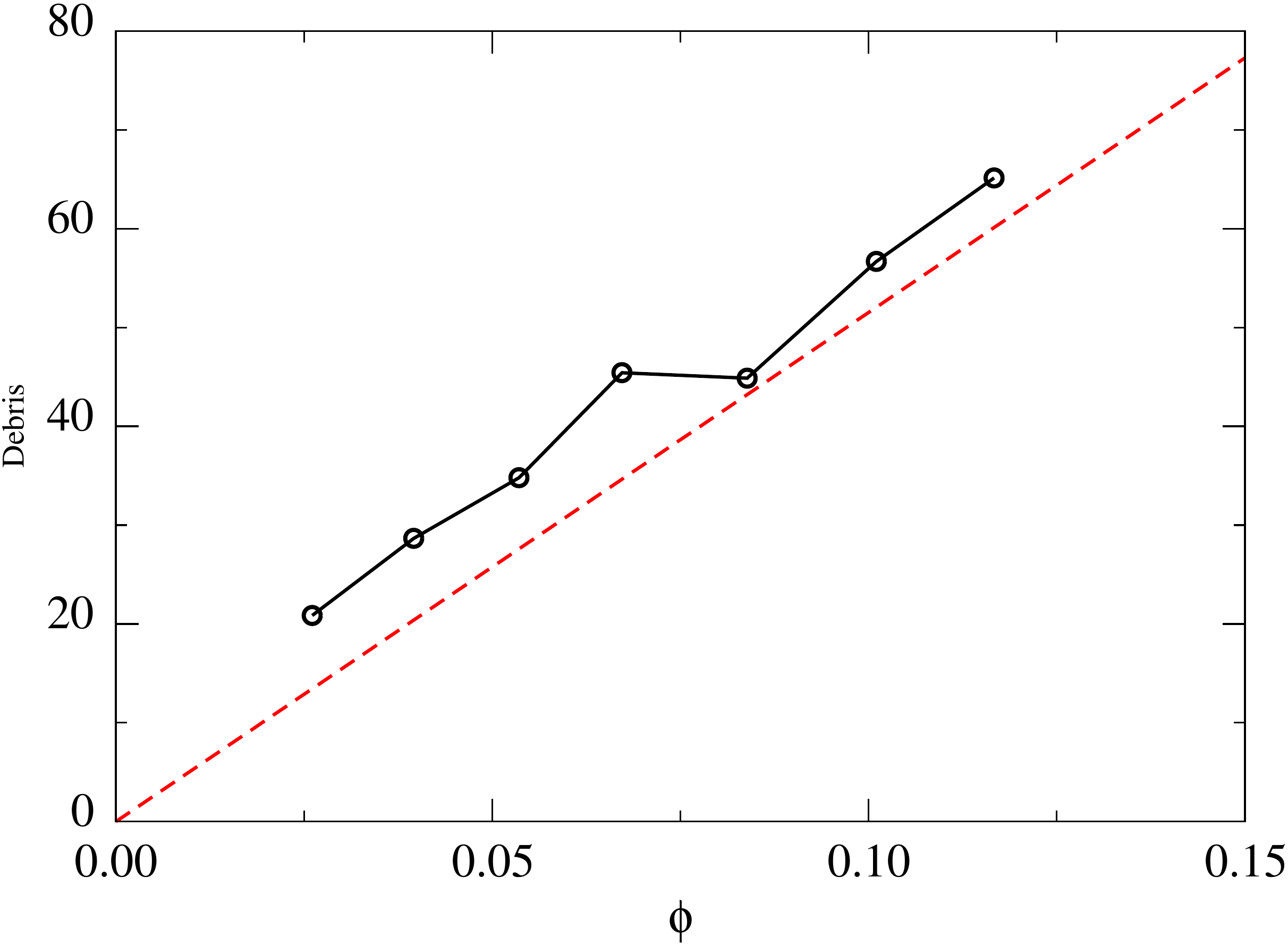}
}
\caption{Debris variation  as function $d$ (left) and $\phi$ (right). The red dashed line is the lower bound 
analytical solution (Eq.~\ref{eq:Debris2})}
\label{fig:closed_yp_stuck_debris_packing_dmean}
\end{figure*}

Supposing the lift obeys Eq.~\ref{eq:lift} where $d^* = d$ and NASP obeys Eq.~\ref{eq:NASP}, 
we suggest the Debris variable is approximated by: 
\begin{equation}
\mbox{Debris} = \frac{3}{2}\frac{A}{\cos(\theta)^2}\frac{\phi}{d^2}
\label{eq:Debris2}
\end{equation}
Figure~\ref{fig:closed_yp_stuck_debris_packing_dmean} (left) shows the Debris as a function of $d$ and 
Fig.~\ref{fig:closed_yp_stuck_debris_packing_dmean} (right) as a function of $\phi$. We show comparisons to our approximate formula (Eq.~\ref{eq:Debris2}).

\subsection{Safety valve lift behavior}
\label{sec:SafetyValveLift}

Many of the behaviors in the safety valve configuration mimic those of the pressure valve. Here we summarize the safety valve data. The used input parameters are resumed in 
Tab.~\ref{tab:spacerangeopenclosednp} for the simulations without particles and Tab.~\ref{tab:spacerangeopenclosedyp} 
for the simulations with particles. 

\begin{figure*}[!ht]
\centerline{
\includegraphics[width=.5\textwidth]{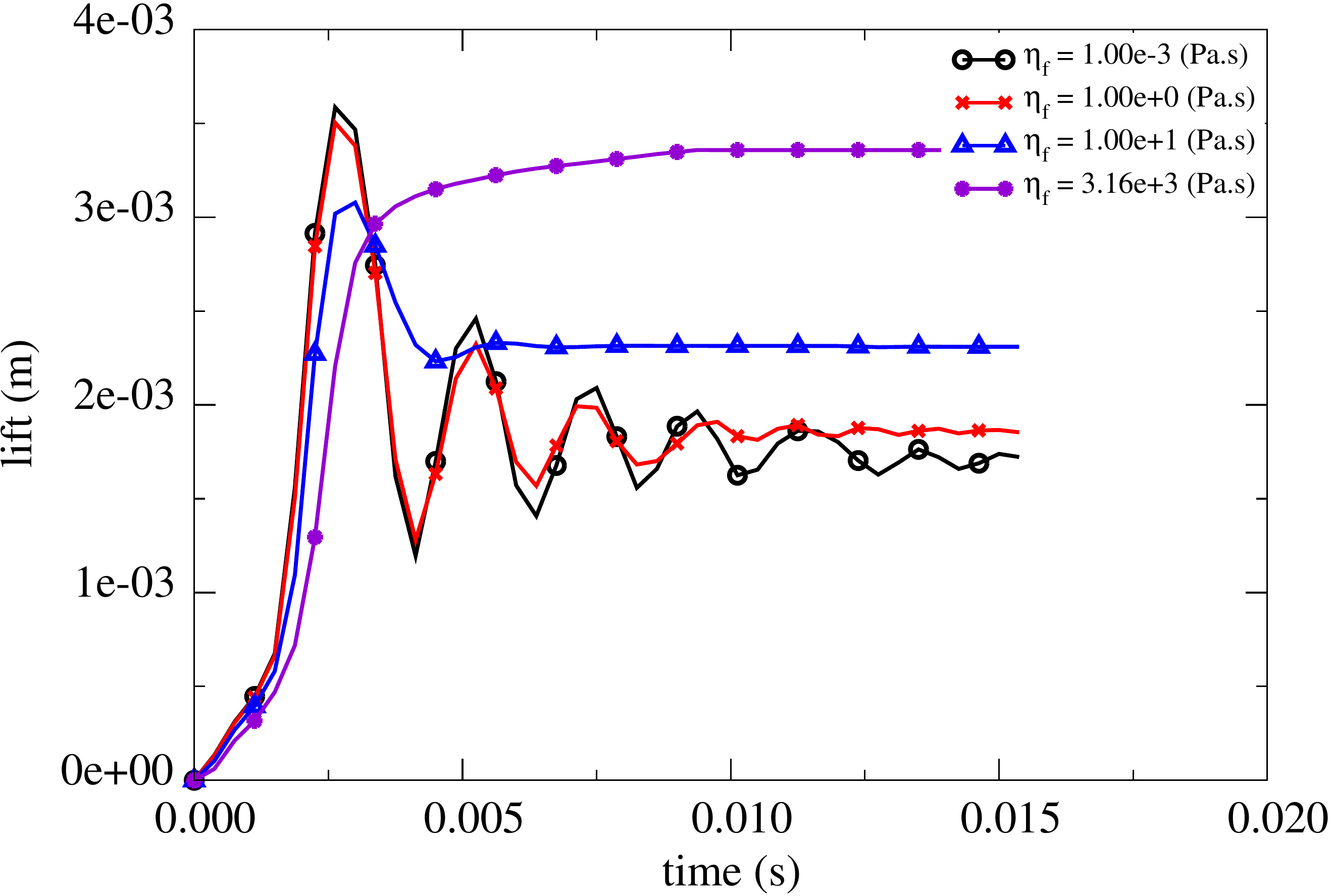}
\includegraphics[width=.5\textwidth]{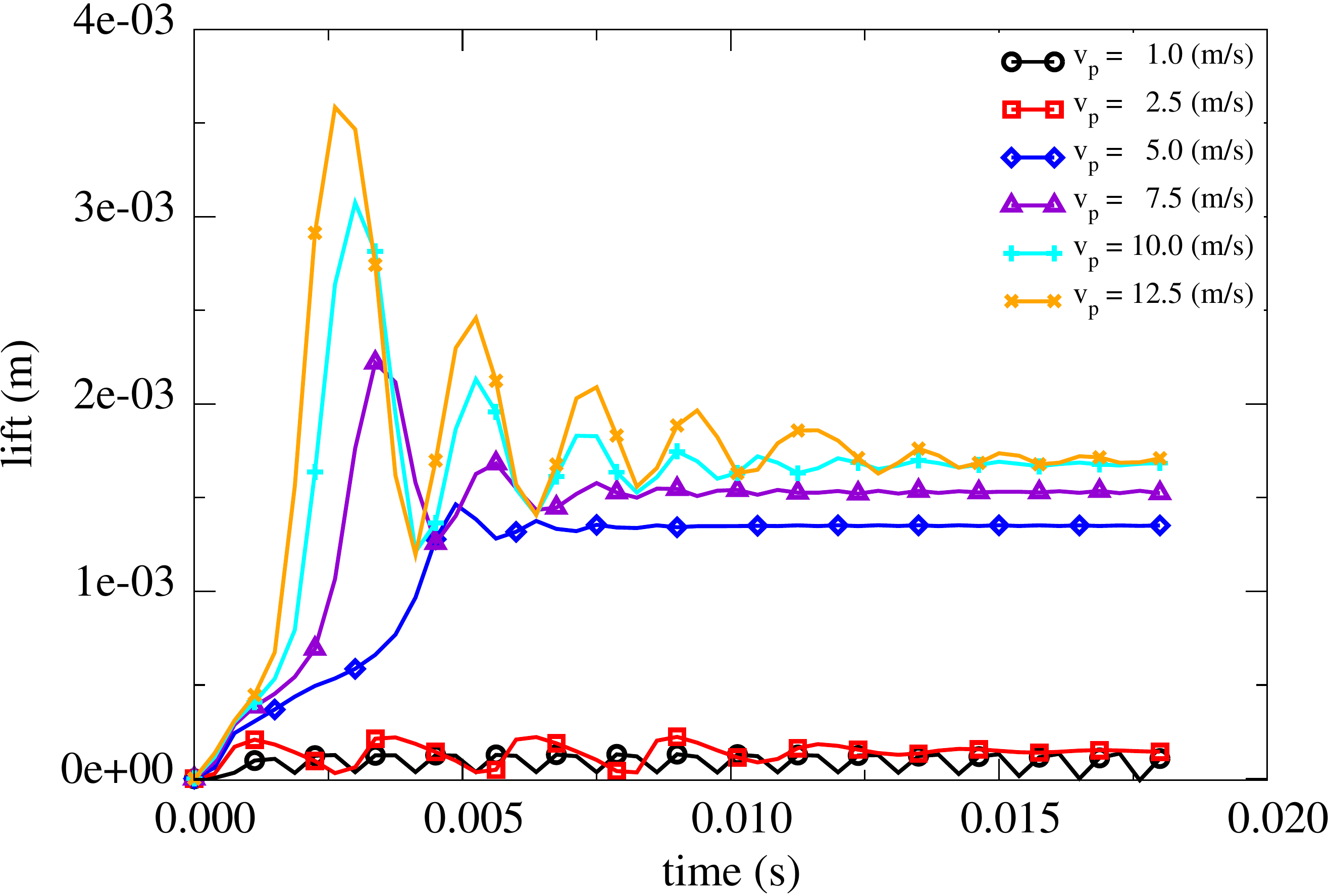}
}
\caption{Valve lift as a function of time for different fluid viscosity (left) and different piston velocity (right) 
(without particles).}
\label{fig:np_opened_lift_eta_v_xxoxxx}
\end{figure*}

Figure~\ref{fig:np_opened_lift_eta_v_xxoxxx} shows the time evolution of the valve lift for different fluid viscosity (left) 
and for different piston velocity $v_p$; both simulations are without particles. The delay at the beginning of the simulation 
is less marked because of the absence of the above valve pressure. The steady lift shows many of the same behaviors as the pressure 
valve, except there can be non-zero $v_p$ (1 m/s and 2.5 m/s) and the valve may not ever open since the open end beneath the valve can prevent the beneath-valve pressure from building up enough to overcome the valve's spring force to open it.

\begin{figure*}[!ht]
\centerline{
\includegraphics[width=.5\textwidth]{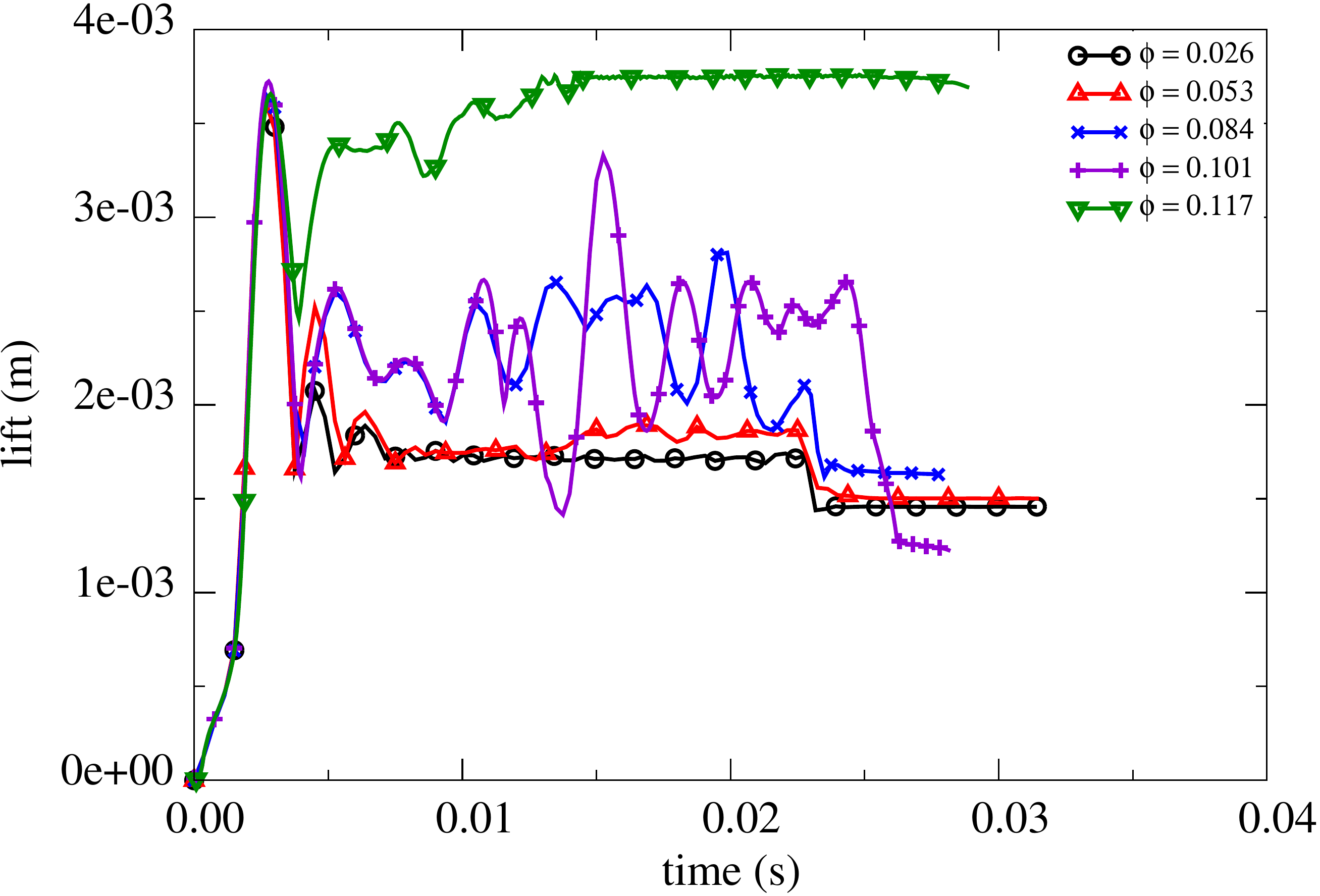}
\includegraphics[width=.5\textwidth]{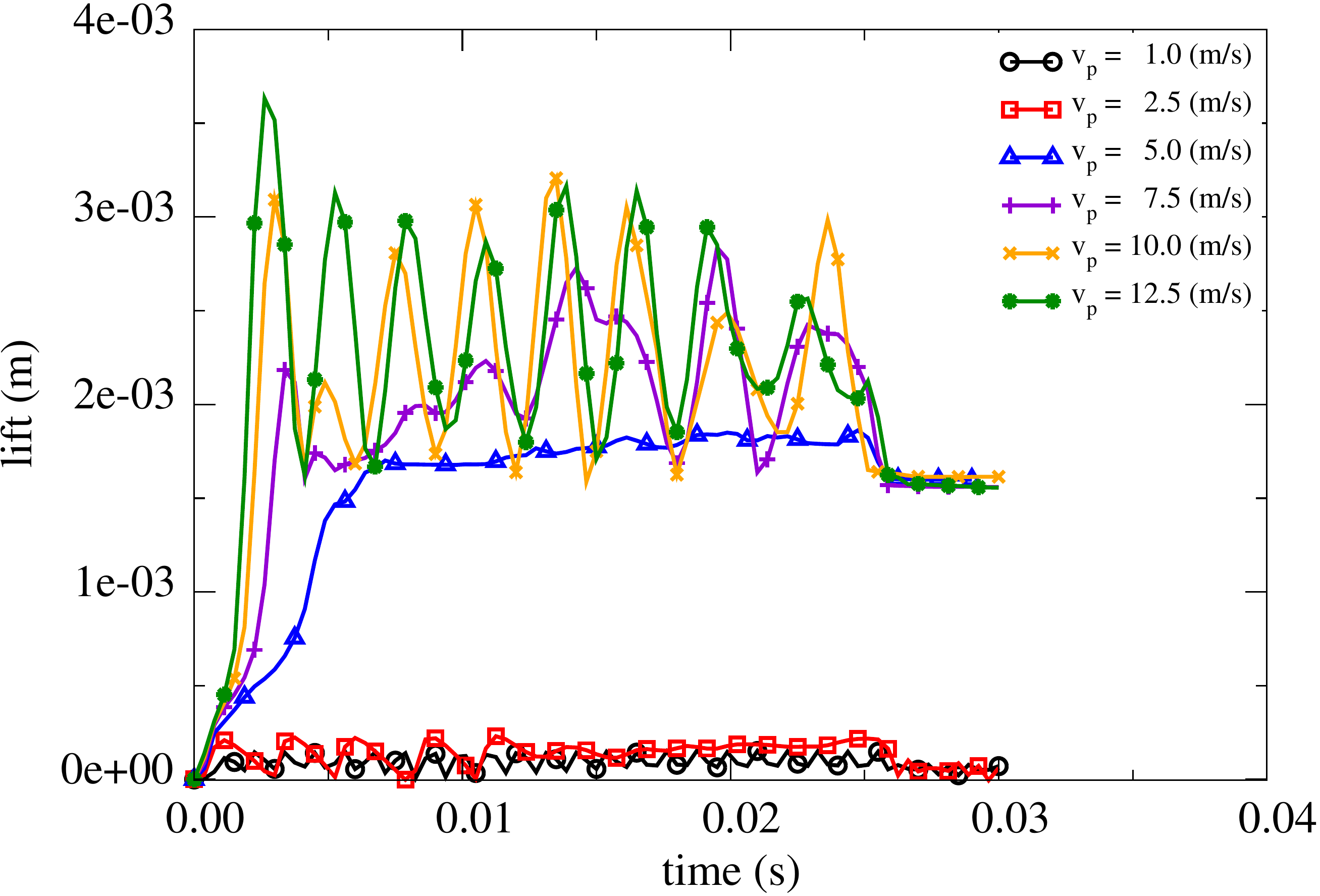}
}
\caption{Valve lift as function of time for different packing fraction (left) and different piston velocity (right).}
\label{fig:opened_yp_lift_v_xxoxxx_sf_xxoxxx}
\end{figure*}

\begin{figure*}[!ht]
\centerline{
\includegraphics[width=.5\textwidth]{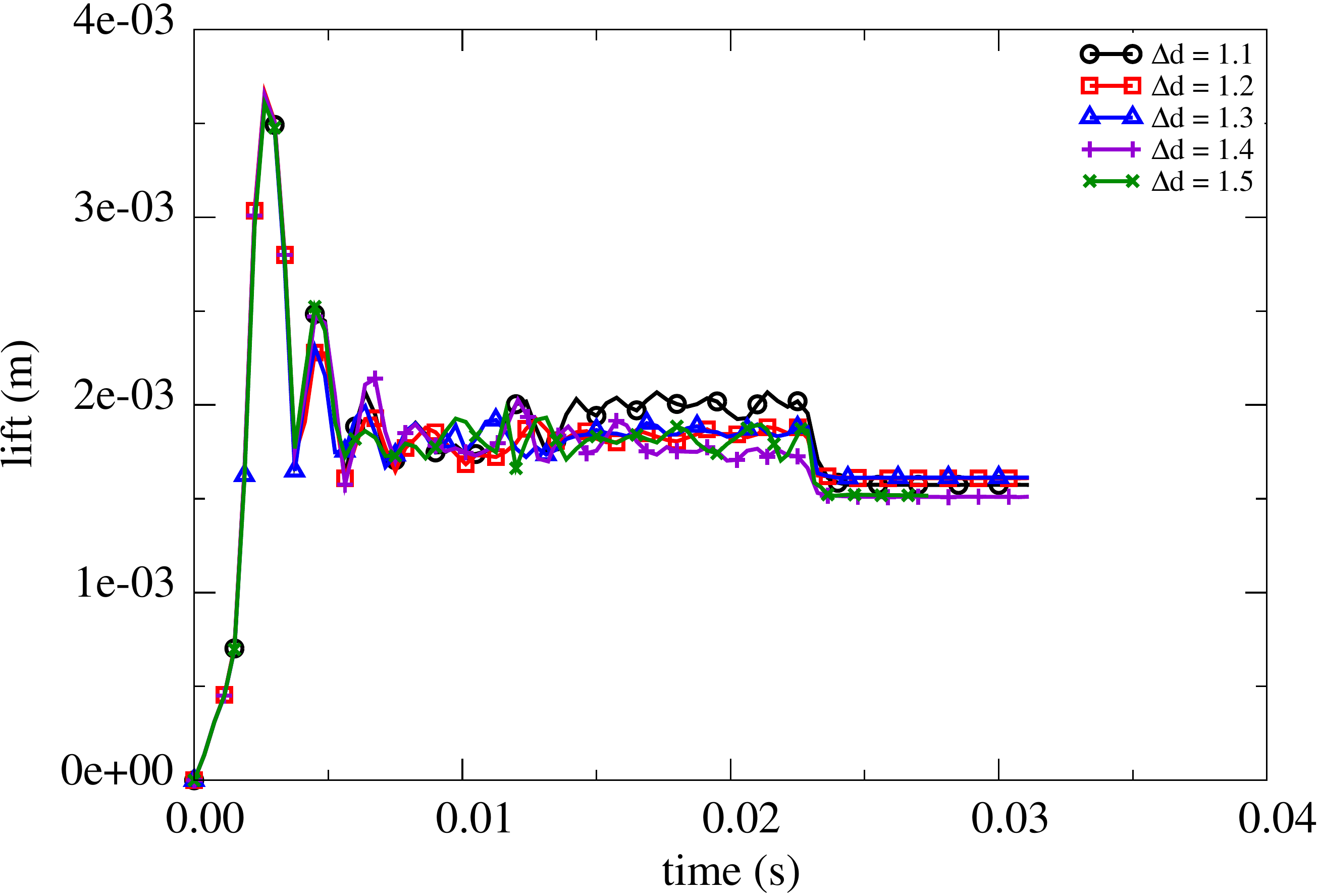}
\includegraphics[width=.5\textwidth]{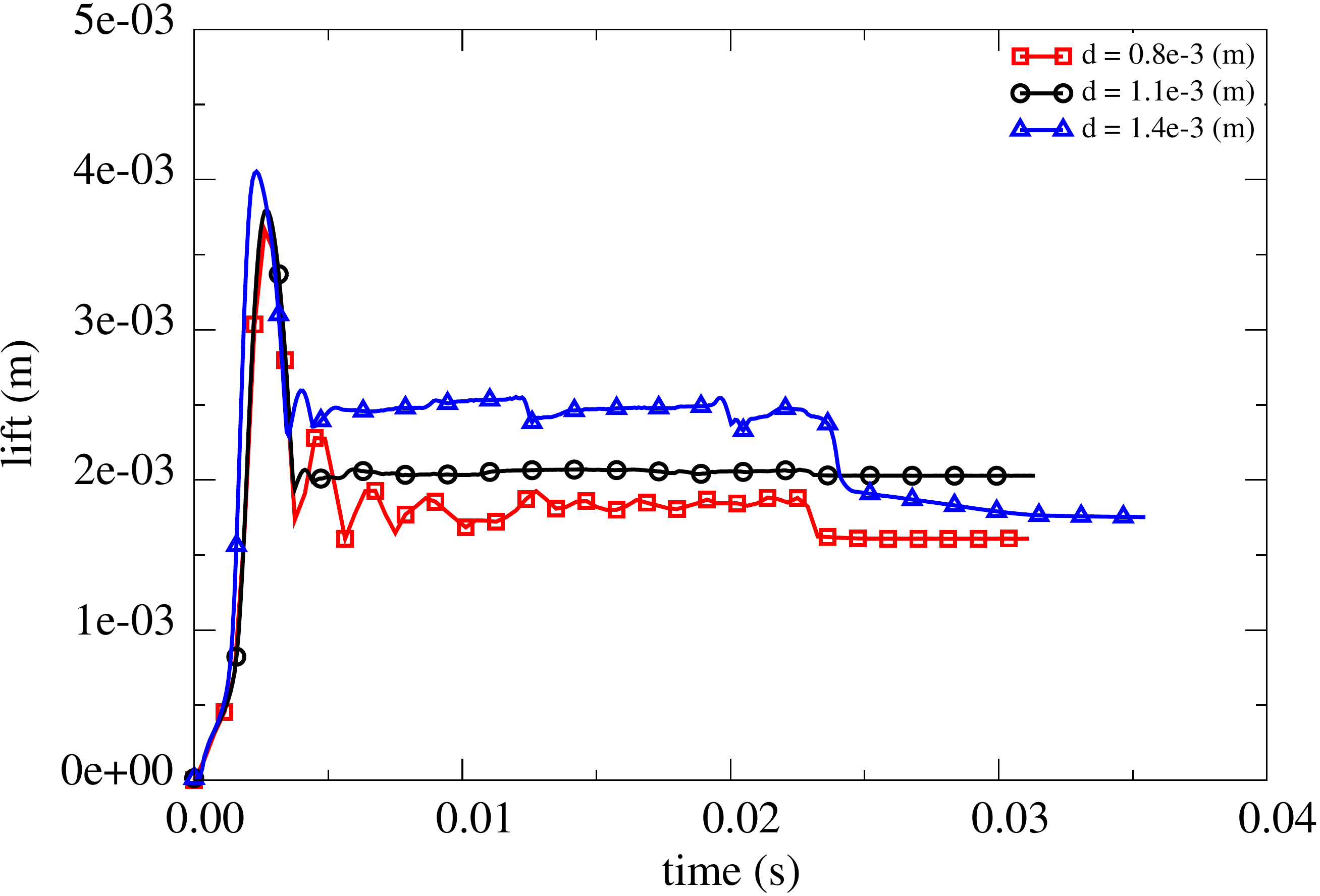}
}
\caption{Valve lift as a function of time for different polydispercity (left) and different mean diameter (right).}
\label{fig:opened_np_lift_d_xxoxxx}
\end{figure*}

The three observed regimes for the pressure valve during the steady lift are also evidenced here as shown in
Fig.~\ref{fig:opened_yp_lift_v_xxoxxx_sf_xxoxxx} for different $\phi$ and $v_p$. The $\varphi _{_l}$ and $\varphi _{_u}$ 
need to be calculated here taking into account the outgoing flux through the modular closure of the safety valve system. 
This then becomes: $\varphi = \varphi _{_{in}} - \varphi _{_{out}}$ where $\varphi _{_{in}}$ is calculated from the input region
and $\varphi _{_{out}}$ is calculated from the flux of outgoing particles through both outlet sections. 
Figure~\ref{fig:opened_np_lift_d_xxoxxx} (left) shows the time evolution of the valve lift for different $\Delta d$ 
where a small deviation in different values of $\Delta d$ is observed. 

\begin{figure}[!ht]
\centerline{
\includegraphics[width=1.\textwidth]{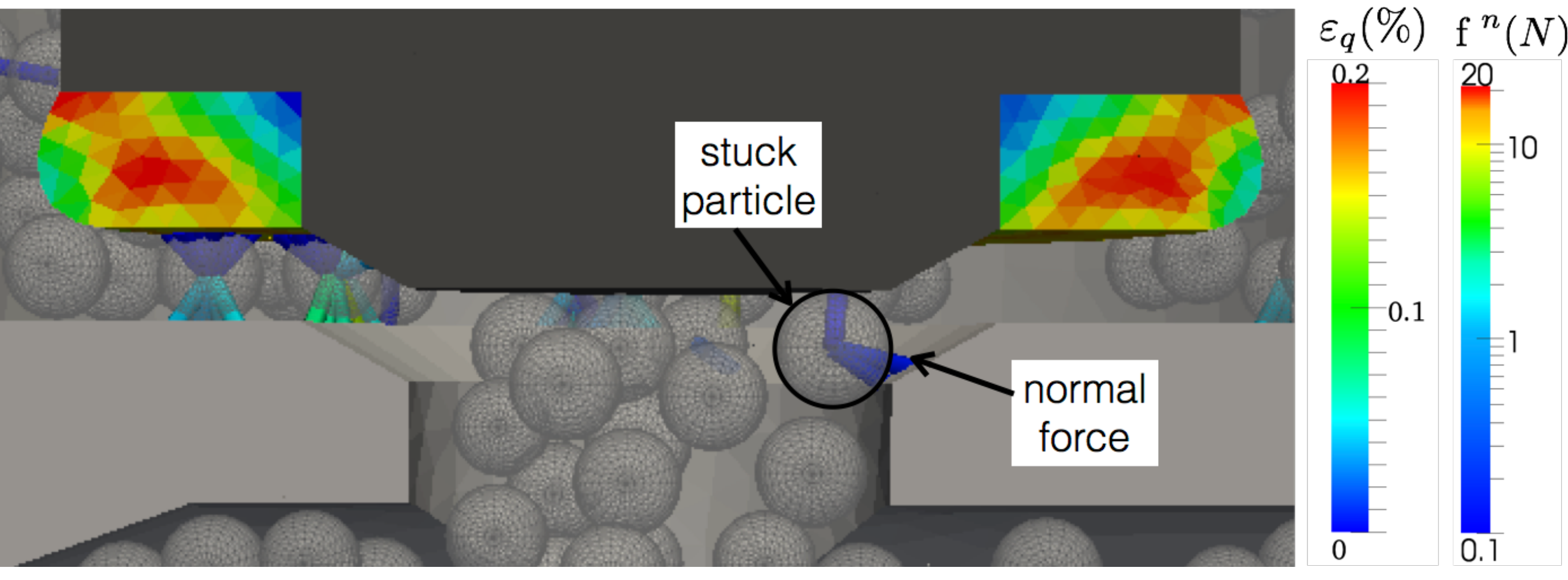}
}
\caption{Final valve lift configuration for $d = 0.0014\ m$ showing (partially) stuck particle in the valve entry and normal forces, $\mbox{f}^{\ n}$, supported by all particles.}
\label{fig:_force_dmax_01o6e03_1_}
\end{figure}

As shown in Fig.~\ref{fig:closing_stuck_particles}, the envelope of the final lift can be predicted by $\text{lift} = d^*/\cos(\theta)$, 
$d_{_{min}}\leqslant d^* \leqslant d_{_{max}}$ however, if the stuck particles are not entirely lodged in the valve channel, this prediction 
is wrong. This is the case in Fig.~\ref{fig:opened_np_lift_d_xxoxxx} (right) for $d = 0.0014\ m$ where the final lift is less than 
the observed lift for $d = 0.0011\ m$. Figure~\ref{fig:_force_dmax_01o6e03_1_} shows that in fact no particles are fully stuck in the
valve channel but there are some stuck in the rubber channel, as indicated by a non-zero normal force $\mbox{f}^{n}$. 
We see in Fig.~\ref{fig:_force_dmax_01o6e03_1_} that even though contact exists between a partially stuck particle and 
the valve channel region, the total force coming from the valve channel is close to $\sim 0.1\ N$ whereas the total force observed in 
the rubber channel is close to $\sim 10\ N$; this means that the main force balancing the valve spring force comes from 
the rubber channel and therefore, the final lift is overpredicted by the previous formula, $\text{lift} = d^*/\cos(\theta)$.

The obtained results for both valves suggest that we may need more investigation with a broader range of particle input parameters to better make a comparison between the safety and the pressure valve geometries.

\section{Conclusion}

In this paper, we have presented a detailed implementation of a 3D DEM-LBM-Rubber coupling in two complex valve geometries. With different 
focused tests, we have validated the implemented methods. The coupling of the three types of materials shows 
a good agreement with our physical predictions. We also have demonstrated the validity of the ZIEB technique, which allowed us 
to run simulations without having to simulate the entire domain

Simulations performed without particles give realistic behaviors. We observe a lubrication effect causing suction that delays 
the opening of the valve after piston motion commences. We find that increasing fluid viscosity increasingly overdamps the valve lift,  reducing or removing the valve oscillations in the quasi-steady regime. We have validated 
a Stokesian pressure drop across the valve channel when the fluid being driven through is sufficiently viscous. 

In the simulations performed with particles in the pressure valve geometry, we found, for the steady lift portion, three qualitative lift behaviors.  
The different regimes appear to be governed by the total particle flux, which combines the imposed piston velocity $v_p$ 
and packing fraction $\phi$ through a flux $\varphi = v_p\ \phi$. The flux variable appears to indicate when the open-valve 
dynamics transition from a steady value of small lift, to an oscillating value that traverses between high and low positions, 
to a steady high lift value. Further investigation may be needed to calibrate the robustness of the $\varphi$ variable in 
determining qualitative valve dynamics. 

The valve closure was also investigated, which occurs when the (virtual) piston motion is stopped. The pressure valve shows a dependency 
of the final lift on particle size and we give a prediction of the lift envelope based on the minimum and maximum particle sizes 
in the polydispersity. We show that if the maximum lift during the open phase does not exceed $d_{_{max}}/\cos(\theta)$, 
the final lift at closure can be less than $d_{_{min}}/\cos(\theta)$ because the particles are not entirely stuck in the valve channel. 

Lastly we demonstrate the robustness of the approach by switching to a safety valve configuration, in which the above-valve 
region is not pressurized and the below-valve region has another exit.  Similar qualitative behaviors are observed as compared 
to the pressure valve, both with and without particles, albeit at different specific values of the piston speed and input particle 
packing fraction.

\begin{acknowledgements}
This work was supported by ARO grant W911 NF-15-1-0598 and Schlumberger Technology Corporation. PM and KK would like to thank J.-Y. Delenne (INRA, UMR IATE Montpellier) for his helpful and useful discussions 
on DEM-LBM coupling and Sachith Dunatunga for his help in streamlining the numerics. Conflict of Interest: The authors declare that they have no conflict of interest.
\end{acknowledgements}

\newpage

\appendix

\section{Safety valve, pressure valve, and their dimensions}
\label{app:dimension}

In this appendix, we give the safety and pressure valve configuration and the dimensions. Units are millimeters $mm$ 
and degrees $^o$.

\begin{figure}[!ht]
\centerline{
\includegraphics[width=1.\textwidth]{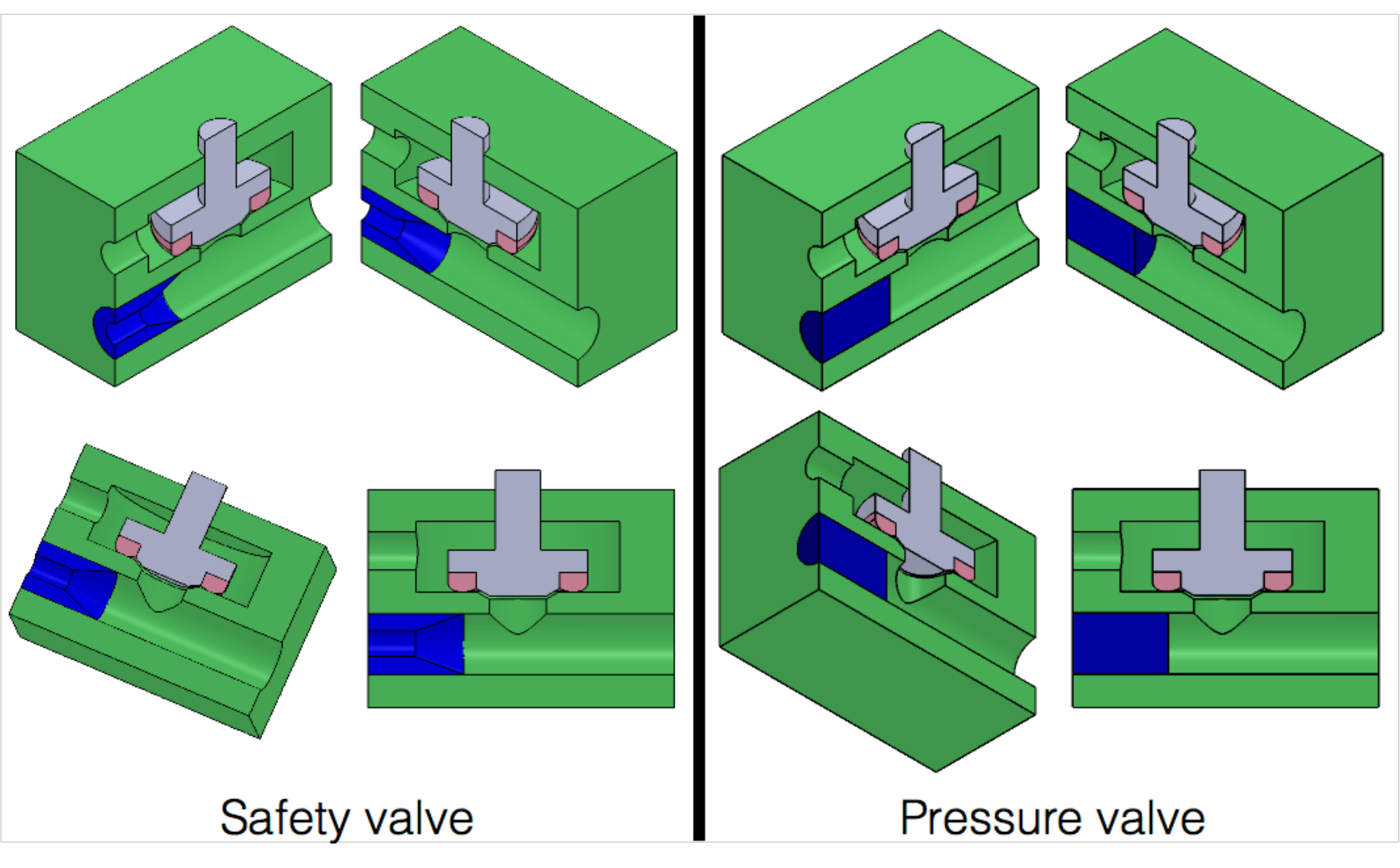}
}
\caption{Safety and pressure valve configuration.}
\label{fig:SafetyPressure}
\end{figure}

\begin{figure}[!ht]
\centerline{
\includegraphics[width=1.\textwidth]{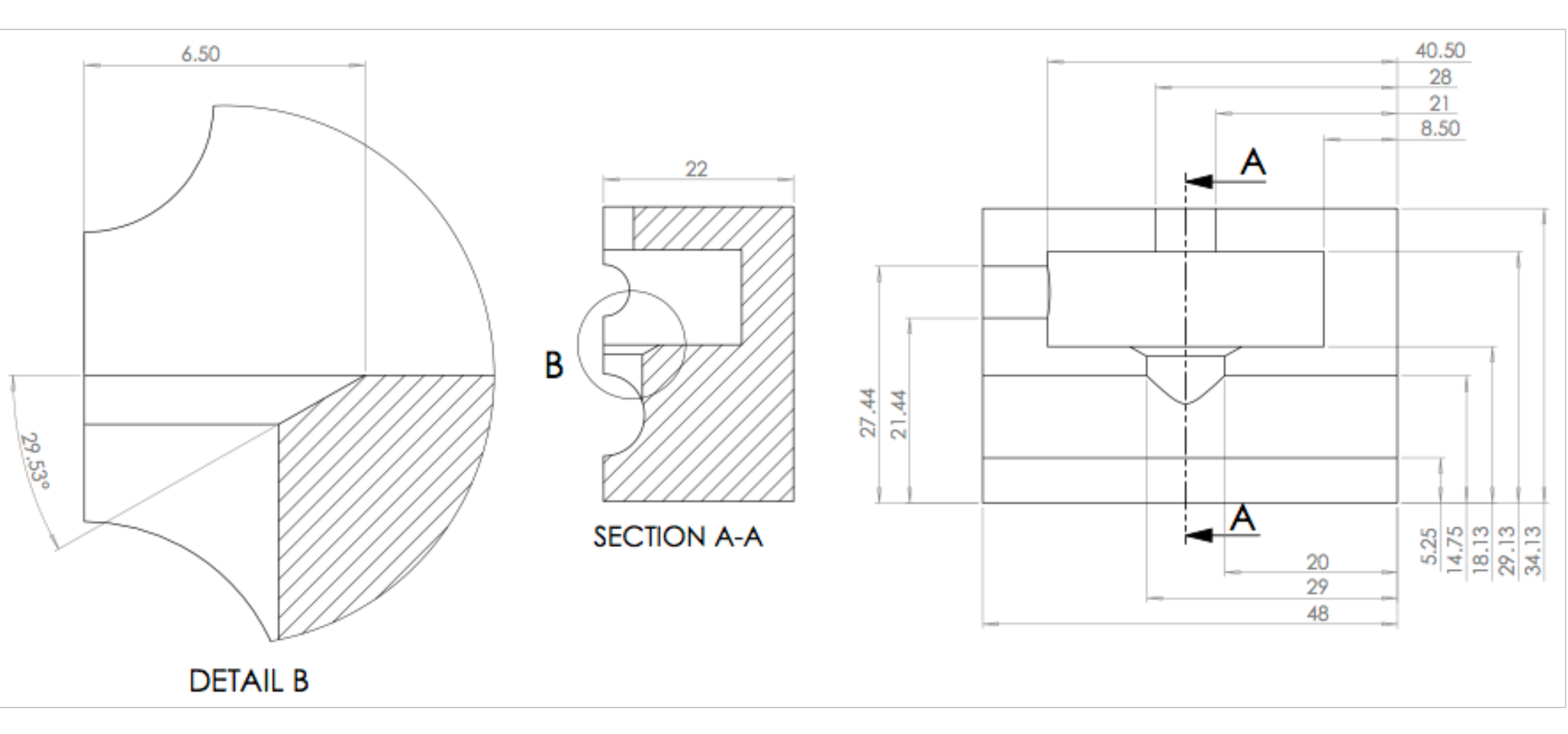}
}
\caption{Frame dimension.}
\label{fig:Seat}
\end{figure}

\begin{figure}[!ht]
\centerline{
\includegraphics[width=1.\textwidth]{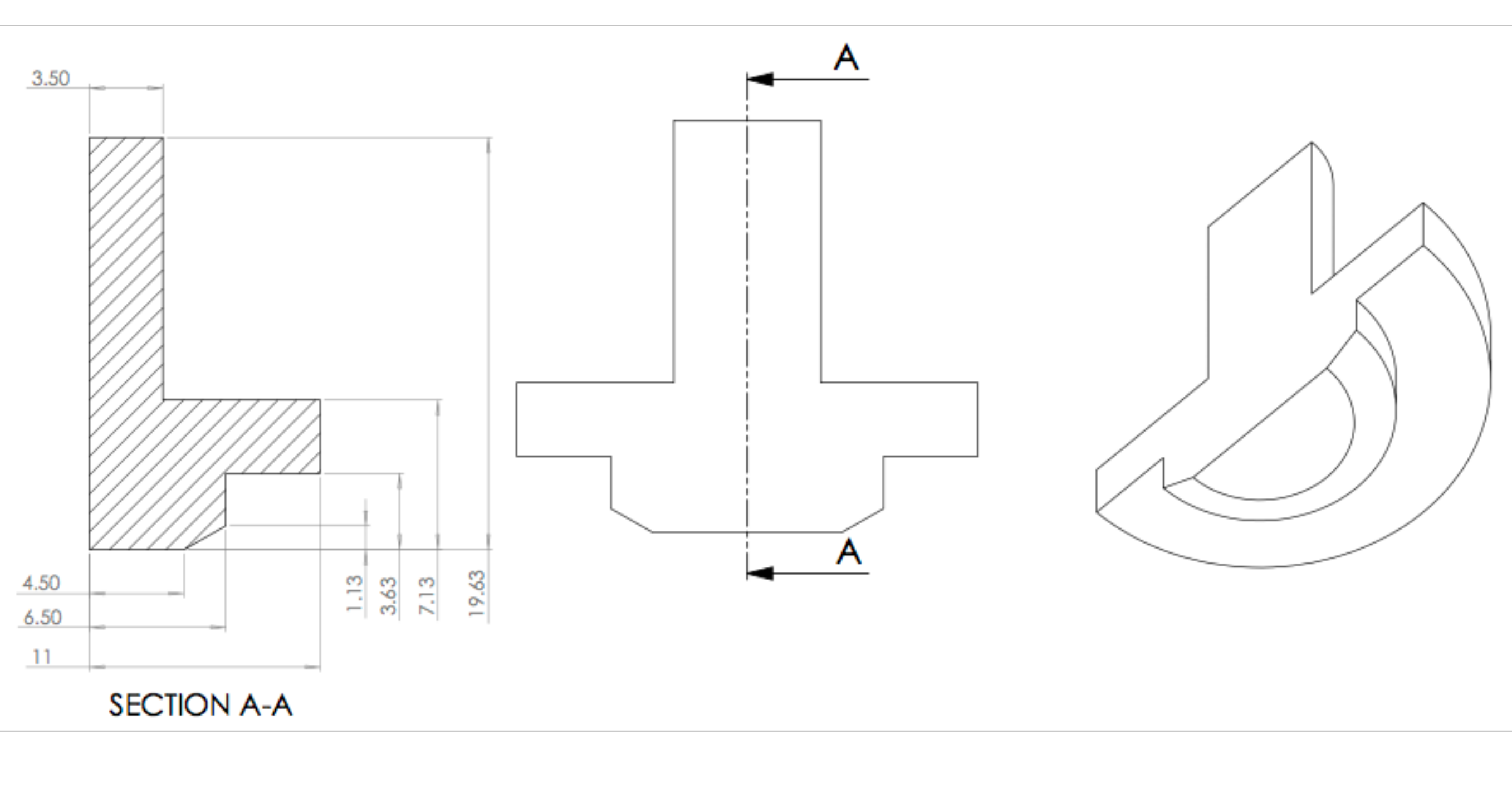}
}
\caption{Valve dimension.}
\label{fig:Valve}
\end{figure}

\begin{figure}[!ht]
\centerline{
\includegraphics[width=1.\textwidth]{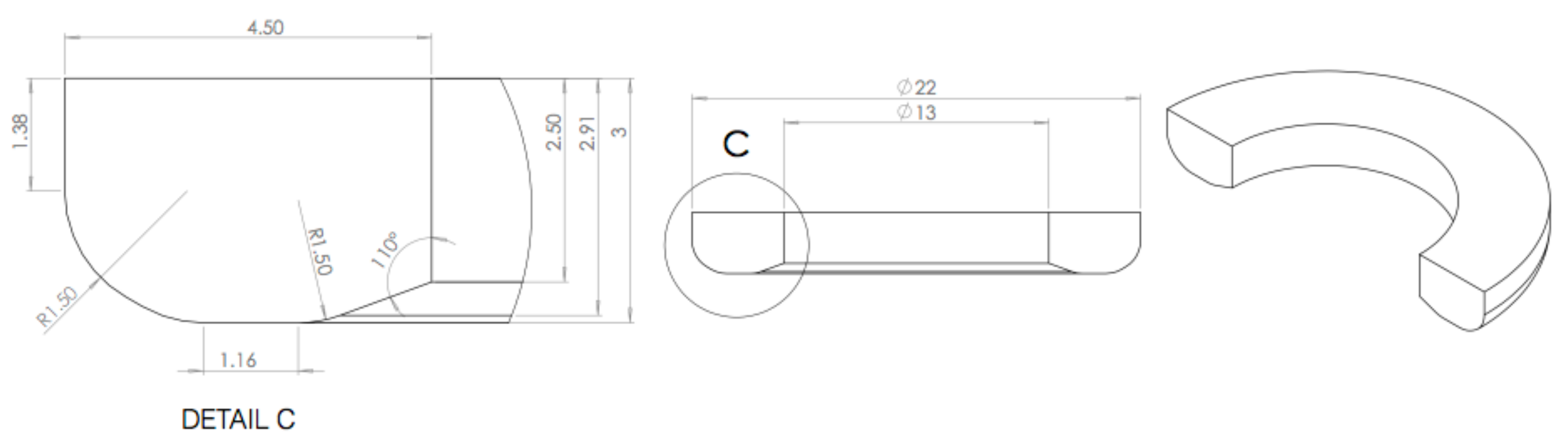}
}
\caption{Rubber dimension.}
\label{fig:Rubber}
\end{figure}

\begin{figure}[!ht]
\centerline{
\includegraphics[width=.65\textwidth]{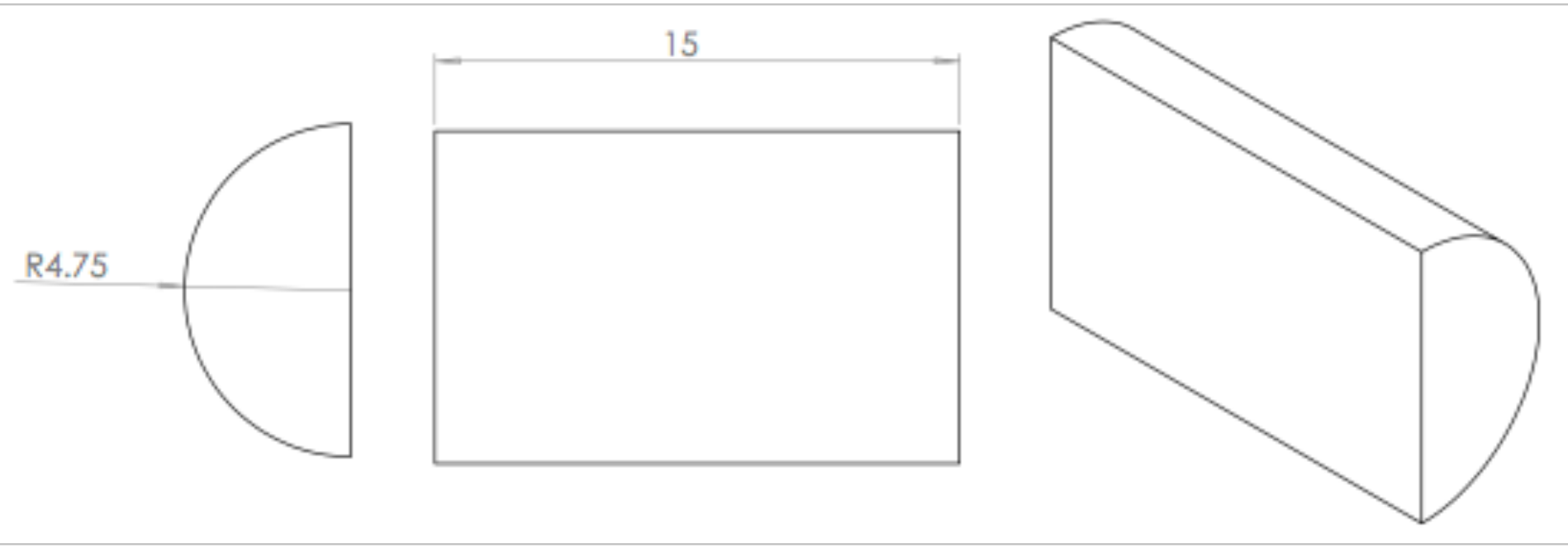}
}
\caption{Pressure Closure dimension.}
\label{fig:PressureClosure}
\end{figure}

\begin{figure}[!ht]
\centerline{
\includegraphics[width=.55\textwidth]{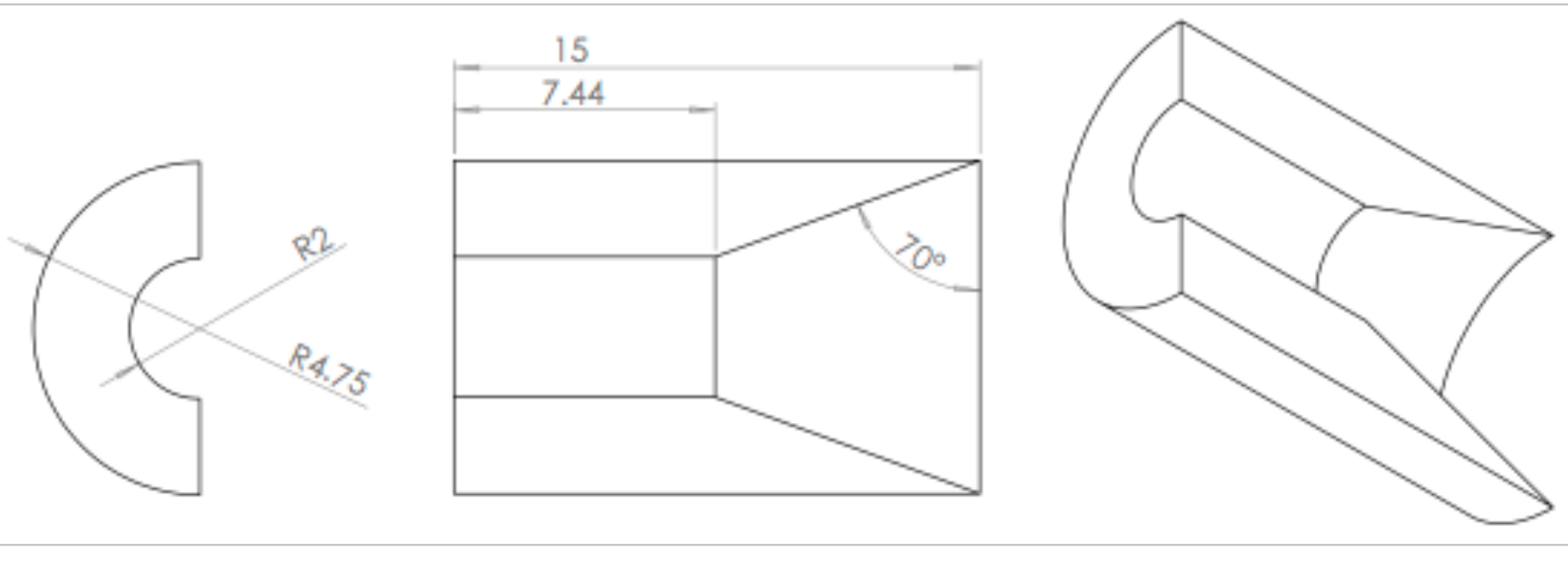}
}
\caption{Safety Closure dimension.}
\label{fig:SafetyClosure}
\end{figure}

\newpage

\bibliographystyle{spphys}

\end{document}